\documentclass[journal=mamobx,manuscript=article]{achemso}

\usepackage[version=3]{mhchem} % Formula subscripts using \ce{}
\usepackage[T1]{fontenc}
\usepackage[utf8]{inputenc}
\usepackage{lmodern}
\usepackage[british]{babel}
\usepackage[numbers,sort&compress]{natbib}
\usepackage{amsmath,amsthm,amssymb, graphicx, multicol, array}
\usepackage{cleveref}
\usepackage{graphicx}
\usepackage{natbib}				%for citations in Harvard style
\usepackage{enumerate}			%to be able to change the enumerate symbols
\usepackage{amsfonts}			% for special set symbols
\usepackage{listings}
\usepackage{a4wide}
\usepackage{afterpage}
\usepackage{color}
\usepackage{comment}
\usepackage{multirow}
\usepackage{xr}
\usepackage{xcolor}
\usepackage[normalem]{ulem}
\makeatletter

\newcommand*{\citen}{}% generate error, if `\citen` is already in use
\DeclareRobustCommand*{\citen}[1]{%
  \begingroup
    \romannumeral-`\x % remove space at the beginning of \setcitestyle
    \setcitestyle{numbers}%
    \cite{#1}%
  \endgroup
}

\newcommand*{\addFileDependency}[1]{% argument=file name and extension
  \typeout{(#1)}
  \@addtofilelist{#1}
  \IfFileExists{#1}{}{\typeout{No file #1.}}
}
\makeatother

\newcommand*{\myexternaldocument}[1]{%
    \externaldocument{#1}%
    \addFileDependency{#1.tex}%
    \addFileDependency{#1.aux}%
}

\myexternaldocument{SI}

\SectionNumbersOn
\newcommand{\add}[1]{\textcolor{black}{#1}}
\newcommand{\maw}[1]{\textcolor{black}{#1}}

\author{Shannon Zhang}
\author{Michael A. Webb}
\email{mawebb@princeton.edu}
\affiliation[Princeton University]
{Department of Chemical and Biological Engineering, Princeton University, Princeton, NJ, 08544 USA}

\title{Asymmetric Effects Underlying Dynamic Heterogeneity in Miscible Blends of Poly(methyl methacrylate) with Poly(ethylene oxide)}

\begin{document}

\begin{abstract} 
The emergence of spatially variable local dynamics, or dynamic heterogeneity, is common in multicomponent polymer systems.
Although often attributed to differences in the intrinsic dynamics of each component, the molecular origin of their coupling and its dependencies remain unclear.
Here, we use molecular dynamics simulations of polyethylene oxide (PEO)/polymethyl methacrylate (PMMA) blends, \maw{across the full range of compositions and multiple thermal regimes, to characterize local fluctuations and sub-chain relaxations for both PEO and PMMA.}
\maw{By constructing probability distributions of local composition and computing entropic measures, we connect nanoscale heterogeneity to differences in mobility between PEO and PMMA, extending beyond mean‐field treatments.}
\maw{While PMMA segmental fluctuations in blends broadly align with $T_\text{g}$‐equivalent neat PMMA systems, PEO exhibits enhanced mobility correlated with increased free volume and broader, more diverse local compositions upon blending.}
Rouse‐mode analysis, used to probe relaxation dynamics over different length scales, shows that \maw{PEO relaxation approaches neat‐like behavior in PEO‐rich domains, whereas PMMA relaxation accelerates uniformly across all mode numbers.
Given the local mobility enhancement of PMMA by PEO, this uniform shift suggests a nanoscale facilitation process that extends PEO’s influence beyond its immediate environment.}
\maw{These findings link the statistics of local compositional heterogeneity to dynamic asymmetry across length scales, provide physical insight into the behavior of this archetypal blend system, and establish a framework for analyzing dynamic coupling in others.}
\end{abstract}

\section{Introduction}
Dynamic heterogeneity refers to spatial and temporal variations in the mobility and relaxation behavior of molecules or particles within a material. In polymer blends, such heterogeneity commonly arises due to differences in the intrinsic segmental dynamics of each component. When blended, interactions between polymers with distinct mobilities create coupled dynamical responses that manifest as perturbations across multiple spatiotemporal scales.\cite{colby_breakdown_1989, roland_dynamical_1991, kamath_dynamic_2003, roland_dynamic_2006, sharma2017, zhang_spatially_2020} Consequently, the emergence of dynamic heterogeneity can strongly influence key material properties such as the glass transition temperature ($T_\text{g}$),\cite{bennemann1999, phan2018, baker2022, ghanekarade2023} viscoelastic behavior,\cite{fredrickson1996, semenov_associating_1995, tanaka_rheological_2000, mogurampelly2016, zhang_role_2022} and ion conductivity.\cite{fenton1973, berthier1983, borodin2006, mogurampelly2016, webb2018, chu2020, deng2021} Therefore, elucidating the molecular origins and characteristic length scales of dynamic heterogeneity is of fundamental scientific interest and may inform the design of polymer blends for targeted applications.

Dynamic heterogeneity is known to be significant in blends of polyethylene oxide (PEO) and polymethyl methacrylate (PMMA). While these polymers are miscible,\cite{shah2023, mu2008, schantz1997, cimmino1989} their glass transition temperatures ($T_\text{g}$) differ by approximately 180 K.\cite{lartigue1997, silva_nanoheterogeneities_2000, R:2022_Habasaki_Atomistic} This dissimilarity gives rise to pronounced differences in segmental dynamics when the polymers are blended, with relaxation times differing by up to 12 orders of magnitude.\cite{lartigue1997, lutz2003} 
Such contrast in mobilities has implications for applications like solid polymer electrolytes,\cite{ghelichi2013, choudhary2018, lim2018, sharon2022, bakar2022} where ionic conductivity is sensitive to local segmental fluctuations.\cite{R:2010_Diddens_Understanding,R:2021_Deng_Role,R:2015_Webb_Chemically_Macromolecules} 
More broadly, PEO/PMMA blends serve as model systems for investigating dynamic coupling between components at the nanoscale, with signatures of heterogeneity accessible in both simulations and experiments.\cite{lutz2003, niedzwiedz2007, garcia_sakai_dynamics_2008, brodeck2010}

Numerous experimental studies have examined the dynamics of PEO and PMMA in blends under varying conditions. Quasi-elastic neutron scattering (QENS) on blends containing up to 30 wt{}\% PEO has shown that the segmental mobility of PMMA, on length scales up to 11~{\AA}, is primarily governed by the temperature difference between the system and the glass transition temperature of the blend, (i.e., $T - T_\text{g}$).\cite{sakai2004, liu2006, maranas2007} 
In contrast, PEO dynamics are strongly influenced by interactions with PMMA, exhibiting distinct behavior at short (<1 nm) and longer (>1 nm) length scales.\cite{genix2005, niedzwiedz2007}
At short length scales, QENS,\cite{sakai2005, garcia_sakai_dynamics_2008} nuclear magnetic resonance (NMR),\cite{lartigue1997, lutz2003} and neutron spin echo (NSE) spectroscopy\cite{farago_collective_2005} reveal narrowing distributions of segmental relaxation times across a range of compositions and temperatures. These effects are attributed to self-concentration phenomena\cite{sakai2005} and local confinement by the stiffer PMMA matrix, with a dependence on local free volume.\cite{lartigue1997, lutz2003, garcia_sakai_dynamics_2008}

On longer length scales, PEO dynamics have been investigated using infrared dichroism and birefringence,\cite{zawada_component_1992} QENS,\cite{genix2005, garcia_sakai_dynamics_2008} and NSE.\cite{niedzwiedz2007} These measurements consistently indicate a pronounced slowdown of PEO segmental motion in blends with low PEO content.\cite{zawada_component_1992, genix2005, niedzwiedz2007, garcia_sakai_dynamics_2008} Unlike the behavior observed at shorter scales, the dynamics at these larger scales are characterized by a broad distribution of relaxation times,\cite{garcia_sakai_dynamics_2008} which is largely attributed to long-range concentration fluctuations.\cite{zawada_component_1992}
Collectively, these observations highlight the presence of dynamic heterogeneity across multiple length scales in PEO/PMMA blends and suggest several underlying mechanisms; however, direct inference of the molecular-level phenomena remains elusive through experimental characterization alone.

To complement experimental observations, several theoretical models have been developed to explain dynamic heterogeneity in PEO/PMMA blends. The Lodge-McLeish (LM) model, also referred to as the chain connectivity model, quantifies the influence of local self-concentration on polymer dynamics.\cite{lodge2000} 
The LM model predicts the segmental dynamics of each polymer component using the concentration within a cooperative volume centered on a monomer. The relevant length scale used to determine the size of the cooperative volume is the Kuhn length of each polymer component. 
It successfully predicts PEO relaxation times in blends containing 10-30 wt{}\% PEO taken from QENS measurements for large spatial scales ($q$=0.69~\r{A}$^{-1}$, approximately 18~\r{A}) but breaks down for smaller spatial scales ($q$=1.3~\r{A}$^{-1}$, approximately 10~\r{A}).\cite{garcia_sakai_dynamics_2008} When the self-concentration is allowed to vary as a fitting parameter, the LM model reasonably fits PEO relaxation times measured by QENS\cite{garcia_sakai_dynamics_2008} and NMR,\cite{lutz2003} as well as terminal PMMA relaxation times.\cite{he_segmental_2003} However, allowing the self-concentration to vary obfuscates the theoretical foundations of and insights from the model. 
\maw{Extensions of the LM model incorporate concentration fluctuations to quantitatively predict relaxation times in polymer blends.\cite{kumar2007, kant2003, shenogin2007, colby2005} These studies demonstrate that concentration fluctuations are necessary to accurately capture both the peak and width of relaxation time spectra, particularly at temperatures below the blend $T_\text{g}$. The correlation volume within which concentration fluctuations predict relaxation times is found to be on the order of the Kuhn length, supporting the assumptions of the LM model.}

Another model uses mesoscale concentration fluctuations with length scales between 1~nm and 1~$\mu$m via the generalized entropy theory of glass formation, the lattice cluster theory of blend thermodynamics, and the Kirkwood-Buff theory of concentration fluctuations to predict structural relaxation times of dynamically asymmetric miscible polymers in blends.\cite{dudowicz_concentration_2014, dudowicz_two_2014} It can qualitatively fit the separated relaxation times of PEO and PMMA in blends but cannot account for the chemical specificity required for fitting quantitative behaviors.\cite{dudowicz_concentration_2014, dudowicz_two_2014} 

Finally, a coupling model\cite{moreno_tests_2007, ngai_interchain_2011, ngai_unified_2013, colmenero_comment_2013, ngai_response_2013} and the generalized Langevin equation framework\cite{colmenero_generalized_2013, colmenero_reply_2013, ngai_comment_2013} have been developed to describe a crossover time that separates single-chain PEO dynamics at short length scales from many-chain coupled PEO dynamics at long length scales. This framework has been used to explain unexpected phenomena and properties that arise specifically in dynamically asymmetric miscible polymer blends, such as different $T_\text{g}$ values for each component and the breakdown of the time-temperature superposition.\cite{ngai_unified_2013} Additionally, the coupling model closely captures PEO segmental dynamics for experimental QENS measurements using a momentum transfer $q$ value between 1 and 2~\AA$^{-1}$ in blends with 10-30 wt{}\% PEO.\cite{garcia_sakai_dynamics_2008} 
These theories qualitatively capture and rationalize trends in blended PEO dynamics up to mesoscale length scales.
However, like with experimental observations,  they do not conclusively illustrate the molecular-level phenomena that contribute to dynamic heterogeneity. 

% Lodge-McLeish model (cite Lodge 2000): cooperative volume centered on a monomer of the polymer of interest within which the concentration of the polymer is enhanced over the bulk concentration due to chain connectivity. Estimates an effective Tg experienced by each component in the blend based on a volume that is proportional to l_k^3. From original LML paper, it has qualitative agreement with effective Tg behavior of some miscible blends. 
% Lutz 2003: attempts to fit 10-30% PEO blend system PEO segmental correlation times taken from NMR to LML model by varying the self-concentration volume fraction (of PEO). While it has reasonable fit to PEO segmental dynamics between ~300K and ~440K, the LML model's calculated self-concentration (0.15) does not match the fitted self-concentration (0.57)
% He 2003: Fits LML model to terminal relaxation times from Zawada (simultaneous measurement of infrared dichroism and birefringence; 20-60% PMMA at T >> Tg). Fails to quantitatively fit PMMA terminal dynamics. Colby successfully fits PMMA terminal dynamics in a blend with more PMMA (80% PMMA) but again, fitted phi_self neq predicted
% Sakai 2008: 10-30% PEO in blends, 308-440K. LM model is predictive for QENS relaxation times taken in spatial range of spherical diameter 18A. Not predictive but can be fit by changing phi_self for smaller spatial scales. Thus, no single constant length scale can be used to control the influence of mixing on dynamics.

Molecular dynamics (MD) simulations have helped to elucidate the microscopic origins of dynamic heterogeneity in polymer blends under certain conditions. Analysis of all-atom simulations with blend compositions of 10-30 wt{}\% PEO suggest the presence of multiple populations of PEO dynamics in blends which may be caused by confinement effects of PEO in a rigid matrix of PMMA. This is deduced from van Hove self-correlation functions of hydrogen atoms from PEO, which show a double-peak structure at 400 K with an unmoving second peak\cite{genix2005} and PEO relaxation times that are broadly distributed.\cite{chen_molecular_2009} 
Monitoring mean-square deviations of atoms in PEO and PMMA also suggests highly disparate local cage sizes, as the magnitude of distance traveled in the ballistic regime is much higher for PEO than PMMA.\cite{genix2005, chen_molecular_2009} 
Additionally, PEO segmental relaxation times have been found to be more stretched upon blending than those of PMMA.\cite{chen_molecular_2009} 
Rouse analyses based on simulations of 20 wt{}\% PEO have illustrated the non-Gaussianity of the distribution of PEO atomic displacements in blends\cite{brodeck_single_2012}, suggesting that local PEO segmental motion for wavelengths on the order of the size of a monomer is not strongly affected by blending but is more strongly affected at larger wavelengths.\cite{brodeck2010} 
These results provide molecular-level insights into dynamic heterogeneity in PEO/PMMA blends, albeit using a varied set of compositions, temperatures, and characterization methods.

In this work, we use atomistic MD simulations to systematically characterize how dynamic heterogeneity manifests at the nanoscale in PEO/PMMA blends. 
\maw{A central objective is to examine how the strength of dynamic coupling between the two components depends on temperature and blend composition. Although prior studies have explored the dynamic behavior of PEO/PMMA blends, inconsistencies in system specifications (e.g., blend composition, molecular weight, and force field) and thermodynamic conditions have made it difficult to identify clear trends. To address this, we comprehensively investigate behavior over the full range of blend compositions at temperatures above, between, and below the simulated apparent apparent $T_\text{g}$ of both polymers. }
Dynamics are characterized by means of both local segmental fluctuations as well as segmental relaxation timescales. 
We find that blending induces asymmetric and composition-dependent changes in both local and collective dynamics, governed by temperature and local composition. 
\add{In particular, the free volume and the diversity in local composition surrounding polymer units are found to correlate strongly with differences in segmental mobility between PEO and PMMA. }
These findings help to clarify what PEO-PMMA interactions influence dynamic heterogeneity and motivate further investigation into their generality and implications for macroscopic properties and functional performance.

\section{Methods}
\subsection{Simulation}
\subsubsection{General simulation protocols} All MD simulations are performed using LAMMPS (ver 29, Sep 2021).\cite{lammps} Systems are modeled using the all-atom optimized potentials for liquid simulations (OPLS-AA)\cite{oplsaa} force field. Real-space non-bonded interactions are truncated at 12~\AA. Long-range electrostatics are handled using the particle-particle-particle-mesh Ewald summation method with a $10^{-4}$ convergence accuracy.\cite{pppmbook, pppmarticle} Equations of motion are integrated using a velocity-Verlet integration scheme and a 1 fs timestep. Periodic boundary conditions are used in all three dimensions. Temperature and pressure are controlled using a Nos\'e-Hoover thermostat and barostat with damping constants of 100 fs and 2000 fs, respectively. 

\subsubsection{Polymer chain generation} \label{sec:genmethod}
Each PEO chain consists of 75 monomers, and each PMMA chain consists of 33 monomers.
\add{Here and throughout the text, the term monomer is used to refer to a single constitutional repeat unit of the polymer.} 
As a result, chains for both polymers possess molecular weights of approximately 3,300 g/mol.
\add{Syndiotactic PMMA was used in all simulations. This choice avoids the need to stochastically generate atactic sequences. Moreover, syndiotactic and atactic PMMA have been previously noted to exhibit similar $T_\text{g}$,\cite{yoshida1982} which may suggest similar dynamical characteristics.
} 

\add{Initial chain configurations are generated to approximate the expected relationship for the mean-squared end-to-end distance in that $\langle R^2 \rangle = L \ell_k$, where $\ell_k$ is the Kuhn length of the polymer and $L$ is the statistical contour length. In practice, rather than $L$ we use $L_\text{ext}\approx\sum_il_i$, where $L_\text{ext}$ is the length of a fully extended polymer chain approximated by the summation of individual bond lengths $l_i$. The individual $l_i$ are obtained from repeat units, following geometry optimization in Avogadro 1.2.0 with the steepest descent algorithm and the UFF force field;\cite{avogadro,avogadro-site} this yields $L$ values of 326.1~\AA\ for PEO and 102.5~\AA\ for PMMA. While this approach will initially yield somewhat more extended conformations than targeted as $L_\text{ext} > L$, this approximation is used only for initialization, and subsequent equilibration \maw{mitigates initial} bias in chain dimensions.} 
The Kuhn lengths used are 8.2~\AA~for PEO and 13.8~\AA\ for PMMA.\cite{hiemenzlodge}
Chains are constructed by sequentially adding monomers, with each new monomer positioned based on a randomly sampled dihedral angle. For PEO, dihedral angles are drawn from a uniform distribution between $-0.65$ and $0.65$ radians; for PMMA, the range is $-0.4$ to $0.4$ radians. For each system described below, this process is repeated until the required number of independent chain configurations within a threshold of 5~\AA\ of the target $\langle R^2 \rangle$ is obtained. These configurations are used as described in Section {\ref{sec:sysprep}}. \maw{The contour length of each polymer is estimated by using geometry optimization in Avogadro 1.2.0 to measure the distance between two monomers on a fully extended chain and multiplying by the total number of monomers. This yields estimated contour lengths of 273~\AA~for PEO and 92~\AA~for PMMA. Dividing these values by the respective Kuhn lengths gives 33.3 and 6.7 Kuhn steps for PEO and PMMA chains, respectively.}

\subsubsection{System preparation}\label{sec:sysprep}
Systems are specified in terms of their composition, given by the fraction of PEO chains relative to all chains in the simulation cell. We use the notation
\begin{equation}
x^\text{(PEO)} \equiv \frac{N^\text{(PEO)}}{N^\text{(PEO)}+N^\text{(PMMA)}}
\end{equation}
where $N^\text{(PEO)}$ and $N^\text{(PMMA)}$ are the number of chains of PEO and PMMA.
Because the PEO and PMMA chains possess similar molecular weights, $x^\text{(PEO)}$ is also comparable to the mass fraction of PEO in the system. 
Condensed-phase systems are then prepared for $x^\text{(PEO)} = $ 0, 0.1, 0.2, 0.3, 0.4, 0.5, 0.6, 0.7, 0.8, 0.9, and 1.0.
Each system contains $N^\text{(PEO)}+N^\text{(PMMA)}$ = 40 total chains.
  For each $x^\text{(PEO)}$, three independent systems are prepared, resulting in a total of $11\times3=33$ systems. 

For each system, an initial configuration is generated by randomly placing each pre-constructed chain into a 60~\AA~$\times$~60~\AA~$\times$~60~\AA\ simulation box with random positions and orientations. A brief energy minimization is then performed where the system is simulated in the microcanonical (NVE) ensemble with a constrained maximum distance of 0.005~\AA\ moved per timestep; this minimization takes place for 0.05~ns.
After, the constrained maximum distance is increased to 0.1~\AA\, and the system is simulated for 0.5~ns. 
This procedure resolves unfavorable atomic overlaps introduced by the random packing procedure. 
Following minimization, 
\add{initial particle velocities are randomly generated from a uniform distribution over the interval $[-0.5,0.5]$, rescaled such that temperature as estimated from the kinetic energy corresponds to 300 K, and finally \maw{shifted} to remove any net linear momentum in the simulation cell}.
The system then undergoes 1 ns of simulation in the NVE ensemble, followed by 1 ns in the canonical (NVT) ensemble at 300~K.
Subsequently, a barostat is introduced to maintain the pressure at 1~atm, and the system is heated from 300~K to 700~K at a rate of 80~K/ns. After reaching 700~K, the system is equilibrated in the isothermal--isobaric (NPT) ensemble for 50~ns. Finally, the system is cooled from 700~K to 100~K at a rate of 10~K/ns at constant pressure.

The cooling trajectories are used for analysis of glass transition temperatures (see Section {\ref{sec:Tg}}).
Simulation configurations are also recorded specifically around the temperatures of 500 K, 360 K, and 220 K. 
From these configurations, additional simulations are performed for 35 ns in the NPT ensemble. 
These simulation trajectories are used to characterize  dynamic heterogeneity by means of local segmental mobilities, as described in Sections {\ref{sec:msf}} and {\ref{sec:fv}}.
\add{Finally, additional simulation is performed for systems at 500~K to enable the Rouse mode analysis described in Section 2.2.5; for $x^\text{(PEO)} = 1.0$, trajectories are extended by 100~ns, and for all other systems, trajectories are extended by 200~ns, with the difference being due to slower relaxation.} 

\add{All blends are prepared well-mixed and remain so for the duration of the simulation. All systems, including neat PEO and PMMA, remain amorphous throughout our simulations. Crystallization is neither observed nor expected on the timescale of our simulations. Partially by consequence, demixing is also not observed across the full composition range. As a result, the data presented are representative of hypothetical miscible systems. The simulations do not establish miscibility for PEO/PMMA blends in this regime but rather explore the consequences of it in a simulation setting.} 

%We note that our system preparation protocol yields amorphous systems across all compositions and temperatures. As such, the following analysis pertains to well-mixed systems without the influence of crystallization, which is not accessible on simulation timescales. While semicrystallinity is an important factor for material properties, its absence here removes a potential confounding factor, allowing a clearer focus on the fundamental dynamical coupling between components. 

\subsection{Analysis}\label{sec:analysis}

\subsubsection{Glass transition temperature}\label{sec:Tg} 

To obtain an apparent glass transition temperature ($T_\text{g}$), we use simulated dilatometry and analyze the temperature dependence of the specific volume, $v(T)$, for each system. An apparent $T_\text{g}$ is identified by a change in the slope of $v(T)$,\cite{buchholz2002, han1994} which reflects a shift in thermophysical relaxation behavior for the monitored quantity; this is associated with transition from a melt to a more glassy state.
During the cooling phase of system preparation (Section~\ref{sec:sysprep}), configurations are recorded at 10~K intervals. For each temperature, we perform an additional 1~ns simulation under isothermal--isobaric (NPT) conditions, starting from the corresponding configuration. The average specific volume is computed from the second half of each trajectory, yielding a set of $(T_i, \bar{v}_i)$ pairs.

The apparent $T_\text{g}$ from this dataset is determined using a previously reported bootstrap resampling procedure.\cite{chu2020,deng2021} We first identify, by visual inspection, a maximum and minimum temperature within the melt regime, $T_\text{max}^\text{m}$ and $T_\text{min}^\text{m}$, and a maximum and minimum temperature within the glassy regime, $T_\text{max}^\text{g}$ and $T_\text{min}^\text{g}$. 
The temperatures $T_\text{min}^\text{m}$ and $T_\text{max}^\text{g}$ define sampling ranges for the melt, $[T_\text{min}^\text{m}, T_\text{max}^\text{m} = T_\text{min}^\text{m} + 100~\text{K}]$, and for the glass, $[T_\text{min}^\text{g} = T_\text{max}^\text{g} - 100~\text{K}, T_\text{max}^\text{g}]$. 
The temperature bounds used for each blend composition are listed in the Supplementary Information, Table S1.
In each resampling iteration, a lower bound $T_\text{lo}^\text{m}$ is randomly selected from the melt range and an upper bound $T_\text{hi}^\text{g}$ from the glass range. Linear regressions are then performed on the data within $[T_\text{lo}^\text{m}, T_\text{max}^\text{m}]$ and $[T_\text{min}^\text{g}, T_\text{hi}^\text{g}]$ to fit the melt and glassy branches, respectively. \add{The intersection of these fits is considered as one $T_\text{g}$ sample.} This process is repeated 10,000 times to generate a distribution of $T_\text{g}$ values. We report the mean of this distribution as the estimated $T_\text{g}$ and its standard deviation as the associated uncertainty.
Simulated $v(T)$ data and representative $T_\text{g}$ values for each blend composition are provided in the Supplementary Information, Figure S1.
 
\subsubsection{Local segmental mobility}\label{sec:msf} 
To characterize local segmental dynamics, we define a segmental mobility parameter, $\mu_{i,\Delta t}$, which relates to the mean-square fluctuation of the positions of a particle over a given observation time $\Delta t$. For a given particle $i$, a mobility is computed as 
\begin{equation}\label{eq:MSF}
    \mu_{i,\Delta t} = \frac{ \langle(\vec{r}_i(t) - \langle\vec{r}_i\rangle_{\Delta t})^2\rangle_{\Delta t}}{\Delta t}
\end{equation}
% \begin{equation}\label{eq:MSF}
%     \text{MSF}_i = \langle(\vec{r}_i(t) - \langle\vec{r}_i\rangle_{\Delta t})^2\rangle_{\Delta t}
% \end{equation}
where $i$ is a backbone carbon, $\vec{r}_i(t)$ is the position of particle $i$ at time $t$, and $\langle \cdot \rangle_{\Delta t}$ denotes an ensemble average over the observation time. 
For the analysis herein, $\Delta t$ = 100 ps, which is substantially shorter than timescales for chain diffusion. Consequently, eq. {\eqref{eq:MSF}} mostly captures local segmental fluctuations. 
This quantity is computed  at 500 K, 360 K, and 220 K from the final 5 ns of the 35-nanosecond trajectories described in Section {\ref{sec:sysprep}}. 
Analysis based on $\mu_{i,\Delta t}$ across compositions and temperatures manifests in two ways.
In Section~\ref{sec:localdyn}, the segmental mobility is computed at the species level by averaging over all backbone carbons of each polymer type. In \add{Section~\ref{sec:local_absenv}}, the segmental mobility is further resolved based on the local environment of each backbone carbon atom to account for compositional heterogeneity introduced by blending.

To characterize the local environment of a particle \add{in Section~\ref{sec:local_absenv}}, we define a normalized self-density parameter, $\tilde{\phi}^{(A)}_i$, which measures the local enrichment of species $A$ around a particle of the same species. For a given backbone carbon atom $i$, this quantity is computed as:
\begin{equation}\label{eq:local}
    \tilde{\phi}^{(A)}_i = \frac{\sum{}_{j=1}^n \omega_{ij}^{(A)} \, \delta_{\alpha_j}^{(A)}}{\left\langle \sum{}_{j=1}^n \omega_{ij}^{(A)} \right\rangle_{x^{(A)} = 1}}
\end{equation}
In eq.~\eqref{eq:local}, the numerator sums over all $n$ monomers in the system, applying position-dependent weights $\omega_{ij}^{(A)}$ and selecting only those monomers of species $A$ via the Kronecker delta $\delta_{\alpha_j}^{(A)}$, where $\alpha_j$ denotes the species identity of monomer $j$. The denominator provides a normalization by the average local density around a particle in a pure $A$ system (subjected to the same weighting coefficients), denoted by $\langle \cdot \rangle_{x^{(A)} = 1}$. 
This normalization provides natural limits then of $\tilde{\phi}^{(A)}_i = 1$ corresponding to a local environment identical to that in a pure system of species $A$, while $\tilde{\phi}^{(A)}_i = 0$ indicates a local environment composed entirely of the other species.

To utilize eq.~\eqref{eq:local}, a scheme for the weighting coefficients must be defined, of which there are many reasonable choices. We choose to define a smoothing kernel of the form
\begin{equation}\label{eq:weights}
\omega_{ij}^{(A)} =
\begin{cases}
1 & r_{ij} \leq \add{r_m^{(A)}} \\
\exp\left(-\dfrac{(r_{ij} - r_m^{(A)})^2}{\sigma^2}\right) & r_{ij} > \add{r_m^{(A)}}
\end{cases}
\end{equation}
where $r_{ij} = |\vec{r}_i - \vec{R}_j|$ is the distance between particle $i$ and the center of mass of monomer $j$ (using the minimum image convention), and $r_m^{(A)}$ and $\sigma$ are parameters that define a smoothing kernel. 
To emphasize spatially local interactions, we set $r_m^{(A)}$ to the radius of gyration of a single \add{constitutional repeat unit} of species $A$, computed from simulations in vacuum at room temperature. This yields $r_m^{\text{(PEO)}} = 1.59$~{\AA} and $r_m^{\text{(PMMA)}} = 2.34$~{\AA}. The smoothing width parameter is set to $\sigma = 12$~{\AA} based on the non-bonded, real-space interaction cutoff.
 
\subsubsection{Free-volume analysis}\label{sec:fv} 
The concept of free volume is often invoked to elucidate facets of polymer dynamics.\cite{rigby1990, putta2001, widmer-cooper_free_2006, mei_local-average_2022}
To quantify the free volume associated with each polymer chain, we implement the following procedure. First, the simulation cell is tessellated using Delaunay triangulation, such that each simplex (tetrahedron) is defined by four atoms. Each chain is associated with a subset of simplices that have at least one vertex belonging to an atom on that chain.
Next, the entire simulation cell is filled with a three-dimensional grid of $n$ equally spaced spherical probes, where $n$ depends on a chosen probe radius. 
Each probe is then classified as occupied or unoccupied based on overlap with any atom in the system, using atomic diameters defined by the $\sigma$ parameters from the OPLS-AA force field. 
Finally, the free volume of a given chain is then computed as the total volume of unoccupied probes that fall within the chain-associated simplices.
Free volume is computed and averaged over the final 5~ns of simulations equilibrated at 500~K, 360~K, and 220~K. 
In the main text, results correspond to a probe radius of 0.5~\AA; additional results for other probe sizes are provided in Figure S2. 

\subsubsection{\add{Distribution of local composition}} \label{sec:distmethod}
\add{To further probe how packing influences polymer dynamics, we perform an analysis akin to that described by \maw{the} Lodge--McLeish (LM) chain connectivity model, which assumes that a cooperative volume of spanned by a Kuhn length $\ell_k$ governs local self-concentration effects.\cite{lodge2000}
\maw{Here, we go beyond mean-field average self-concentrations and calculate} the distribution of local compositions surrounding PEO and PMMA backbone carbons within spheres of radius equal to the Kuhn length, $\ell_k$.\cite{liu2009} For PEO, we use $\ell_k$ = 8.2~\AA; for PMMA, we use $\ell_k$ = 13.8~\AA.\cite{hiemenzlodge}}

\add{The calculation proceeds similarly to that described in Section {\ref{sec:fv}}. 
\maw{First, the simulation cell is filled $n$ spherical probes distributed on a simple cubic lattice; $n$ depends on the probe radius, which is 0.5~\AA~for our analysis.} 
\maw{Each probe is subsequently classified as PEO, PMMA, or unoccupied based on whether the center-to-center distance between a probe and a particle is less than the atomic diameter; atomic diameters are based on the values of $\sigma$ as prescribed by the employed force field;} probes are also assigned to a given polymer chain based on which atom overlaps with the probe.} 

\add{For each backbone carbon of polymer type $A$, we then quantify three local composition measures within the $\ell_k$ sphere: the intramolecular volume fraction $\phi_\text{intra}^\text{(A)}$ (i.e., contributions from the same chain), the intermolecular PEO volume fraction $\phi_\text{inter,PEO}^\text{(A)}$ (i.e., contributions from surrounding PEO chains), and the intermolecular PMMA volume fraction $\phi_\text{inter,PMMA}^\text{(A)}$ (i.e., contributions from surrounding PMMA chains). 
\maw{These quantities are calculated as the sum of volumes of appropriately classified probes divided by the total sphere volume.
The total sphere volume is defined as the sum of intramolecular, intermolecular PEO, intermolecular PMMA, and free volume contributions within the relevant sphere. This normalization is slightly different from prior analyses} of a similar nature, where normalization was done on a species-specific basis.\cite{liu2009} 
\maw{Distributions of volume fractions} are computed and averaged over the final 5 ns of trajectories equilibrated at 500 K, 360 K, and 220 K.}
%H(xPEO;phi_kappa,B^(A))
% write as integral over P(phi)log2P(phi)dphi

\add{To further characterize these distributions, we calculate their Shannon entropy: 
\begin{equation}\label{eq:shannon}
H(x^\text{(PEO)};\phi_\kappa ^\text{(A)}) = \int P(\phi_\kappa^\text{(A)}) \log_2 \left[P(\phi_{\kappa}^\text{(A)})\right] d\phi_{\kappa}^\text{(A)}
\end{equation}
where $P(\phi_{\kappa}^\text{(A)})$ is the probability distribution of local volume fraction of atoms near a backbone carbon of type $A$ under condition $\kappa$ (intramolecular, intermolecular PEO, or intermolecular PMMA). For interpretability, eq.~\eqref{eq:shannon} is normalized by the Shannon entropy of a uniform distribution to yield $\tilde{H}(\cdot)$. This normalization sets $\tilde{H} = 0$ for a Dirac delta distribution and $\tilde{H} = 1$ for a uniform distribution. Thus, broader and more heterogeneous local environments drive $\tilde{H}$ towards unity, while narrower or more uniform environments drive it towards zero.}
%\add{In this way, our approach allows both a direct comparison to the predicted self-concentration values from the LM model and also a flexible description that resolves full distributions of local environments.}
%As shown below, while the LM framework predicts intramolecular fractions of $\sim 0.22$--0.23 for both species, our simulations reveal sharply peaked values near $\sim 0.06$, which are shown in Figure \ref{fig:intra_si}. This deviation highlights the limitations of applying the LM model to atomistic simulations with spatial and temporal constraints. Indeed, the LM model has been found to quantitatively agree with experimentally measured segmental dynamics for large spatial scales ($q=0.69$~\AA$^{-1}$) but fails for smaller scales ($q=1.3$~\AA$^{-1}$).\cite{garcia_sakai_dynamics_2008} However, this deviation also illustrates how this framework extends the LM concept by resolving full distributions rather than relying on a single averaged cooperative volume.\cite{liu2009}}

\subsubsection{Rouse mode analysis}\label{sec:Rouse} 

To investigate how dynamic heterogeneity manifests in collective polymer dynamics, we perform a Rouse mode analysis to extract characteristic relaxation times associated with different statistical segment lengths of the polymer chains. 
For a polymer comprised of $N$ monomers, Rouse mode coordinates are computed by
\begin{equation}
\vec{X}_p(t) = \sqrt{\frac{c_p}{2}} \sum_{i=0}^{N-1} \vec{r}_i(t) \cos \left[\frac{p \pi }{N} \left(i + \frac{1}{2} \right) \right]
\end{equation}
where $p$ = 0, $\dots$, $N-1$ indicates the mode index, $\vec{r}_i(t)$ is the position of the center of mass of the $i$th monomer on the chain at time $t$ , and $c_p$ is a $p$-dependent constant, such that $c_1 = 1$ and $c_p = 2$ for all other modes.
The zeroth mode corresponds to the behavior of the center of mass of a chain, while all other modes roughly correspond to the collective behavior of sub-chains of $\frac{(N-1)}{p}$ segments.

The Rouse mode coordinates are used to compute time autocorrelation functions (ACFs), which characterize the relaxation timescales of subchains of varying segment lengths. Each ACF is well-described by a stretched exponential function:\cite{brodeck2010, moreno_entangledlike_2008}
\begin{equation} \label{eqacf}
    \frac{\langle \vec{X}_p(t) \cdot \vec{X}_p(0) \rangle}{ \langle \vec{X}_p(0) \cdot \vec{X}_p(0) \rangle } = \exp\left[ -\left( \frac{t}{\tau_p} \right)^{\beta_p} \right]
\end{equation}
where $\tau_p$ and $\beta_p$ are fitting parameters. Rather than fitting ACFs to eq.~\eqref{eqacf}, however, ACFs are fit to a linearized form of eq.~\eqref{eqacf} for simplicity. Taking the natural logarithm yields 
\begin{equation} \label{eqacf_lin}
    \ln \left[ -\ln \left( \frac{\langle \vec{X}_p(t) \cdot \vec{X}_p(0) \rangle}{\langle \vec{X}_p(0) \cdot \vec{X}_p(0) \rangle} \right) \right] = \beta_p \ln(t) - \beta_p \ln(\tau_p)
\end{equation}
allowing extraction of $\beta_p$ and $\tau_p$ via linear regression. An effective relaxation time is then calculated as
\begin{equation} \label{eqefftau}
    \tau_{p}^{\text{eff}} = \int_0^{\infty} \exp\left[ -\left( \frac{t}{\tau_p} \right)^{\beta_p} \right] dt = \tau_p^{1/\beta_p} \Gamma \left( \frac{\beta_p+1}{\beta_p} \right)
\end{equation}
where $\Gamma(\cdot)$ denotes the gamma function. 
% This procedure is used directly for PMMA.
% For PEO, the fitting is simplified by linearizing eq.~\eqref{eqacf}. Taking the natural logarithm yields
% \begin{equation} \label{eqacf_lin}
%     \ln \left[ -\ln \left( \frac{\langle \vec{X}_p(t) \cdot \vec{X}_p(0) \rangle}{\langle \vec{X}_p(0) \cdot \vec{X}_p(0) \rangle} \right) \right] = \beta_p \ln(t) - \beta_p \ln(\tau_p)
% \end{equation}
% allowing extraction of $\beta_p$ and $\tau_p$ via linear regression. The distinction in fitting approaches reflects the stronger non-ideal behavior observed in PMMA compared to PEO, as length scale of a monomer is generally shorter than the Kuhn length.
\add{To perform the fitting, the ACFs were computed over time intervals of 8 ns for PEO, \maw{while for PMMA, intervals were 125 ns, 75 ns, and 20 ns for $x^\text{(PEO)} =$ 0.0, $x^\text{(PEO)} =0.1$–0.2, and $x^\text{(PEO)} =0.9$–0.3, respectively.} These intervals were selected to ensure that the average normalized ACF of the largest Rouse mode for each species decayed to at least 0.2 \maw{at 500 K, to facilitate reliable fitting}. Representative ACFs and fitted curves are shown in the Supplementary Information (S10 and S11).} 

\section{Results \& Discussion}

\subsection{Composition-dependent glass transition temperatures} \label{sec:tganalysis}

\begin{figure}[htb] 
    \includegraphics{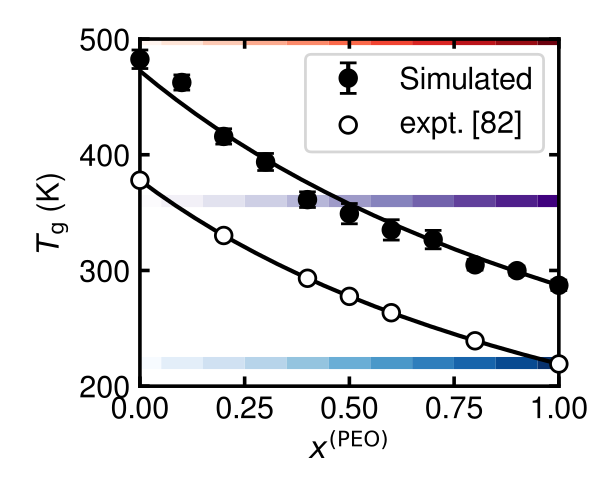}
    \caption{Dependence of \maw{apparent} glass transition temperature ($T_\text{g}$) on blend composition. Markers represent simulated (filled) and experimental (empty) $T_\text{g}$ values for blends of PEO and PMMA. Experimental results are from Ref. \citen{wu1987}. Solid black lines are fits to the Fox equation. Error bars reflect standard errors calculated from three independent system configurations. Horizontal, colored bands provide visual reference to 500 K (red), 360 K (purple), and 220 K (blue), which are examined in subsequent figures. The color gradation within each band distinguishes blends at the same temperature but different compositions.}
    \label{figtg}
\end{figure}

We begin by briefly comparing experimental and simulated trends regarding $T_\text{g}$ for these PEO/PMMA systems as a function of composition.
Figure \ref{figtg} shows that both the simulated apparent $T_\text{g}$ as well as the experimental $T_\text{g}$ values\cite{wu1987} follow the Fox equation
\begin{equation}\label{eqfox}
    \frac{1}{T_\text{g}} = \frac{w^{\text{(PEO)}}}{T_\text{g}^{\text{(PEO)}}} + \frac{w^\text{(PMMA)}}{T_\text{g}^{\text{(PMMA)}}}
\end{equation}
where $w^{(i)}$ is the mass fraction of species $i$ and $T_\text{g}^{(i)}$ is the $T_\text{g}$ of a pure neat system of species $i$. 
\add{The PMMA used in the experimental measurements of $T_\text{g}$ are of unspecified tacticity.\cite{wu1987}}
The apparent $T_\text{g}$ values extracted from simulation are systematically higher than experimental values by approximately 100~K. This offset is roughly consistent across the range of examined compositions and is generally expected due to the much faster cooling rates and shorter observation times inherent to molecular simulations.\cite{tgrate,R:2020_Hung_Forecasting} 
Given this systematic disparity, the similarity in trends and alignment  with the Fox equation suggests that the employed force field captures the essential physics governing blend dynamics and responds appropriately to changes in composition.

\add{Experiments using differential scanning calorimetry (DSC) on PEO/PMMA blends have reported both single\cite{wu1987,fernandes1986,liberman1984} and dual\cite{lodge2006,silva_nanoheterogeneities_2000} $T_\text{g}$ signals at intermediate compositions, with the former attributed to broad, overlapping peaks arising from local composition fluctuations.\cite{lodge2006} 
In general, rapid quench rates and finite system sizes inherent to simulations may obscure multiple signatures observed experimentally.
\maw{Our simulations yield a single apparent $T_\text{g}$ based on the systematic fitting procedure described in the Supplementary Information (Section S1)}
\maw{However, recent simulation work\cite{R:2022_Habasaki_Atomistic} suggested two $T_\text{g}$ values from density–temperature data using hyperbolic fitting.}
\maw{Our data} (Figure S1) illustrate a broad transition window at intermediate compositions than could support hyperbolic fitting with two distinct inflection points.}  
\add{Regardless of whether the simulated dilatometry yields one or two apparent $T_\text{g}$, the key outcome is that the \maw{simulations produce a reasonable composition dependence, suggesting it can capture} asymmetries in PEO and PMMA dynamics.}

Figure \ref{figtg} also highlights three temperatures that are of specific interest in the following sections. These temperatures are selected to span distinct thermal regimes: (i) 220~K lies below the apparent $T_\text{g}$ of both pure components and therefore below that of any blend; (ii) 360~K falls between the $T_\text{g}$ values of pure PEO and PMMA, such that some blends are above and others below their respective $T_\text{g}$; and (iii) 500~K exceeds the $T_\text{g}$ of both pure components and all blends. This temperature range enables the examination of how interspecies dynamical coupling depends not only on the different $T_\text{g}$ values of PEO and PMMA but also on the absolute temperature with respect to these $T_\text{g}$ values.

\subsection{Characterization of species-dependent local dynamics} \label{sec:localdyn}

\afterpage{
\begin{figure}[htb]
    \centering
    \includegraphics{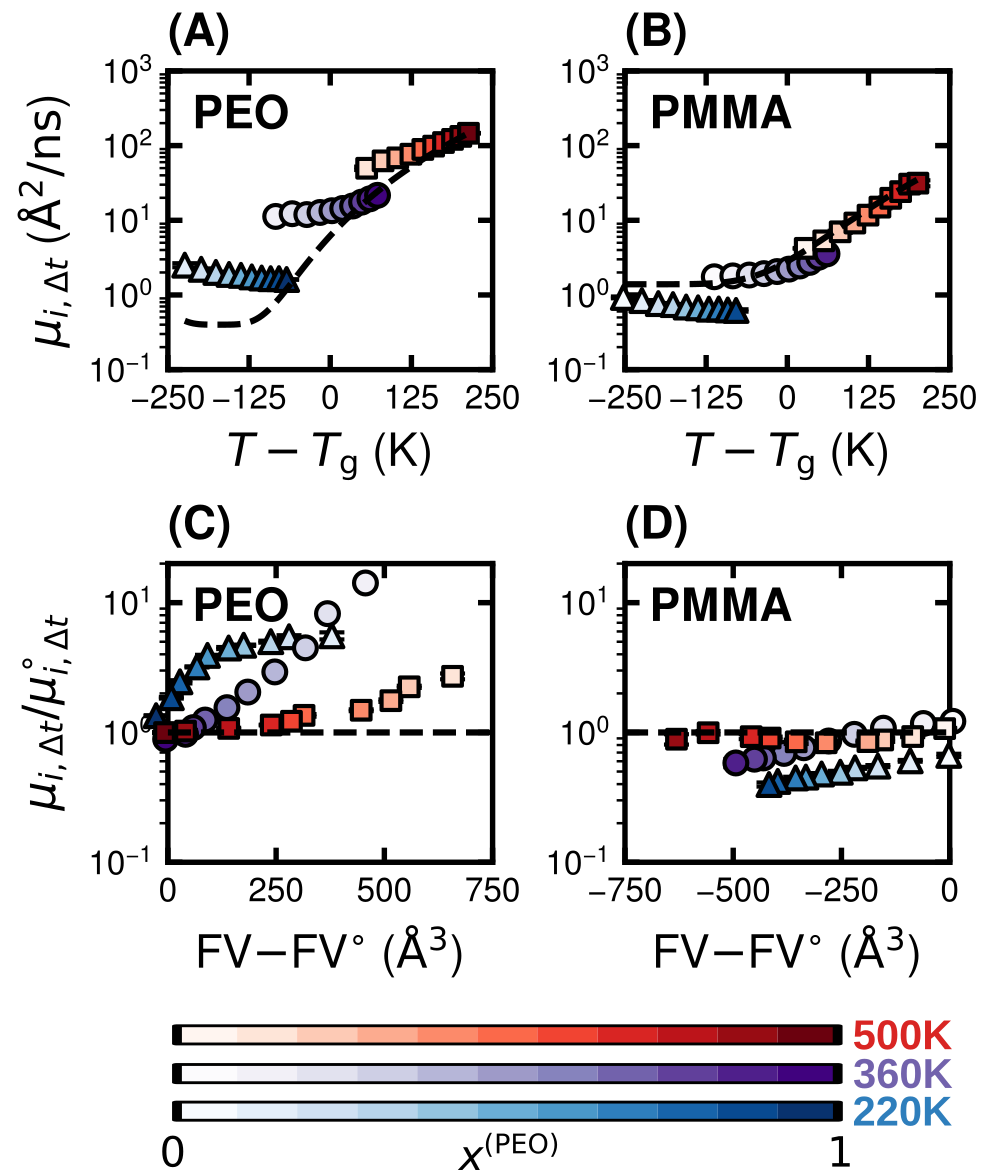}
    \caption{Analysis of local segmental mobilities across blend compositions and temperatures.
    The local segmental mobility, $\mu_{i,\Delta t}$ as a function of temperature relative to the apparent $T_\text{g}$ for (A) PEO and (B) PMMA. \add{Data points at 220 K, 360 K, and 500 K are represented as blue triangles, purple circles, and red squares, respectively.}
    The dashed black lines represent fits to the neat polymer reference, $\mu_{i,\Delta t}^\circ$.
    Segmental mobility as a function of free volume (FV) for (C) PEO and (D) PMMA. The FV${}^\circ$ denotes reference to the FV of the neat polymer. Error bars reflecting standard errors from three independent systems are generally smaller than the symbol size. }
    \label{figMSFrelt}
\end{figure}
}

As an initial characterization of nanoscale dynamic heterogeneity, we examine how blending influences the average local dynamics of PEO and PMMA compared to their behavior in neat systems. Specifically, we analyze species-resolved segmental mobilities ($\mu_{i,\Delta t}$) of polymer segments across varying blend compositions at temperatures below (220~K), between (360~K), and above (500~K) the apparent $T_\text{g}$ values of the pure components. 

Figures~\ref{figMSFrelt}A and \ref{figMSFrelt}B show the $\mu_{i,\Delta t}$ as a function of $T - T_\text{g}$, where $T_\text{g}$ varies with composition. 
The dynamics for PEO in blends (Figure~\ref{figMSFrelt}A) exhibit significant deviations from neat behavior at the same distance from $T_\text{g}$ (dashed black line) across most compositions but particularly in PMMA-rich blends (low $x^\text{(PEO)}$). This indicates that $T - T_\text{g}$ is not a reliable predictor of local segmental dynamics for PEO.
In contrast, PMMA (Figure~\ref{figMSFrelt}B) dynamics in blends closely follow the behavior of neat PMMA, albeit with less strong correlation at temperatures below $T_\text{g}$. This trend is consistent with previous findings suggesting that PMMA dynamics are effectively governed by the temperature difference from the $T_\text{g}$ of the blend\cite{sakai2004, liu2006, maranas2007} and also reveal an asymmetry in dynamical coupling between PEO and PMMA.
\add{\maw{The mobilities} of both PEO and PMMA at 220 K (below $T_\text{g}$ of both pure components) decrease with increasing $x^\text{(PEO)}$.
This behavior is both opposite to the behavior at 360 K and 500 K and also possibly unexpected because PEO is the higher-mobility species in the blend.}

To elucidate the prior results, we examine variations in species-dependent packing behavior, which is expected to manifest in different free volumes of the chains.
Figures~\ref{figMSFrelt}C and \ref{figMSFrelt}D show the normalized segmental mobility as a function of a change in free volume (FV) from the neat system (denoted by `$\circ$'). 
The rationale for this comparison derives from considerations involving free-volume theory, which may suggest that $\log [\mu_{i,\Delta t}/\mu_{i,\Delta t}^\circ] \propto (\text{FV}-\text{FV}^\circ)$.
The data indeed possess roughly linear behaviors in the limit of smaller perturbations in FV.   
The positive trends in Figures~\ref{figMSFrelt}C and \ref{figMSFrelt}D across all temperatures indicate that positive deviations in segmental mobility from neat polymer behavior generally correlate with increases in local free volume upon blending and vice versa. 

\add{\maw{Increases in FV also explain} the apparent enhancement in mobility of both PEO and PMMA at 220 K with increasing mole fraction of the lower mobility species.
In particular, both species are effectively glassy at 220 K, with low mobilities of similar magnitude ($\mu^\circ_{\text{PEO}, \Delta t}=0.16$~\AA/ns and $\mu^\circ_{\text{PMMA}, \Delta t}=0.09$~\AA/ns), such that the relative mobility advantage of PEO is diminished (Figure S4). Consequently, in this regime, we suggest that the dynamics are mostly controlled by variations in free volume. 
\maw{Although PMMA is canonically slower, it is also bulkier, such that its addition leads to increases FV, and thus less effective packing.
By contrast, }PEO-rich blends exhibit lower mobilities due to more efficient packing of chains. The less effective packing of PMMA relative to PEO is supported by the observation that the free volume of neat PMMA surpasses that of neat PEO at the same $T-T_\text{g}$ (\maw{Figure S3}).}

\add{An exception to the observation that dynamics are controlled by free volume relates to PMMA at high temperatures. In this case, mobility in blends \maw{resembles that of} a $T_\text{g}$-equivalent neat polymer reference, despite variations in FV (Figure~\ref{figMSFrelt}D).} 
These trends support the notion that packing effects contribute to dynamic heterogeneity, but they again highlight an asymmetry in coupling. While increased free volume tends to correlate with enhanced mobility for PEO, a reduction in free volume does not universally imply suppressed segmental dynamics in PMMA by the same magnitude. This asymmetry reflects the influence of other factors beyond free volume in controlling relative enhancement/suppression of polymer dynamics in blends.

\subsection{Influence of local environment on segmental mobility} \label{sec:lenlocaldyn}
\subsubsection{\maw{Analysis of local mobility-composition coupling}} \label{sec:local_absenv}
While previous results focused on local dynamics at the species level, we now explicitly consider variations due to the local environment of individual polymer segments. The central hypothesis is that a PEO segment surrounded entirely by other PEO segments should behave similarly to one in a pure PEO system, with minimal influence from PMMA, and vice versa.
However, as the local environment becomes enriched in the opposite species, there will be interaction-based coupling that will lead to deviations.
To test this, we analyze segmental mobilities as a function of a normalized self-density parameter $\tilde{\phi}^{(A)}_i$, which is approximately unity when the environment is similar to that of the neat system and approaches zero when surrounded completely by the other species.

\begin{figure}[htb!]
    \centering
    \includegraphics{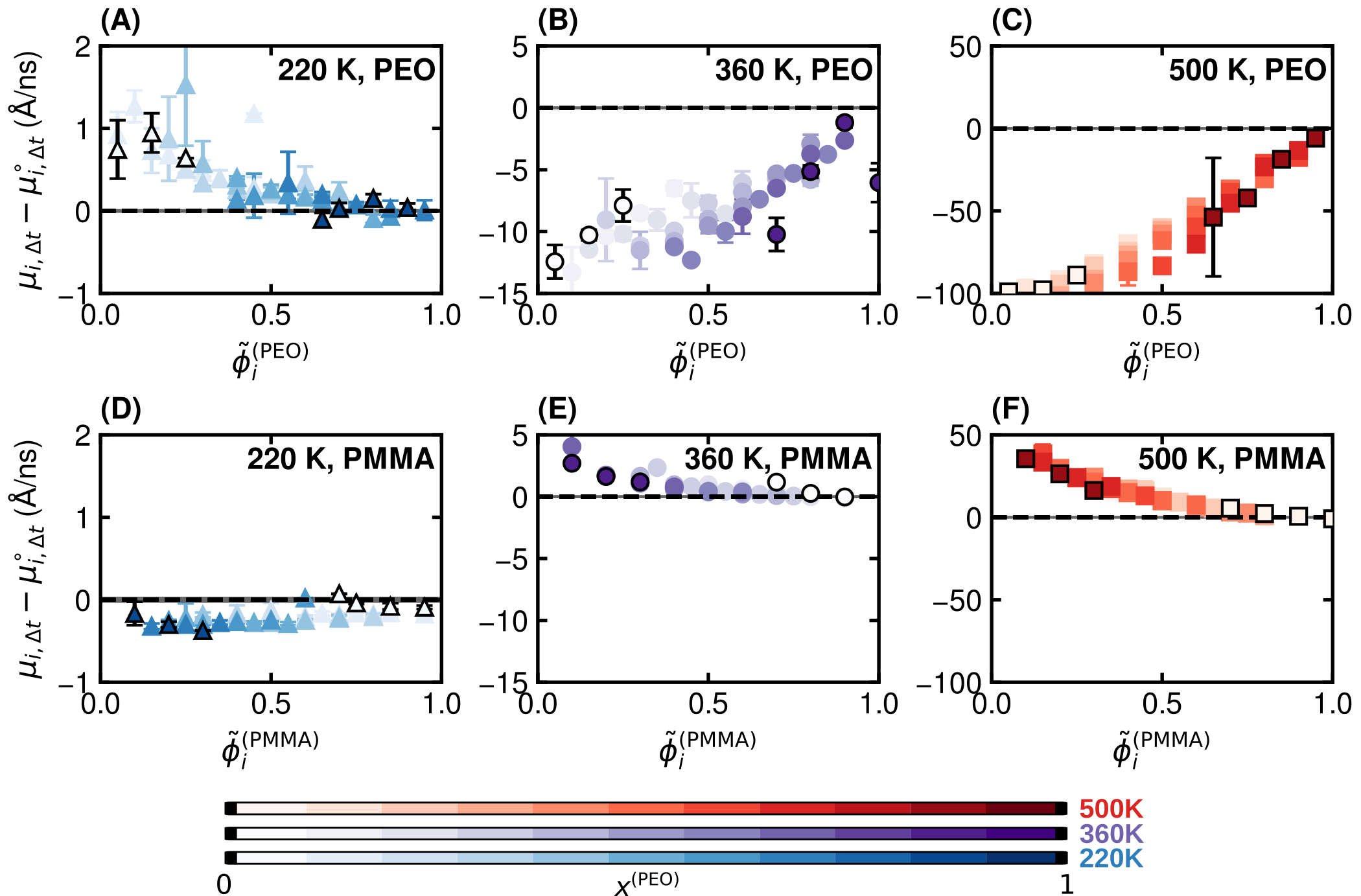}
    \caption{Variation in relative segmental mobility based on local environment. Deviations for neat-polymer mobility  for PEO at (A) 220 K, (B) 360 K, and (C) 500 K and for PMMA at (D) 220 K, (E) 360 K, and (F) 500 K. 
    Data is shown for all blend compositions, with gradation from light (PMMA-rich) to dark (PEO-rich), as indicated by the color bars. 
    Results for chains in blends with the most extreme compositions ($x^\text{(PEO)}=0.1$ and $0.9$) are outlined in black for visual clarity. 
    Error bars reflect standard errors from three independent systems. Horizontal dashed lines provide a guide to the eye for the neat-polymer mobility. The gray shaded area around the dashed lines reflect standard deviations calculated from three independent neat systems. }
    \label{figMSFsp}
\end{figure}

Figure \ref{figMSFsp} shows that the influence of local environment is asymmetric between species and strongly temperature-dependent. 
\add{The temperature-dependence is  first apparent by comparing the results for PEO where $\mu_{i,\Delta t}$ is enhanced by blending at a temperature below both component $T_\text{g}$ (Figure \ref{figMSFsp}A) whereas it is suppressed upon blending at higher temperatures (Figures \ref{figMSFsp}B,C)}.
\add{Meanwhile, that the influence is asymmetric between species is evident by contrast with the PMMA results  (Figures \ref{figMSFsp}D-F), which display nominally the opposite behavior of enhancement upon blending at high temperatures and minor suppression upon blending at low temperatures.}
\add{The observations at 360 K and 500 K are physically intuitive, as PEO-rich environments tend to exhibit larger mobilities, while PMMA-rich environments are slower.}
\add{The observations at 220 K where mobilities appear enhanced in PMMA-rich environments are likely due to the free-volume effects described in Section \ref{sec:localdyn}.}

\add{The dependence of segmental mobility on local composition also displays intriguing differences between suppression and enhancement effects in blends.}
In cases where $\mu_{i,\Delta t}$ is enhanced upon blending (above the dashed line), $\mu_{i,\Delta t}$ gradually approaches the neat reference, $\mu_{i,\Delta t}^\circ$, as the local environment becomes enriched in the same species; the notion of gradual in this context reflects a vanishing of the first derivative, $\frac{\partial \mu_{i,\Delta t}}{\partial \tilde{\phi}_i}$.
Where $\mu_{i,\Delta t}$ is suppressed (below the dashed line), $\mu_{i,\Delta t}$ approaches $\mu_{i,\Delta t}^\circ$ more abruptly. 
While all dynamics tend to the neat reference in the limit of that the local environment is enriched in that species, 
these behaviors, which are species-agnostic, reveal that suppression effects are more readily evident than enhancements. In other words, the local coupling of polymers with the opposing species has a much stronger magnitude of effect whereby the mobility of faster-moving chains is reduced more than that of slower-moving chains is enhanced in a blend. Upon investigation of the behavior of $\mu_{i,\Delta t} - \mu_{i,\Delta t}^\circ$ normalized by $\mu_{i,\Delta t}^\circ$ shown in Figure S5, it is evident, however, that the relative change in $\mu_{i,\Delta t}$ is actually much larger for species with enhanced mobility than suppressed mobility. 

There is also disparity in the  composition-dependence of these observations between the two polymer species.
In PEO, Figures \ref{figMSFsp}B and \ref{figMSFsp}C show that trends in segmental mobility with respect to $\tilde\phi_i^{\text{(PEO)}}$ differ based on $x^\text{(PEO)}$. We observe that the extent of dynamical coupling is weaker in blends with more PMMA (i.e., for a given $\tilde\phi_i^{\text{(PEO)}}$, $\mu_{i,\Delta t}-\mu_{i,\Delta t}^\circ$ is smaller  in systems with lower $x^\text{(PEO)}$).
By contrast, in PMMA, Figures \ref{figMSFsp}E and \ref{figMSFsp}F show how $\mu_{i,\Delta t}-\mu_{i,\Delta t}^\circ$ collapses onto a single  curve for all compositions. 
This can be accounted for by the composition-dependent packing behavior.
Namely, the suppression of PEO mobility due to the presence of more PMMA in the local environment is negated in part by the larger free volume, which would tend to enhance mobility of chains in PMMA-rich blends (Figure \ref{figMSFrelt}). This effect is largely absent for PMMA at all temperatures, once again reflecting the asymmetrical nature of dynamic heterogeneity in these blends.

\subsubsection{\maw{Analysis of local composition distributions}}\label{sec:distenv}

\add{While Figure~\ref{figMSFsp} demonstrates how local composition leads to deviations in segmental dynamics from neat-polymer behavior, \maw{further understanding blend properties} requires knowledge of how frequently such environments occur and how they are distributed. 
To this end, we characterize the local environments sampled by PEO and PMMA with the objective to relate these to deviations from $T_\text{g}$-equivalent neat dynamics. 
We compute probability distributions of local intermolecular PMMA volume fractions surrounding particles from PEO, $\phi_{\text{inter,PMMA}}^\text{(PEO)}$, and PMMA, $\phi_{\text{inter,PMMA}}^\text{(PMMA)}$, and characterize their breadth using a normalized Shannon entropy, $\tilde{H}$, where lower $\tilde{H}$ indicates more homogeneous environments and higher values indicates environments are more heterogeneously distributed. 
}

\begin{figure}[htb!]
    \centering
    \includegraphics{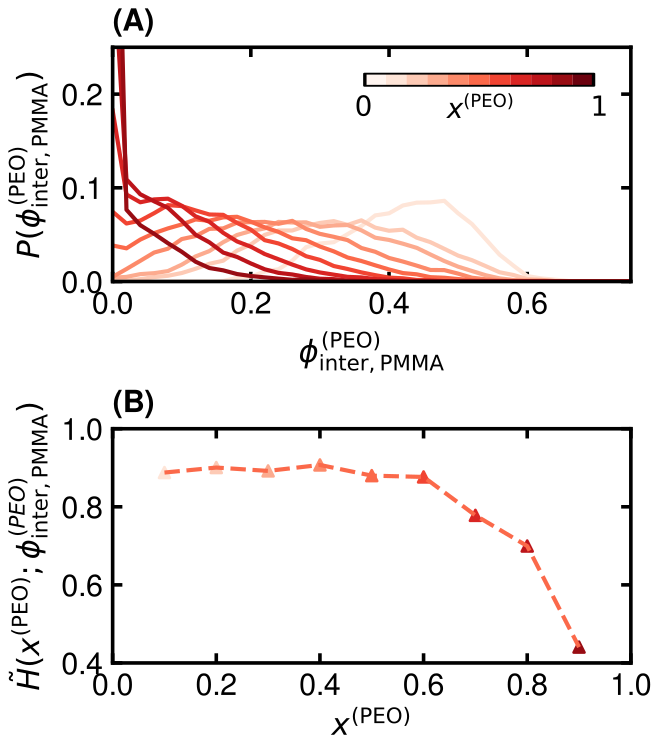}
    \caption{Characterization of local composition distributions around PEO. (A) Probability distributions of intermolecular PMMA volume fraction around PEO and (B) corresponding normalized Shannon entropies. Results are shown for systems at 500 K. The color gradient corresponds to a gradient in blend composition containing the most PEO (dark) to the least PEO (light). The dashed red line is provided as a guide to the eye. }
    \label{figdistpmma_peo}
\end{figure}

\add{Figure~\ref{figdistpmma_peo}A shows how the distributions of intermolecular PMMA volume fractions around PEO units vary with blend composition at 500~K; analogous results at 220~K and 360~K are provided in the Supplementary Information (Figure S8). For PEO-rich blends, the distributions peak near $\phi_{\text{inter,PMMA}}^{(\text{PEO})} = 0$, indicating that most PEO atoms experience negligible local PMMA content, consistent with the overall scarcity of PMMA. As $x^{(\text{PEO})}$ decreases and there is more PMMA present in the blend, the distributions shift toward higher $\phi_{\text{inter,PMMA}}^{(\text{PEO})}$.
The distributions also broaden, revealing that local environments become increasingly heterogeneous as the blend composition becomes more balanced. 
This trend is reflected quantitatively in the normalized Shannon entropy, $\tilde{H}$ (Figure~\ref{figdistpmma_peo}B), which rises with decreasing $x^{(\text{PEO})}$, signifying a wider range of local PMMA fractions sampled by PEO atoms. 
The entropy plateaus for $x^{(\text{PEO})} \lesssim 0.6$, indicating that the extent of heterogeneity becomes composition-independent once sufficient PMMA is present in the blend.}

\begin{figure}[htb!]
    \centering
    \includegraphics{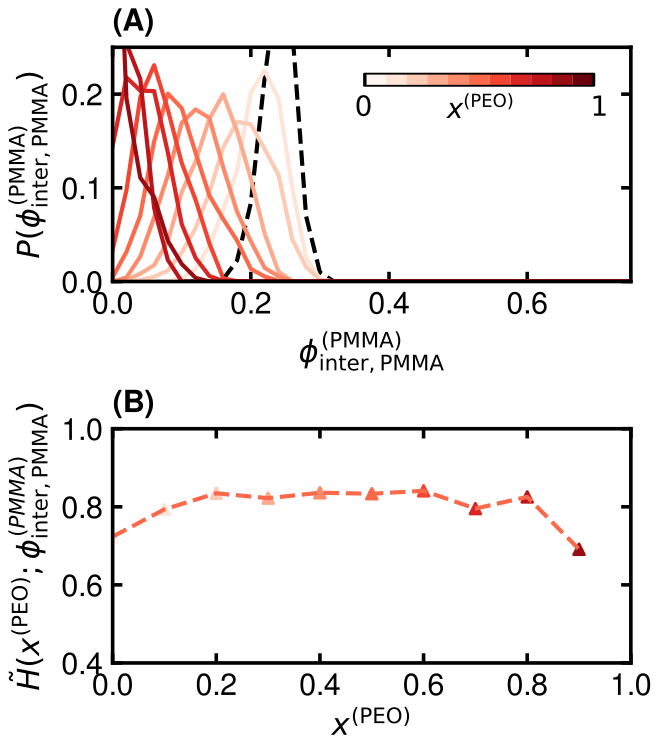}
    \caption{Characterization of local composition distributions for PMMA. (A) Probability distributions of intermolecular PMMA volume fraction around PMMA and (B) corresponding normalized Shannon entropies. Results are shown for systems at 500 K. The color gradient corresponds to a gradient in blend composition containing the most PEO (dark) to the least PEO (light). The black dashed line in (A) is the distribution of neat PMMA. The red dashed line in (B) is provided as a guide to the eye. }
    \label{figdistpmma_pmma}
\end{figure}

\add{Figure~\ref{figdistpmma_pmma}A shows how the distributions of intermolecular PMMA volume fractions around PMMA units vary with blend composition. As in Figure~\ref{figdistpmma_peo}A, the distributions shift to higher $\phi_{\text{inter,PMMA}}^{(\text{PMMA})}$ with decreasing $x^{(\text{PEO})}$, reflecting the overall increase in PMMA content. However, unlike the distributions around PEO, their shapes remain comparatively narrow and composition-invariant, indicating that the local environments of PMMA change in a more uniform and predictable manner. This behavior is corroborated by the normalized Shannon entropy in Figure~\ref{figdistpmma_pmma}B, which remains nearly constant across compositions and is systematically lower than for PEO, signifying more homogeneous local surroundings. 
Together, these results highlight another asymmetry between PEO and PMMA upon blending. Namely, while the local environment of PMMA changes in a consistent way that can be tracked  by the mean composition, as in the Lodge–McLeish framework,\cite{lodge2000}  the local environment of PEO cannot because the nature of the distribution itself changes significantly with composition.
In addition, PEO \maw{segments experience} overall \maw{more} local compositional heterogeneity.}

\add{We attribute the dynamical response of PEO in the blends to the increased heterogeneity and composition dependence of its local environment with respect to surrounding PMMA. 
These features correlate strongly with the deviations of local segmental mobilities from those of $T_\text{g}$-equivalent neat systems (Figure~\ref{figMSFrelt}). 
For PEO, the growing diversity of local intermolecular PMMA environments with decreasing $x^{(\text{PEO})}$ (Figure~\ref{figdistpmma_peo}A) coincides with its marked departure from the neat reference behavior. 
In contrast, the local environments of PMMA evolve more uniformly with composition (Figure~\ref{figdistpmma_pmma}), resulting in dynamics that follow expectations based on the overall reduction in $T_\text{g}$ with PEO incorporation. 
Distributions of local intramolecular self-contributions (Figure S6) are comparatively insensitive to composition, as expected.
Meanwhile, the distributions of local intermolecular PEO contributions (Figure S7) again change simply in response to composition for PMMA, and for PEO, there is a modest increase in heterogeneity at low PMMA blending fractions, but this is less pronounced than the strong compositional diversity observed for surrounding PMMA (Figure S8).
However, the distributions of local environments show minimal temperature dependence (Figures S6-S9), indicating that increased compositional heterogeneity alone does not account for the deviations observed well below $T_\text{g}$ (220~K). 
Instead, at such temperatures, the relative dynamics are better explained by differences in free-volume between the probed systems and their references (Figures~\ref{figMSFrelt}C,D).
}

\subsection{Rouse mode analysis of collective dynamics in the melt state} \label{sec:rouseanalysis}

\begin{figure}[htb]
    \centering
    \includegraphics{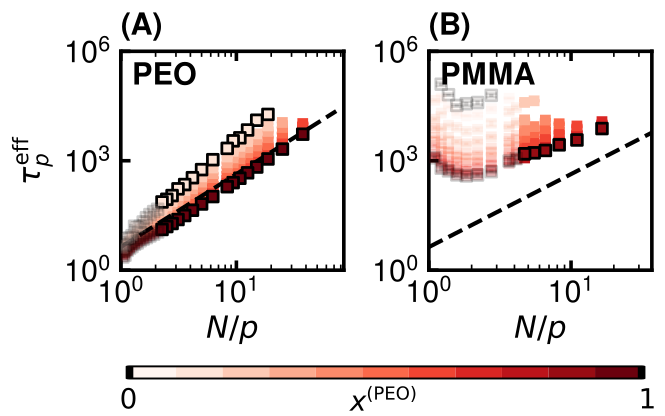}
    \caption{Rouse mode analysis at 500 K for chains in blends of varying composition. The effective Rouse relaxation time $\tau^{\text{eff}}_p$ as a function of sub-chain length $N/p$ for (A) PEO and (B) PMMA.
    For $p>8$, symbols for data are only shown for every third value of $p$ for visual clarity. 
    Dashed black lines are a guide to the eye to indicate the expected ideal scaling of $\tau_p \sim p^{-2}$. The position of the line is the same across panels and is set to align with the behavior of neat PEO. 
    Results for chains in blends with the most extreme compositions are outlined in black for visual clarity of trends. Error bars reflect standard errors from three independent systems and are generally smaller than the symbol size. 
    \add{Transparent markers are used for $p$ corresponding to sub-chains equal to or less than an estimated Kuhn length.
    Deviations from the ideal $p^{-2}$ scaling are expected for modes shorter than the Rouse bead size, which is generally larger than a Kuhn length.}
}
    \label{figrouse}
\end{figure}

To gain insight into how blending affects \add{collective dynamics of PEO and PMMA, we perform a Rouse mode analysis of chains of both species. This approach shifts the focus from local segmental fluctuations to collective sub-chain dynamics, enabling us to probe relaxation across longer times and length scales. 
In this way, our aim is not to test whether the dynamics are strictly Rouse-like but rather use Rouse coordinates as a convenient collective-variable framework to assess how relaxation times vary with sub-chain length and composition.}
Given the computational challenges of equilibration and convergence as well as the dramatic increase in relaxation times expected at temperatures below $T_\text{g}$, \add{our analysis is restricted \maw{to} systems at 500~K (Figure \ref{figrouse}).} 

Figure \ref{figrouse}A shows that the characteristic relaxation times, $\tau^\text{eff}_p$, of PEO in PEO/PMMA blends deviates from those of pure PEO,  across all blend compositions and for all mode numbers. 
Pure PEO approximately follows the expected Rouse scaling (dashed black line), $\tau_p \sim p^{-2}$,\cite{brodeck2009} \add{albeit} when using $\tau^\text{eff}_p$. 
As the mode number increases ($p \to N$), $\tau^\text{eff}_p$ for all blends \add{begins to} converge; \add{this implies that the relaxation behavior of the smallest sub-chains of PEO are similar, whether in blends or in neat melts.}

\add{It should be noted that
the classic Rouse scaling of $\tau_p \sim p^{-2}$ is only expected down to the Rouse bead size,
which is distinct from the Kuhn length and can span multiple chemical monomers.\cite{agapov2010} }
\add{Here, the apparent convergence in relaxation times occurs for sub-chains likely smaller than a canonical Rouse bead.}
\add{Meanwhile,} deviations from pure PEO behavior become more pronounced at lower $p$ as the blend becomes more PMMA-rich (decreasing $x^\text{(PEO)}$). 
\maw{Furthermore, since $\tau_p^{\mathrm{eff}}$ represents the mean relaxation time extracted from stretched exponential decays, apparent deviations from ideal scaling may reflect both physical departures\cite{maitra2007,R:1998_Paul_Chain, R:2002_Krushev_Role} from Rouse dynamics as well as non-exponential relaxation.}
\add{All together, these results intuitively suggest that chain relaxation at shorter length scales is less affected by the surrounding melt environment than at longer length scales.} 

\add{Figure~\ref{figrouse}B shows that \maw{variations in relaxation behavior of PMMA chain segments with $N/p$ are functionally similar across blend compositions, resulting in systematic vertical offsets} in relaxation times. This indicates that blending primarily shifts the overall timescale of PMMA dynamics while preserving their scaling, suggesting that the data \maw{might} be collapsed by a composition-dependent rescaling factor, governed by the distance from $T_\text{g}$ (see also Figure S15).}
\add{\maw{Furthermore,} the relaxation times for PMMA deviate from classical Rouse scaling across all mode numbers and compositions, exhibiting an upturn at short sub-chain lengths. This anomalous slowdown likely reflects that the relatively stiff PMMA chains contain only a few effective Rouse beads, such that the shortest modes are dominated by local constraints rather than entropic relaxation.}
\add{This contrasts with Figure~\ref{figrouse}A, where PEO relaxation times \maw{scale differently with $N/p$ as a function of blend composition} yet converge for short sub-chains, indicating that PEO dynamics can recover neat-like behavior at sufficiently small length scales. In comparison, PMMA dynamics remain uniformly perturbed by PEO across all scales.}

%For the present systems, the Rouse bead size is estimated to be approximately one Kuhn length. To avoid artifacts from modes shorter than this length scale, we restrict the analysis to modes $p \leq 32$ for PEO and $p \leq 6$ for PMMA. Relaxation times for modes outside these ranges are shown as transparent data points. Within these constraints, the Rouse framework provides a consistent way to examine how collective dynamics evolve upon blending.In this study, we estimate that the Rouse bead size is approximately one Kuhn length for both PEO and PMMA.\cite{agapov2010}  To avoid confounding results with unexpected behaviors of the chains in the Rouse framework, we restrict the Rouse analysis to Rouse modes $p$ that are representative of subchains of at least the Rouse bead size. For PEO, the analysis is restricted to modes $p\leq32$ and for PMMA, the analysis is restricted to modes $p\leq6$. Relaxation times and stretching parameters for Rouse modes that fall outside of this range are shown as transparent data points. While classic Rouse scaling of $\tau_p \sim p^{-2}$ is only expected down to the Rouse bead size, which generally exceeds a single chemical monomer and is distinct from the Kuhn length,\cite{agapov2010} our analysis at the chemical monomer scale still provides a useful collective-variable framework. Here, the focus is not on confirming ideal Rouse behavior, but on leveraging Rouse coordinates to examine how relaxation times vary across different sub-chain length scales.

\subsection{Interplay Between Local Mobility and Relaxation Timescales} \label{sec:interplay}

At this stage, we remark on an apparent contradiction between the results shown in Figures~\ref{figrouse}A and \ref{figrouse}B and those in Figures~\ref{figMSFsp}C and \ref{figMSFsp}F. Figure~\ref{figMSFsp} indicates that PEO segmental mobility is suppressed even in the presence of a small number of nearby PMMA segments, whereas PMMA mobility is only notably affected when the local environment becomes significantly enriched in PEO ($\tilde{\phi}_i^\text{(PMMA)}$ < 0.5). In contrast, the Rouse mode analysis suggests a strong overall influence of PEO on PMMA \add{because relaxation times for PMMA do not converge to the neat reference, even for short sub-chains\maw{, whereas those for PEO do}.}

Two factors help reconcile this discrepancy. 
First, although PEO appears more affected in absolute terms in Figure~\ref{figMSFsp}, normalizing by the neat-polymer reference reveals that PMMA experiences a much larger relative enhancement in dynamics in PEO-rich environments (Figure S5). 
Second, there is also a fundamental distinction between the two measured quantities.
Figure~\ref{figMSFsp} relates to a segmental mobility, based on short-time fluctuations over a 100~ps interval, and Figure~\ref{figrouse} relates to segmental relaxation, which reflects structural decorrelation, which occurs at much longer timescales. 
Thus, while local dynamics may appear similar over short times,  this does not necessarily imply similar behavior in longer-time relaxation.
We suspect that results between these analyses would begin to align if the time interval for assessing segmental mobility was substantially increased.
This underscores a key nuance of dynamic heterogeneity, in that its effects depend on the timescale of the process being observed.

\section{Conclusions} \label{sec:conclusions}

\maw{The seemingly straightforward dependence of PMMA dynamics on a shifted $T_\text{g}$ in contrast to the composition-dependent dynamics of PEO in PEO/PMMA blends have been the subject of several prior works.\cite{sakai2004, liu2006, maranas2007, brodeck2010} 
Here, using atomistic molecular dynamics simulations, we extend these observations by systematically examining how overall and local composition influence segmental mobility and collective relaxation, across blend compositions, multiple length scales, and thermal regimes.
Analyses of both local fluctuations and sub-chain relaxation dynamics reveal a coherent physical picture linking free-volume effects and local compositional heterogeneity to deviations from neat-polymer behavior.}

Local environments were found to asymmetrically affect segmental mobilities. \maw{At short times, suppression by ``slow'' environments was stronger than enhancement by ``fast'' ones; nonetheless, PEO accelerated PMMA dynamics far more than PMMA suppressed PEO.} 
Accounting for $T_\text{g}$ differences, PEO segments \maw{in blends} exhibited enhanced dynamics relative to neat PEO, whereas PMMA segments \maw{in blends} remained comparable to $T_\text{g}$-equivalent neat PMMA systems. 
\maw{These effects were directly connected to variations in local composition. 
PEO dynamics increased with local free volume, which is heterogenously distributed when blending with PMMA, yielding a response that differs from mean-field expectations. 
Meanwhile, PMMA dynamics varied more uniformly with blend composition, as does the local compositional variations of PMMA segments.}

Rouse-mode analysis at 500~K revealed complementary signatures at collective length scales. \add{PEO relaxation approached neat-like behavior at short sub-chain lengths or in PEO-rich blends, whereas PMMA relaxation was uniformly accelerated across all modes with increasing PEO content. This behavior suggests a more idealized, composition-dependent response for PMMA compared to the strongly heterogeneous PEO dynamics.}
\add{In tandem with our observations on the local enhancement of PMMA dynamics in the vicinity of PEO, we suggest this arises from a nanoscale facilitation mechanism whereby locally softened PMMA near PEO propagates enhanced mobility to neighboring regions.}
\add{This interpretation resonates with the dynamic coupling framework proposed by Ngai and Roland,\cite{R:1993_Ngai_Chemical} in which mobility in the faster component propagates through intermolecular interactions to accelerate relaxation of the slower matrix. }

Overall, these findings provide a molecular-level framework for understanding dynamic asymmetry in polymer blends. 
\add{We note that our results pertain to amorphous systems of syndiotactic PMMA. Given the similar $T_\text{g}$ and segmental dynamics of atactic PMMA, we expect our findings to be broadly transferable to amorphous blends containing atactic PMMA as well. The same analytical approaches could be applied to systems with crystallinity or other tacticities to probe how packing and morphology shape asymmetric dynamics. Similar mechanisms are anticipated in other flexible–rigid polymer pairs such as PDMS/PMMA or PDMS/PS, offering a basis to test the generality of this physical picture. More broadly, the results suggest that the influence of one species emerges only at certain local compositions and evolves nonlinearly with environment. This insight may guide strategies for tuning viscoelasticity or transport by adjusting the composition and morphology of dynamically distinct phases.}

\section*{Acknowledgments}\label{sec:acknowledgement}
This research was supported by the National Science Foundation under Grant No. 2237470.
S.Z. acknowledges additional support by Grant No. 2022398 from the United States-Israel Binational Science Foundation (BSF).
Calculations were performed using resources from Princeton Research Computing at Princeton University, which is a consortium led by the Princeton Institute for Computational Science and Engineering (PICSciE) and Office of Information Technology's Research Computing.

\section*{Data availability}
The data underlying this study are available in the published article and its Supporting Information. Additional files facilitating reproduction by performing simulations are available at 
\url{https://github.com/webbtheosim/md-simulation-files/tree/main/2025-peo-pmma-blends}.

\section*{Supporting Information}
%The Supporting Information is available free of charge at \url{}. 

Calculation of apparent $T_\text{g}$; calculation of free volume; comparison of neat PEO and neat PMMA free volume; relative segmental mobility enhancement as a function of temperature; influence of local environment on normalized segmental mobility; influence of the distribution of local environment on segmental mobility; and Rouse mode analysis. Input files for performing simulations at various blend compositions.
%A list of sections. Section 2. Section 3. Section 4.

%%%%%%%%%%%%%%%%%%%%%%%%%%%%%%%%%%%%%%%%%%%%%%%%%%%%%%%%%%%%%%%%%%%%%
\bibliography{ref}

\providecommand{\latin}[1]{#1}
\makeatletter
\providecommand{\doi}
  {\begingroup\let\do\@makeother\dospecials
  \catcode`\{=1 \catcode`\}=2 \doi@aux}
\providecommand{\doi@aux}[1]{\endgroup\texttt{#1}}
\makeatother
\providecommand*\mcitethebibliography{\thebibliography}
\csname @ifundefined\endcsname{endmcitethebibliography}
  {\let\endmcitethebibliography\endthebibliography}{}
\begin{mcitethebibliography}{94}
\providecommand*\natexlab[1]{#1}
\providecommand*\mciteSetBstSublistMode[1]{}
\providecommand*\mciteSetBstMaxWidthForm[2]{}
\providecommand*\mciteBstWouldAddEndPuncttrue
  {\def\EndOfBibitem{\unskip.}}
\providecommand*\mciteBstWouldAddEndPunctfalse
  {\let\EndOfBibitem\relax}
\providecommand*\mciteSetBstMidEndSepPunct[3]{}
\providecommand*\mciteSetBstSublistLabelBeginEnd[3]{}
\providecommand*\EndOfBibitem{}
\mciteSetBstSublistMode{f}
\mciteSetBstMaxWidthForm{subitem}{(\alph{mcitesubitemcount})}
\mciteSetBstSublistLabelBeginEnd
  {\mcitemaxwidthsubitemform\space}
  {\relax}
  {\relax}

\bibitem[Colby(1989)]{colby_breakdown_1989}
Colby,~R.~H. Breakdown of time-temperature superposition in miscible polymer
  blends. \emph{Polymer} \textbf{1989}, \emph{30}, 1275--1278, DOI:
  \doi{10.1016/0032-3861(89)90048-7}\relax
\mciteBstWouldAddEndPuncttrue
\mciteSetBstMidEndSepPunct{\mcitedefaultmidpunct}
{\mcitedefaultendpunct}{\mcitedefaultseppunct}\relax
\EndOfBibitem
\bibitem[Roland and Ngai(1991)Roland, and Ngai]{roland_dynamical_1991}
Roland,~C.~M.; Ngai,~K.~L. Dynamical heterogeneity in a miscible polymer blend.
  \emph{Macromolecules} \textbf{1991}, \emph{24}, 2261--2265, DOI:
  \doi{10.1021/ma00009a021}, Publisher: American Chemical Society\relax
\mciteBstWouldAddEndPuncttrue
\mciteSetBstMidEndSepPunct{\mcitedefaultmidpunct}
{\mcitedefaultendpunct}{\mcitedefaultseppunct}\relax
\EndOfBibitem
\bibitem[Kamath \latin{et~al.}(2003)Kamath, Colby, and
  Kumar]{kamath_dynamic_2003}
Kamath,~S.; Colby,~R.~H.; Kumar,~S.~K. Dynamic {Heterogeneity} in {Miscible}
  {Polymer} {Blends} with {Stiffness} {Disparity}: {Computer} {Simulations}
  {Using} the {Bond} {Fluctuation} {Model}. \emph{Macromolecules}
  \textbf{2003}, \emph{36}, 8567--8573, DOI: \doi{10.1021/ma034682x},
  Publisher: American Chemical Society\relax
\mciteBstWouldAddEndPuncttrue
\mciteSetBstMidEndSepPunct{\mcitedefaultmidpunct}
{\mcitedefaultendpunct}{\mcitedefaultseppunct}\relax
\EndOfBibitem
\bibitem[Roland \latin{et~al.}(2006)Roland, McGrath, and
  Casalini]{roland_dynamic_2006}
Roland,~C.~M.; McGrath,~K.~J.; Casalini,~R. Dynamic {Heterogeneity} in
  {Poly}(vinyl methyl ether)/{Poly}(2-chlorostyrene) {Blends}.
  \emph{Macromolecules} \textbf{2006}, \emph{39}, 3581--3587, DOI:
  \doi{10.1021/ma060315k}\relax
\mciteBstWouldAddEndPuncttrue
\mciteSetBstMidEndSepPunct{\mcitedefaultmidpunct}
{\mcitedefaultendpunct}{\mcitedefaultseppunct}\relax
\EndOfBibitem
\bibitem[Sharma and Green(2017)Sharma, and Green]{sharma2017}
Sharma,~R.~P.; Green,~P.~F. Component Dynamics in Polymer/Polymer Blends: Role
  of Spatial Compositional Heterogeneity. \emph{Macromolecules} \textbf{2017},
  \emph{50}, 6617--6630, DOI: \doi{10.1021/acs.macromol.7b00092}\relax
\mciteBstWouldAddEndPuncttrue
\mciteSetBstMidEndSepPunct{\mcitedefaultmidpunct}
{\mcitedefaultendpunct}{\mcitedefaultseppunct}\relax
\EndOfBibitem
\bibitem[Zhang \latin{et~al.}(2020)Zhang, Rocha, Lu, Yuan, Uji-i, Floudas,
  Müllen, Xiao, Hofkens, and Debroye]{zhang_spatially_2020}
Zhang,~G.; Rocha,~S.; Lu,~G.; Yuan,~H.; Uji-i,~H.; Floudas,~G.~A.; Müllen,~K.;
  Xiao,~L.; Hofkens,~J.; Debroye,~E. Spatially and {Temporally} {Resolved}
  {Heterogeneities} in a {Miscible} {Polymer} {Blend}. \emph{ACS Omega}
  \textbf{2020}, \emph{5}, 23931--23939, DOI:
  \doi{10.1021/acsomega.0c03173}\relax
\mciteBstWouldAddEndPuncttrue
\mciteSetBstMidEndSepPunct{\mcitedefaultmidpunct}
{\mcitedefaultendpunct}{\mcitedefaultseppunct}\relax
\EndOfBibitem
\bibitem[Bennemann \latin{et~al.}(1999)Bennemann, Donati, Baschnagel, and
  Glotzer]{bennemann1999}
Bennemann,~C.; Donati,~C.; Baschnagel,~J.; Glotzer,~S.~C. Growing range of
  correlated motion in a polymer melt on cooling towards the glass transition.
  \emph{Nature} \textbf{1999}, \emph{399}, 246--249, DOI:
  \doi{10.1038/20406}\relax
\mciteBstWouldAddEndPuncttrue
\mciteSetBstMidEndSepPunct{\mcitedefaultmidpunct}
{\mcitedefaultendpunct}{\mcitedefaultseppunct}\relax
\EndOfBibitem
\bibitem[Phan and Schweizer(2018)Phan, and Schweizer]{phan2018}
Phan,~A.~D.; Schweizer,~K.~S. Elastically Collective Nonlinear Langevin
  Equation Theory of Glass-Forming Liquids: Transient Localization,
  Thermodynamic Mapping, and Cooperativity. \emph{The Journal of Physical
  Chemistry B} \textbf{2018}, \emph{122}, 8451--8461, DOI:
  \doi{10.1021/acs.jpcb.8b04975}\relax
\mciteBstWouldAddEndPuncttrue
\mciteSetBstMidEndSepPunct{\mcitedefaultmidpunct}
{\mcitedefaultendpunct}{\mcitedefaultseppunct}\relax
\EndOfBibitem
\bibitem[Baker \latin{et~al.}(2022)Baker, Reynolds, Masurel, Olmsted, and
  Mattsson]{baker2022}
Baker,~D.~L.; Reynolds,~M.; Masurel,~R.; Olmsted,~P.~D.; Mattsson,~J.
  Cooperative Intramolecular Dynamics Control the Chain-Length-Dependent Glass
  Transition in Polymers. \emph{Physical Review X} \textbf{2022}, \emph{12},
  021047, DOI: \doi{10.1103/PhysRevX.12.021047}\relax
\mciteBstWouldAddEndPuncttrue
\mciteSetBstMidEndSepPunct{\mcitedefaultmidpunct}
{\mcitedefaultendpunct}{\mcitedefaultseppunct}\relax
\EndOfBibitem
\bibitem[Ghanekarade and Simmons(2023)Ghanekarade, and
  Simmons]{ghanekarade2023}
Ghanekarade,~A.; Simmons,~D.~S. Combined Mixing and Dynamical Origins of Tg
  Alterations Near Polymer–Polymer Interfaces. \emph{Macromolecules}
  \textbf{2023}, \emph{56}, 379--392, DOI:
  \doi{10.1021/acs.macromol.2c01621}\relax
\mciteBstWouldAddEndPuncttrue
\mciteSetBstMidEndSepPunct{\mcitedefaultmidpunct}
{\mcitedefaultendpunct}{\mcitedefaultseppunct}\relax
\EndOfBibitem
\bibitem[Fredrickson(1996)]{fredrickson1996}
Fredrickson,~G.~H. The theory of polymer dynamics. \emph{Current Opinion in
  Solid State \& Materials Science} \textbf{1996}, \emph{1}, 812--816\relax
\mciteBstWouldAddEndPuncttrue
\mciteSetBstMidEndSepPunct{\mcitedefaultmidpunct}
{\mcitedefaultendpunct}{\mcitedefaultseppunct}\relax
\EndOfBibitem
\bibitem[Semenov \latin{et~al.}(1995)Semenov, Joanny, and
  Khokhlov]{semenov_associating_1995}
Semenov,~A.~N.; Joanny,~J.-F.; Khokhlov,~A.~R. Associating polymers:
  {Equilibrium} and linear viscoelasticity. \emph{Macromolecules}
  \textbf{1995}, \emph{28}, 1066--1075, DOI: \doi{10.1021/ma00108a038},
  Publisher: American Chemical Society\relax
\mciteBstWouldAddEndPuncttrue
\mciteSetBstMidEndSepPunct{\mcitedefaultmidpunct}
{\mcitedefaultendpunct}{\mcitedefaultseppunct}\relax
\EndOfBibitem
\bibitem[Tanaka \latin{et~al.}(2000)Tanaka, Takahara, and
  Kajiyama]{tanaka_rheological_2000}
Tanaka,~K.; Takahara,~A.; Kajiyama,~T. Rheological {Analysis} of {Surface}
  {Relaxation} {Process} of {Monodisperse} {Polystyrene} {Films}.
  \emph{Macromolecules} \textbf{2000}, \emph{33}, 7588--7593, DOI:
  \doi{10.1021/ma000406w}, Publisher: American Chemical Society\relax
\mciteBstWouldAddEndPuncttrue
\mciteSetBstMidEndSepPunct{\mcitedefaultmidpunct}
{\mcitedefaultendpunct}{\mcitedefaultseppunct}\relax
\EndOfBibitem
\bibitem[Mogurampelly \latin{et~al.}(2016)Mogurampelly, Sethuraman, Pryamitsyn,
  and Ganesan]{mogurampelly2016}
Mogurampelly,~S.; Sethuraman,~V.; Pryamitsyn,~V.; Ganesan,~V. Influence of
  nanoparticle-ion and nanoparticle-polymer interactions on ion transport and
  viscoelastic properties of polymer electrolytes. \emph{The Journal of
  Chemical Physics} \textbf{2016}, \emph{144}, 154905, DOI:
  \doi{10.1063/1.4946047}\relax
\mciteBstWouldAddEndPuncttrue
\mciteSetBstMidEndSepPunct{\mcitedefaultmidpunct}
{\mcitedefaultendpunct}{\mcitedefaultseppunct}\relax
\EndOfBibitem
\bibitem[Zhang \latin{et~al.}(2022)Zhang, Moore, Zhao, Govind, Wolf, Jin,
  Walsh, Riggleman, and Fakhraai]{zhang_role_2022}
Zhang,~A.; Moore,~A.~R.; Zhao,~H.; Govind,~S.; Wolf,~S.~E.; Jin,~Y.;
  Walsh,~P.~J.; Riggleman,~R.~A.; Fakhraai,~Z. The role of intramolecular
  relaxations on the structure and stability of vapor-deposited glasses.
  \emph{The Journal of Chemical Physics} \textbf{2022}, \emph{156}, 244703,
  DOI: \doi{10.1063/5.0087600}\relax
\mciteBstWouldAddEndPuncttrue
\mciteSetBstMidEndSepPunct{\mcitedefaultmidpunct}
{\mcitedefaultendpunct}{\mcitedefaultseppunct}\relax
\EndOfBibitem
\bibitem[Fenton \latin{et~al.}(1973)Fenton, Parker, and Wright]{fenton1973}
Fenton,~D.~E.; Parker,~J.~M.; Wright,~P.~V. Complexes of alkali metal ions with
  poly(ethylene oxide). \emph{Polymer} \textbf{1973}, \emph{14}, 589, DOI:
  \doi{10.1016/0032-3861(73)90146-8}\relax
\mciteBstWouldAddEndPuncttrue
\mciteSetBstMidEndSepPunct{\mcitedefaultmidpunct}
{\mcitedefaultendpunct}{\mcitedefaultseppunct}\relax
\EndOfBibitem
\bibitem[Berthier \latin{et~al.}(1983)Berthier, Gorecki, Minier, Armand,
  Chabagno, and Rigaud]{berthier1983}
Berthier,~C.; Gorecki,~W.; Minier,~M.; Armand,~M.~B.; Chabagno,~J.~M.;
  Rigaud,~P. Microscopic investigation of ionic conductivity in alkali metal
  salts-poly(ethylene oxide) adducts. \emph{Solid State Ionics} \textbf{1983},
  \emph{11}, 91--95, DOI: \doi{10.1016/0167-2738(83)90068-1}\relax
\mciteBstWouldAddEndPuncttrue
\mciteSetBstMidEndSepPunct{\mcitedefaultmidpunct}
{\mcitedefaultendpunct}{\mcitedefaultseppunct}\relax
\EndOfBibitem
\bibitem[Borodin and Smith(2006)Borodin, and Smith]{borodin2006}
Borodin,~O.; Smith,~G.~D. Mechanism of Ion Transport in Amorphous Poly(ethylene
  oxide)/LiTFSI from Molecular Dynamics Simulations. \emph{Macromolecules}
  \textbf{2006}, \emph{39}, 1620--1629, DOI:
  \doi{https://doi.org/10.1021/ma052277v}\relax
\mciteBstWouldAddEndPuncttrue
\mciteSetBstMidEndSepPunct{\mcitedefaultmidpunct}
{\mcitedefaultendpunct}{\mcitedefaultseppunct}\relax
\EndOfBibitem
\bibitem[Webb \latin{et~al.}(2018)Webb, Yamamoto, Savoie, Wang, and
  Miller]{webb2018}
Webb,~M.~A.; Yamamoto,~U.; Savoie,~B.~M.; Wang,~Z.-G.; Miller,~T. F.~I.
  Globally Suppressed Dynamics in Ion-Doped Polymers. \emph{ACS Macro Letters}
  \textbf{2018}, \emph{7}, 734--738, DOI:
  \doi{10.1021/acsmacrolett.8b00237}\relax
\mciteBstWouldAddEndPuncttrue
\mciteSetBstMidEndSepPunct{\mcitedefaultmidpunct}
{\mcitedefaultendpunct}{\mcitedefaultseppunct}\relax
\EndOfBibitem
\bibitem[Chu \latin{et~al.}(2020)Chu, Webb, Deng, Col\'on, Kambe, Krishnan,
  Nealey, and de~Pablo]{chu2020}
Chu,~W.; Webb,~M.~A.; Deng,~C.; Col\'on,~Y.~J.; Kambe,~Y.; Krishnan,~S.;
  Nealey,~P.~F.; de~Pablo,~J.~J. Understanding Ion Mobility in
  P2VP/NMP\textsuperscript{+}I\textsuperscript{-} Polymer Electrolytes: A
  Combined Simulation and Experimental Study. \emph{Macromolecules}
  \textbf{2020}, \emph{53}, 2783--2792, DOI:
  \doi{10.1021/acs.macromol.9b02329}\relax
\mciteBstWouldAddEndPuncttrue
\mciteSetBstMidEndSepPunct{\mcitedefaultmidpunct}
{\mcitedefaultendpunct}{\mcitedefaultseppunct}\relax
\EndOfBibitem
\bibitem[Deng \latin{et~al.}(2021)Deng, Webb, Bennington, Sharon, Nealey,
  Patel, and de~Pablo]{deng2021}
Deng,~C.; Webb,~M.~A.; Bennington,~P.; Sharon,~D.; Nealey,~P.~F.; Patel,~S.~N.;
  de~Pablo,~J.~J. Role of Molecular Architecture on Ion Transport in Ethylene
  oxide-Based Polymer Electrolytes. \emph{Macromolecules} \textbf{2021},
  \emph{54}, 2266--2276, DOI: \doi{10.1021/acs.macromol.0c02424}\relax
\mciteBstWouldAddEndPuncttrue
\mciteSetBstMidEndSepPunct{\mcitedefaultmidpunct}
{\mcitedefaultendpunct}{\mcitedefaultseppunct}\relax
\EndOfBibitem
\bibitem[Shah \latin{et~al.}(2023)Shah, He, Gao, and Balsara]{shah2023}
Shah,~N.~J.; He,~L.; Gao,~K.~W.; Balsara,~N.~P. Thermodynamics and Phase
  Behavior of Poly(ethylene oxide)/Poly(methyl methacrylate)/Salt Blend
  Electrolytes Studied by Small-Angle Neutron Scattering. \emph{Macromolecules}
  \textbf{2023}, \emph{56}, 2889--2898, DOI:
  \doi{https://doi.org/10.1021/acs.macromol.2c02533}\relax
\mciteBstWouldAddEndPuncttrue
\mciteSetBstMidEndSepPunct{\mcitedefaultmidpunct}
{\mcitedefaultendpunct}{\mcitedefaultseppunct}\relax
\EndOfBibitem
\bibitem[Mu \latin{et~al.}(2008)Mu, Huang, Lu, and Sun]{mu2008}
Mu,~D.; Huang,~X.-R.; Lu,~Z.-Y.; Sun,~C.-C. Computer simulation study on the
  compatibility of poly(ethylene oxide)/poly(methyl methacrylate) blends.
  \emph{Chemical Physics} \textbf{2008}, \emph{348}, 122--129, DOI:
  \doi{10.1016/j.chemphys.2008.03.015}\relax
\mciteBstWouldAddEndPuncttrue
\mciteSetBstMidEndSepPunct{\mcitedefaultmidpunct}
{\mcitedefaultendpunct}{\mcitedefaultseppunct}\relax
\EndOfBibitem
\bibitem[Schantz(1997)]{schantz1997}
Schantz,~S. Structure and Mobility in Poly(ethylene oxide)/Poly(methyl
  methacrylate) Blends Investigated by 13C Solid-State NMR.
  \emph{Macromolecules} \textbf{1997}, \emph{30}, 1419--1425, DOI:
  \doi{10.1021/ma961538l}\relax
\mciteBstWouldAddEndPuncttrue
\mciteSetBstMidEndSepPunct{\mcitedefaultmidpunct}
{\mcitedefaultendpunct}{\mcitedefaultseppunct}\relax
\EndOfBibitem
\bibitem[Cimmino \latin{et~al.}(1989)Cimmino, Martuscelli, and
  Silvestre]{cimmino1989}
Cimmino,~S.; Martuscelli,~E.; Silvestre,~C. Miscibility prediction based on the
  corresponding states theory: poly(ethylene oxide)/atactic poly(methyl
  methacrylate) system. \emph{Polymer} \textbf{1989}, \emph{30}, 393--398, DOI:
  \doi{https://doi.org/10.1016/0032-3861(89)90003-7}\relax
\mciteBstWouldAddEndPuncttrue
\mciteSetBstMidEndSepPunct{\mcitedefaultmidpunct}
{\mcitedefaultendpunct}{\mcitedefaultseppunct}\relax
\EndOfBibitem
\bibitem[Lartigue \latin{et~al.}(1997)Lartigue, Guillermo, and
  Cohen-Addad]{lartigue1997}
Lartigue,~C.; Guillermo,~A.; Cohen-Addad,~J.~P. Proton NMR investigation of the
  local dynamics in {PEO/PMMA} blends. \emph{Journal of Polymer Science Part B:
  Polymer Physics} \textbf{1997}, \emph{35}, 1059--1105, DOI:
  \doi{10.1002/(SICI)1099-0488(199705)35:7<1095::AID-POLB8>3.0.CO;2-V}\relax
\mciteBstWouldAddEndPuncttrue
\mciteSetBstMidEndSepPunct{\mcitedefaultmidpunct}
{\mcitedefaultendpunct}{\mcitedefaultseppunct}\relax
\EndOfBibitem
\bibitem[Silva \latin{et~al.}(2000)Silva, Machado, Song, and
  Hourston]{silva_nanoheterogeneities_2000}
Silva,~G.~G.; Machado,~J.~C.; Song,~M.; Hourston,~D.~J. Nanoheterogeneities in
  {PEO}/{PMMA} blends: {A} modulated differential scanning calorimetry
  approach. \emph{Journal of Applied Polymer Science} \textbf{2000}, \emph{77},
  2034--2043, DOI:
  \doi{10.1002/1097-4628(20000829)77:9<2034::AID-APP20>3.0.CO;2-Q}, \_eprint:
  https://onlinelibrary.wiley.com/doi/pdf/10.1002/1097-4628\%2820000829\%2977\%3A9\%3C2034\%3A\%3AAID-APP20\%3E3.0.CO\%3B2-Q\relax
\mciteBstWouldAddEndPuncttrue
\mciteSetBstMidEndSepPunct{\mcitedefaultmidpunct}
{\mcitedefaultendpunct}{\mcitedefaultseppunct}\relax
\EndOfBibitem
\bibitem[Habasaki(2022)]{R:2022_Habasaki_Atomistic}
Habasaki,~J. Atomistic molecular dynamics in polyethylene oxide and polymethyl
  methacrylate blends having significantly different glass transition
  temperatures. \emph{International Journal of Applied Glass Science}
  \textbf{2022}, \emph{13}, 347--358, DOI: \doi{10.1111/ijag.16553}\relax
\mciteBstWouldAddEndPuncttrue
\mciteSetBstMidEndSepPunct{\mcitedefaultmidpunct}
{\mcitedefaultendpunct}{\mcitedefaultseppunct}\relax
\EndOfBibitem
\bibitem[Lutz \latin{et~al.}(2003)Lutz, He, Ediger, Cao, Lin, and
  Jones]{lutz2003}
Lutz,~T.~R.; He,~Y.; Ediger,~M.~D.; Cao,~H.; Lin,~G.; Jones,~A.~A. Rapid
  {P}oly(ethylene oxide) {S}egmental {D}ynamics in {B}lends with {P}oly(methyl
  methacrylate). \emph{Macromolecules} \textbf{2003}, \emph{36}, 1724--1730,
  DOI: \doi{10.1021/ma021634o}\relax
\mciteBstWouldAddEndPuncttrue
\mciteSetBstMidEndSepPunct{\mcitedefaultmidpunct}
{\mcitedefaultendpunct}{\mcitedefaultseppunct}\relax
\EndOfBibitem
\bibitem[Ghelichi \latin{et~al.}(2013)Ghelichi, Qazvini, Jafari, Khonakdar,
  Farajollahi, and Scheffler]{ghelichi2013}
Ghelichi,~M.; Qazvini,~N.~T.; Jafari,~S.~H.; Khonakdar,~H.~A.; Farajollahi,~Y.;
  Scheffler,~C. Conformational, thermal, and ionic conductivity behavior of PEO
  in PEO/PMMA miscible blend: Investigating the effect of lithium salt.
  \emph{Journal of Applied Polymer Science} \textbf{2013}, \emph{129},
  1868--1874, DOI: \doi{10.1002/app.38897}\relax
\mciteBstWouldAddEndPuncttrue
\mciteSetBstMidEndSepPunct{\mcitedefaultmidpunct}
{\mcitedefaultendpunct}{\mcitedefaultseppunct}\relax
\EndOfBibitem
\bibitem[Choudhary(2018)]{choudhary2018}
Choudhary,~S. Effects of amorphous silica nanoparticles and polymer blend
  compositions on the structural, thermal and dielectric properties of
  PEO–PMMA blend based polymer nanocomposites. \emph{Journal of Polymer
  Research} \textbf{2018}, \emph{25}, 116, DOI:
  \doi{10.1007/s10965-018-1510-x}\relax
\mciteBstWouldAddEndPuncttrue
\mciteSetBstMidEndSepPunct{\mcitedefaultmidpunct}
{\mcitedefaultendpunct}{\mcitedefaultseppunct}\relax
\EndOfBibitem
\bibitem[Lim \latin{et~al.}(2018)Lim, Jung, and Hwang]{lim2018}
Lim,~Y.~S.; Jung,~H.-A.; Hwang,~H. Fabrication of PEO-PMMA-LiClO4-Based Solid
  Polymer Electrolytes Containing Silica Aerogel Particles for All-Solid-State
  Lithium Batteries. \emph{Energies} \textbf{2018}, \emph{11}, 2559, DOI:
  \doi{10.3390/en11102559}\relax
\mciteBstWouldAddEndPuncttrue
\mciteSetBstMidEndSepPunct{\mcitedefaultmidpunct}
{\mcitedefaultendpunct}{\mcitedefaultseppunct}\relax
\EndOfBibitem
\bibitem[Sharon \latin{et~al.}(2022)Sharon, Deng, Bennington, Webb, Patel,
  de~Pablo, and Nealey]{sharon2022}
Sharon,~D.; Deng,~C.; Bennington,~P.; Webb,~M.~A.; Patel,~S.~N.;
  de~Pablo,~J.~J.; Nealey,~P.~F. Critical Percolation Threshold for
  Solvation-Site Connectivity in Polymer Electrolyte Mixtures.
  \emph{Macromolecules} \textbf{2022}, \emph{55}, 7212--7221, DOI:
  \doi{10.1021/acs.macromol.2c00988}\relax
\mciteBstWouldAddEndPuncttrue
\mciteSetBstMidEndSepPunct{\mcitedefaultmidpunct}
{\mcitedefaultendpunct}{\mcitedefaultseppunct}\relax
\EndOfBibitem
\bibitem[Bakar \latin{et~al.}(2022)Bakar, Darvishi, Li, Han, Aydemir,
  Nizamoglu, Hong, and Senses]{bakar2022}
Bakar,~R.; Darvishi,~S.; Li,~T.; Han,~M.; Aydemir,~U.; Nizamoglu,~S.; Hong,~K.;
  Senses,~E. Effect of Polymer Topology on Microstructure, Segmental Dynamics,
  and Ionic Conductivity in PEO/PMMA-Based Solid Polymer Electrolytes.
  \emph{ACS Appl. Polym. Mater.} \textbf{2022}, \emph{4}, 179--190, DOI:
  \doi{10.1021/acsapm.1c01178}\relax
\mciteBstWouldAddEndPuncttrue
\mciteSetBstMidEndSepPunct{\mcitedefaultmidpunct}
{\mcitedefaultendpunct}{\mcitedefaultseppunct}\relax
\EndOfBibitem
\bibitem[Diddens \latin{et~al.}(2010)Diddens, Heuer, and
  Borodin]{R:2010_Diddens_Understanding}
Diddens,~D.; Heuer,~A.; Borodin,~O. Understanding the Lithium Transport within
  a Rouse-Based Model for a {PEO}/{LiTFSI} Polymer Electrolyte.
  \emph{Macromolecules} \textbf{2010}, \emph{43}, 2028--2036, DOI:
  \doi{10.1021/ma901893h}\relax
\mciteBstWouldAddEndPuncttrue
\mciteSetBstMidEndSepPunct{\mcitedefaultmidpunct}
{\mcitedefaultendpunct}{\mcitedefaultseppunct}\relax
\EndOfBibitem
\bibitem[Deng \latin{et~al.}(2021)Deng, Webb, Bennington, Sharon, Nealey,
  Patel, and de~Pablo]{R:2021_Deng_Role}
Deng,~C.; Webb,~M.~A.; Bennington,~P.; Sharon,~D.; Nealey,~P.~F.; Patel,~S.~N.;
  de~Pablo,~J.~J. Role of Molecular Architecture on Ion Transport in Ethylene
  oxide-Based Polymer Electrolytes. \emph{Macromolecules} \textbf{2021},
  \emph{54}, 2266--2276, DOI: \doi{10.1021/acs.macromol.0c02424}\relax
\mciteBstWouldAddEndPuncttrue
\mciteSetBstMidEndSepPunct{\mcitedefaultmidpunct}
{\mcitedefaultendpunct}{\mcitedefaultseppunct}\relax
\EndOfBibitem
\bibitem[Webb \latin{et~al.}(2015)Webb, Savoie, Wang, and {Miller
  III}]{R:2015_Webb_Chemically_Macromolecules}
Webb,~M.~A.; Savoie,~B.~M.; Wang,~Z.-G.; {Miller III},~T.~F. Chemically
  Specific Dynamic Bond Percolation Model for Ion Transport in Polymer
  Electrolytes. \emph{Macromolecules} \textbf{2015}, \emph{48}, 7346--7358,
  DOI: \doi{10.1021/acs.macromol.5b01437}\relax
\mciteBstWouldAddEndPuncttrue
\mciteSetBstMidEndSepPunct{\mcitedefaultmidpunct}
{\mcitedefaultendpunct}{\mcitedefaultseppunct}\relax
\EndOfBibitem
\bibitem[Niedzwiedz \latin{et~al.}(2007)Niedzwiedz, Wischnewski, Monkenbusch,
  Richter, Genix, Arbe, Colmenero, Strauch, and Straube]{niedzwiedz2007}
Niedzwiedz,~K.; Wischnewski,~A.; Monkenbusch,~M.; Richter,~D.; Genix,~A.-C.;
  Arbe,~A.; Colmenero,~J.; Strauch,~M.; Straube,~E. Polymer Chain Dynamics in a
  Random Environment: Heterogeneous Mobilities. \emph{Physical Review Letters}
  \textbf{2007}, \emph{98}, 168301, DOI:
  \doi{10.1103/PhysRevLett.98.168301}\relax
\mciteBstWouldAddEndPuncttrue
\mciteSetBstMidEndSepPunct{\mcitedefaultmidpunct}
{\mcitedefaultendpunct}{\mcitedefaultseppunct}\relax
\EndOfBibitem
\bibitem[García~Sakai \latin{et~al.}(2008)García~Sakai, Maranas, Peral, and
  Copley]{garcia_sakai_dynamics_2008}
García~Sakai,~V.; Maranas,~J.~K.; Peral,~I.; Copley,~J. R.~D. Dynamics of
  {PEO} in {Blends} with {PMMA}: {Study} of the {Effects} of {Blend}
  {Composition} via {Quasi}-{Elastic} {Neutron} {Scattering}.
  \emph{Macromolecules} \textbf{2008}, \emph{41}, 3701--3710, DOI:
  \doi{10.1021/ma0714870}, Publisher: American Chemical Society\relax
\mciteBstWouldAddEndPuncttrue
\mciteSetBstMidEndSepPunct{\mcitedefaultmidpunct}
{\mcitedefaultendpunct}{\mcitedefaultseppunct}\relax
\EndOfBibitem
\bibitem[Brodeck \latin{et~al.}(2010)Brodeck, Alvarez, Moreno, Colmenero, and
  Richter]{brodeck2010}
Brodeck,~M.; Alvarez,~F.; Moreno,~A.~J.; Colmenero,~J.; Richter,~D. Chain
  Motion in Nonentangled Dynamically Asymmetric Polymer Blends: Comparison
  between Atomistic Simulations of PEO/PMMA and a Generic Bead--Spring Model.
  \emph{Macromolecules} \textbf{2010}, \emph{43}, 3036--3051, DOI:
  \doi{10.1021/ma902820a}\relax
\mciteBstWouldAddEndPuncttrue
\mciteSetBstMidEndSepPunct{\mcitedefaultmidpunct}
{\mcitedefaultendpunct}{\mcitedefaultseppunct}\relax
\EndOfBibitem
\bibitem[Sakai \latin{et~al.}(2004)Sakai, Chen, Maranas, and
  Chowdhuri]{sakai2004}
Sakai,~V.~G.; Chen,~C.; Maranas,~J.~K.; Chowdhuri,~Z. Effect of Blending with
  Poly(ethylene oxide) on the Dynamics of Poly(methyl methacrylate): A
  Quasi-Elastic Neutron Scattering Approach. \emph{Macromolecules}
  \textbf{2004}, \emph{37}, 9975--9983, DOI: \doi{10.1021/ma0497355}\relax
\mciteBstWouldAddEndPuncttrue
\mciteSetBstMidEndSepPunct{\mcitedefaultmidpunct}
{\mcitedefaultendpunct}{\mcitedefaultseppunct}\relax
\EndOfBibitem
\bibitem[Liu \latin{et~al.}(2006)Liu, Sakai, and Maranas]{liu2006}
Liu,~J.; Sakai,~V.~G.; Maranas,~J.~K. Composition Dependence of Segmental
  Dynamics of Poly(methyl methacrylate) in Miscible Blends with Poly(ethylene
  oxide). \emph{Macromolecules} \textbf{2006}, \emph{39}, 2866--2874, DOI:
  \doi{10.1021/ma052136t}\relax
\mciteBstWouldAddEndPuncttrue
\mciteSetBstMidEndSepPunct{\mcitedefaultmidpunct}
{\mcitedefaultendpunct}{\mcitedefaultseppunct}\relax
\EndOfBibitem
\bibitem[Maranas(2007)]{maranas2007}
Maranas,~J.~K. The effect of environment on local dynamics of macromolecules.
  \emph{Current Opinion in Colloid \& Interface Science} \textbf{2007},
  \emph{12}, 29--42, DOI: \doi{10.1016/j.cocis.2007.03.003}\relax
\mciteBstWouldAddEndPuncttrue
\mciteSetBstMidEndSepPunct{\mcitedefaultmidpunct}
{\mcitedefaultendpunct}{\mcitedefaultseppunct}\relax
\EndOfBibitem
\bibitem[Genix \latin{et~al.}(2005)Genix, Arbe, Alvarez, Colmenero, Willner,
  and Richter]{genix2005}
Genix,~A.-C.; Arbe,~A.; Alvarez,~F.; Colmenero,~J.; Willner,~L.; Richter,~D.
  Dynamics of poly(ethylene oxide) in a blend with poly(methyl methacrylate): A
  quasielastic neutron scattering and molecular dynamics simulations study.
  \emph{Physical Review E} \textbf{2005}, \emph{72}, 031808, DOI:
  \doi{10.1103/PhysRevE.72.031808}\relax
\mciteBstWouldAddEndPuncttrue
\mciteSetBstMidEndSepPunct{\mcitedefaultmidpunct}
{\mcitedefaultendpunct}{\mcitedefaultseppunct}\relax
\EndOfBibitem
\bibitem[Sakai \latin{et~al.}(2005)Sakai, Maranas, Chowdhuri, Peral, and
  Copley]{sakai2005}
Sakai,~V.~G.; Maranas,~J.~K.; Chowdhuri,~Z.; Peral,~I.; Copley,~J. R.~D.
  Miscible blend dynamics and the length scale of local compositions.
  \emph{Journal of Polymer Science Part B: Polymer Physics} \textbf{2005},
  \emph{43}, 2914--2923, DOI: \doi{10.1002/polb.20562}\relax
\mciteBstWouldAddEndPuncttrue
\mciteSetBstMidEndSepPunct{\mcitedefaultmidpunct}
{\mcitedefaultendpunct}{\mcitedefaultseppunct}\relax
\EndOfBibitem
\bibitem[Farago \latin{et~al.}(2005)Farago, Chen, Maranas, Kamath, Colby,
  Pasquale, and Long]{farago_collective_2005}
Farago,~B.; Chen,~C.; Maranas,~J.~K.; Kamath,~S.; Colby,~R.~H.;
  Pasquale,~A.~J.; Long,~T.~E. Collective motion in {Poly}(ethylene
  oxide)/poly(methylmethacrylate) blends. \emph{Physical Review E}
  \textbf{2005}, \emph{72}, 031809, DOI: \doi{10.1103/PhysRevE.72.031809},
  Publisher: American Physical Society\relax
\mciteBstWouldAddEndPuncttrue
\mciteSetBstMidEndSepPunct{\mcitedefaultmidpunct}
{\mcitedefaultendpunct}{\mcitedefaultseppunct}\relax
\EndOfBibitem
\bibitem[Zawada \latin{et~al.}(1992)Zawada, Ylitalo, Fuller, Colby, and
  Long]{zawada_component_1992}
Zawada,~J.~A.; Ylitalo,~C.~M.; Fuller,~G.~G.; Colby,~R.~H.; Long,~T.~E.
  Component relaxation dynamics in a miscible polymer blend: poly(ethylene
  oxide)/poly(methyl methacrylate). \emph{Macromolecules} \textbf{1992},
  \emph{25}, 2896--2902, DOI: \doi{10.1021/ma00037a017}\relax
\mciteBstWouldAddEndPuncttrue
\mciteSetBstMidEndSepPunct{\mcitedefaultmidpunct}
{\mcitedefaultendpunct}{\mcitedefaultseppunct}\relax
\EndOfBibitem
\bibitem[Lodge and McLeish(2000)Lodge, and McLeish]{lodge2000}
Lodge,~T.~P.; McLeish,~T. C.~B. Self-Concentrations and Effective Glass
  Transition Temperatures in Polymer Blends. \emph{Macromolecules}
  \textbf{2000}, \emph{33}, 5278--5284, DOI:
  \doi{https://doi.org/10.1021/ma9921706}\relax
\mciteBstWouldAddEndPuncttrue
\mciteSetBstMidEndSepPunct{\mcitedefaultmidpunct}
{\mcitedefaultendpunct}{\mcitedefaultseppunct}\relax
\EndOfBibitem
\bibitem[He \latin{et~al.}(2003)He, Lutz, and Ediger]{he_segmental_2003}
He,~Y.; Lutz,~T.~R.; Ediger,~M.~D. Segmental and terminal dynamics in miscible
  polymer mixtures: {Tests} of the {Lodge}–{McLeish} model. \emph{The Journal
  of Chemical Physics} \textbf{2003}, \emph{119}, 9956--9965, DOI:
  \doi{10.1063/1.1615963}\relax
\mciteBstWouldAddEndPuncttrue
\mciteSetBstMidEndSepPunct{\mcitedefaultmidpunct}
{\mcitedefaultendpunct}{\mcitedefaultseppunct}\relax
\EndOfBibitem
\bibitem[Kumar \latin{et~al.}(2007)Kumar, Shenogin, and Colby]{kumar2007}
Kumar,~S.~K.; Shenogin,~S.; Colby,~R.~H. Dynamics of Miscible Polymer Blends:
  Role of Concentration Fluctuations on Characteristic Segmental Relaxation
  Times. \emph{Macromolecules} \textbf{2007}, \emph{40}, 5759--5766, DOI:
  \doi{https://doi.org/10.1021/ma070502y}\relax
\mciteBstWouldAddEndPuncttrue
\mciteSetBstMidEndSepPunct{\mcitedefaultmidpunct}
{\mcitedefaultendpunct}{\mcitedefaultseppunct}\relax
\EndOfBibitem
\bibitem[Kant \latin{et~al.}(2003)Kant, Kumar, and Colby]{kant2003}
Kant,~R.; Kumar,~S.~K.; Colby,~R.~H. What Length Scales Control the Dynamics of
  Miscible Polymer Blends? \emph{Macromolecules} \textbf{2003}, \emph{36},
  10087--10094, DOI: \doi{https://doi.org/10.1021/ma0347215}\relax
\mciteBstWouldAddEndPuncttrue
\mciteSetBstMidEndSepPunct{\mcitedefaultmidpunct}
{\mcitedefaultendpunct}{\mcitedefaultseppunct}\relax
\EndOfBibitem
\bibitem[Shenogin \latin{et~al.}(2007)Shenogin, Kant, Colby, and
  Kumar]{shenogin2007}
Shenogin,~S.; Kant,~R.; Colby,~R.~H.; Kumar,~S.~K. Dynamics of Miscible Polymer
  Blends: Predicting the Dielectric Response. \emph{Macromolecules}
  \textbf{2007}, \emph{40}, 5767--5775, DOI:
  \doi{https://doi.org/10.1021/ma070503q}\relax
\mciteBstWouldAddEndPuncttrue
\mciteSetBstMidEndSepPunct{\mcitedefaultmidpunct}
{\mcitedefaultendpunct}{\mcitedefaultseppunct}\relax
\EndOfBibitem
\bibitem[Colby and Lipson(2005)Colby, and Lipson]{colby2005}
Colby,~R.~H.; Lipson,~J. E.~G. Modeling the Segmental Relaxation Time
  Distribution of Miscible Polymer Blends: Polyisoprene/Poly(vinylethylene).
  \emph{Macromolecules} \textbf{2005}, \emph{38}, 4919--4928, DOI:
  \doi{https://doi.org/10.1021/ma0500741}\relax
\mciteBstWouldAddEndPuncttrue
\mciteSetBstMidEndSepPunct{\mcitedefaultmidpunct}
{\mcitedefaultendpunct}{\mcitedefaultseppunct}\relax
\EndOfBibitem
\bibitem[Dudowicz \latin{et~al.}(2014)Dudowicz, Freed, and
  Douglas]{dudowicz_concentration_2014}
Dudowicz,~J.; Freed,~K.~F.; Douglas,~J.~F. Concentration fluctuations in
  miscible polymer blends: {Influence} of temperature and chain rigidity.
  \emph{The Journal of Chemical Physics} \textbf{2014}, \emph{140}, 194901,
  DOI: \doi{10.1063/1.4875345}\relax
\mciteBstWouldAddEndPuncttrue
\mciteSetBstMidEndSepPunct{\mcitedefaultmidpunct}
{\mcitedefaultendpunct}{\mcitedefaultseppunct}\relax
\EndOfBibitem
\bibitem[Dudowicz \latin{et~al.}(2014)Dudowicz, Douglas, and
  Freed]{dudowicz_two_2014}
Dudowicz,~J.; Douglas,~J.~F.; Freed,~K.~F. Two glass transitions in miscible
  polymer blends? \emph{The Journal of Chemical Physics} \textbf{2014},
  \emph{140}, 244905, DOI: \doi{10.1063/1.4884123}\relax
\mciteBstWouldAddEndPuncttrue
\mciteSetBstMidEndSepPunct{\mcitedefaultmidpunct}
{\mcitedefaultendpunct}{\mcitedefaultseppunct}\relax
\EndOfBibitem
\bibitem[Moreno and Colmenero(2007)Moreno, and Colmenero]{moreno_tests_2007}
Moreno,~A.~J.; Colmenero,~J. Tests of mode coupling theory in a simple model
  for two-component miscible polymer blends. \emph{Journal of Physics:
  Condensed Matter} \textbf{2007}, \emph{19}, 466112, DOI:
  \doi{10.1088/0953-8984/19/46/466112}\relax
\mciteBstWouldAddEndPuncttrue
\mciteSetBstMidEndSepPunct{\mcitedefaultmidpunct}
{\mcitedefaultendpunct}{\mcitedefaultseppunct}\relax
\EndOfBibitem
\bibitem[Ngai and Wang(2011)Ngai, and Wang]{ngai_interchain_2011}
Ngai,~K.~L.; Wang,~L.-M. Interchain coupled chain dynamics of poly(ethylene
  oxide) in blends with poly(methyl methacrylate): {Coupling} model analysis.
  \emph{The Journal of Chemical Physics} \textbf{2011}, \emph{135}, 194902,
  DOI: \doi{10.1063/1.3662130}\relax
\mciteBstWouldAddEndPuncttrue
\mciteSetBstMidEndSepPunct{\mcitedefaultmidpunct}
{\mcitedefaultendpunct}{\mcitedefaultseppunct}\relax
\EndOfBibitem
\bibitem[Ngai and Capaccioli(2013)Ngai, and Capaccioli]{ngai_unified_2013}
Ngai,~K.~L.; Capaccioli,~S. Unified explanation of the anomalous dynamic
  properties of highly asymmetric polymer blends. \emph{The Journal of Chemical
  Physics} \textbf{2013}, \emph{138}, 054903, DOI:
  \doi{10.1063/1.4789585}\relax
\mciteBstWouldAddEndPuncttrue
\mciteSetBstMidEndSepPunct{\mcitedefaultmidpunct}
{\mcitedefaultendpunct}{\mcitedefaultseppunct}\relax
\EndOfBibitem
\bibitem[Colmenero(2013)]{colmenero_comment_2013}
Colmenero,~J. Comment on “{Unified} explanation of the anomalous dynamic
  properties of highly asymmetric polymer blends” [{J}. {Chem}. {Phys}. 138,
  054903 (2013)]. \emph{The Journal of Chemical Physics} \textbf{2013},
  \emph{138}, 197101, DOI: \doi{10.1063/1.4804624}\relax
\mciteBstWouldAddEndPuncttrue
\mciteSetBstMidEndSepPunct{\mcitedefaultmidpunct}
{\mcitedefaultendpunct}{\mcitedefaultseppunct}\relax
\EndOfBibitem
\bibitem[Ngai and Capaccioli(2013)Ngai, and Capaccioli]{ngai_response_2013}
Ngai,~K.~L.; Capaccioli,~S. Response to “{Comment} on ‘{Unified}
  explanation of the anomalous dynamic properties of highly asymmetric polymer
  blends’” [{J}. {Chem}. {Phys}. 138, 197101 (2013)]. \emph{The Journal of
  Chemical Physics} \textbf{2013}, \emph{138}, 197102, DOI:
  \doi{10.1063/1.4804625}\relax
\mciteBstWouldAddEndPuncttrue
\mciteSetBstMidEndSepPunct{\mcitedefaultmidpunct}
{\mcitedefaultendpunct}{\mcitedefaultseppunct}\relax
\EndOfBibitem
\bibitem[Colmenero(2013)]{colmenero_generalized_2013}
Colmenero,~J. A {Generalized} {Rouse} {Incoherent} {Scattering} {Function} for
  {Chain} {Dynamics} of {Unentangled} {Polymers} in {Dynamically} {Asymmetric}
  {Blends}. \emph{Macromolecules} \textbf{2013}, \emph{46}, 5363--5370, DOI:
  \doi{10.1021/ma400309c}, Publisher: American Chemical Society\relax
\mciteBstWouldAddEndPuncttrue
\mciteSetBstMidEndSepPunct{\mcitedefaultmidpunct}
{\mcitedefaultendpunct}{\mcitedefaultseppunct}\relax
\EndOfBibitem
\bibitem[Colmenero(2013)]{colmenero_reply_2013}
Colmenero,~J. Reply to “{Comment} on ‘{A} {Generalized} {Rouse}
  {Incoherent} {Scattering} {Function} for {Chain} {Dynamics} of {Unentangled}
  {Polymers} in {Dynamically} {Asymmetric} {Blends}’”.
  \emph{Macromolecules} \textbf{2013}, \emph{46}, 8056--8058, DOI:
  \doi{10.1021/ma4017983}, Publisher: American Chemical Society\relax
\mciteBstWouldAddEndPuncttrue
\mciteSetBstMidEndSepPunct{\mcitedefaultmidpunct}
{\mcitedefaultendpunct}{\mcitedefaultseppunct}\relax
\EndOfBibitem
\bibitem[Ngai and Capaccioli(2013)Ngai, and Capaccioli]{ngai_comment_2013}
Ngai,~K.~L.; Capaccioli,~S. Comment on “{A} {Generalized} {Rouse}
  {Incoherent} {Scattering} {Function} for {Chain} {Dynamics} of {Unentangled}
  {Polymers} in {Dynamically} {Asymmetric} {Blends}”. \emph{Macromolecules}
  \textbf{2013}, \emph{46}, 8054--8055, DOI: \doi{10.1021/ma401732c},
  Publisher: American Chemical Society\relax
\mciteBstWouldAddEndPuncttrue
\mciteSetBstMidEndSepPunct{\mcitedefaultmidpunct}
{\mcitedefaultendpunct}{\mcitedefaultseppunct}\relax
\EndOfBibitem
\bibitem[Chen and Maranas(2009)Chen, and Maranas]{chen_molecular_2009}
Chen,~C.; Maranas,~J.~K. A {Molecular} {View} of {Dynamic} {Responses} {When}
  {Mixing} {Poly}(ethylene oxide) and {Poly}(methyl methacrylate).
  \emph{Macromolecules} \textbf{2009}, \emph{42}, 2795--2805, DOI:
  \doi{10.1021/ma802183h}, Publisher: American Chemical Society\relax
\mciteBstWouldAddEndPuncttrue
\mciteSetBstMidEndSepPunct{\mcitedefaultmidpunct}
{\mcitedefaultendpunct}{\mcitedefaultseppunct}\relax
\EndOfBibitem
\bibitem[Brodeck \latin{et~al.}(2012)Brodeck, Alvarez, Colmenero, and
  Richter]{brodeck_single_2012}
Brodeck,~M.; Alvarez,~F.; Colmenero,~J.; Richter,~D. Single {Chain} {Dynamic}
  {Structure} {Factor} of {Poly}(ethylene oxide) in {Dynamically} {Asymmetric}
  {Blends} with {Poly}(methyl methacrylate). {Neutron} {Scattering} and
  {Molecular} {Dynamics} {Simulations}. \emph{Macromolecules} \textbf{2012},
  \emph{45}, 536--542, DOI: \doi{10.1021/ma2016634}, Publisher: American
  Chemical Society\relax
\mciteBstWouldAddEndPuncttrue
\mciteSetBstMidEndSepPunct{\mcitedefaultmidpunct}
{\mcitedefaultendpunct}{\mcitedefaultseppunct}\relax
\EndOfBibitem
\bibitem[Thompson \latin{et~al.}(2022)Thompson, Aktulga, Berger, Bolintineanu,
  Brown, Crozier, in't Veld, Kohlmeyer, Moore, Nguyen, Shan, Stevens,
  Tranchida, Trott, and Plimpton]{lammps}
Thompson,~A.~P.; Aktulga,~H.~M.; Berger,~R.; Bolintineanu,~D.~S.; Brown,~W.~M.;
  Crozier,~P.~S.; in't Veld,~J.~J.; Kohlmeyer,~A.; Moore,~S.~G.; Nguyen,~T.~D.;
  Shan,~R.; Stevens,~M.~J.; Tranchida,~J.; Trott,~C.; Plimpton,~S.~J. LAMMPS -
  a flexible simulation tool for particle-based materials modeling at the
  atomic, meso, and continuum scales. \emph{Computer Physics Communications}
  \textbf{2022}, \emph{271}, 108171, DOI: \doi{10.1016/j.cpc.2021.108171}\relax
\mciteBstWouldAddEndPuncttrue
\mciteSetBstMidEndSepPunct{\mcitedefaultmidpunct}
{\mcitedefaultendpunct}{\mcitedefaultseppunct}\relax
\EndOfBibitem
\bibitem[William L.~Jorgensen and Tirado-Rives(1996)William L.~Jorgensen, and
  Tirado-Rives]{oplsaa}
William L.~Jorgensen,~D. S.~M.; Tirado-Rives,~J. Development and Testing of the
  OPLS All-Atom Force Field on Conformational Energetics and Properties of
  Organic Liquids. \emph{Journal of the American Chemical Society}
  \textbf{1996}, \emph{118}, 11225--11236, DOI: \doi{10.1021/ja9621760}\relax
\mciteBstWouldAddEndPuncttrue
\mciteSetBstMidEndSepPunct{\mcitedefaultmidpunct}
{\mcitedefaultendpunct}{\mcitedefaultseppunct}\relax
\EndOfBibitem
\bibitem[Hockney and Eastwood(1988)Hockney, and Eastwood]{pppmbook}
Hockney,~R.~W.; Eastwood,~J.~W. \emph{Computer simulation using particles};
  1988\relax
\mciteBstWouldAddEndPuncttrue
\mciteSetBstMidEndSepPunct{\mcitedefaultmidpunct}
{\mcitedefaultendpunct}{\mcitedefaultseppunct}\relax
\EndOfBibitem
\bibitem[Pollock and Glosli(1996)Pollock, and Glosli]{pppmarticle}
Pollock,~E.~L.; Glosli,~J. Comments on {P3M}, {FMM}, and the {E}wald method for
  large periodic {C}oulombic systems. \emph{Computer Physics Communications}
  \textbf{1996}, \emph{95}, 93--110, DOI:
  \doi{10.1016/0010-4655(96)00043-4}\relax
\mciteBstWouldAddEndPuncttrue
\mciteSetBstMidEndSepPunct{\mcitedefaultmidpunct}
{\mcitedefaultendpunct}{\mcitedefaultseppunct}\relax
\EndOfBibitem
\bibitem[Yoshida and Kobayashi(1982)Yoshida, and Kobayashi]{yoshida1982}
Yoshida,~H.; Kobayashi,~Y. Effect of Tacticity on Relaxation Process with
  Sub-Tg Annealing in Poly(methyl methacrylate). \emph{Polymer Journal}
  \textbf{1982}, \emph{14}, 925--926, DOI:
  \doi{https://doi.org/10.1295/polymj.14.925}\relax
\mciteBstWouldAddEndPuncttrue
\mciteSetBstMidEndSepPunct{\mcitedefaultmidpunct}
{\mcitedefaultendpunct}{\mcitedefaultseppunct}\relax
\EndOfBibitem
\bibitem[Hanwell \latin{et~al.}(2012)Hanwell, Curtis, Lonie, Vandermeersch,
  Zurek, and Hutchinson]{avogadro}
Hanwell,~M.~D.; Curtis,~D.~E.; Lonie,~D.~C.; Vandermeersch,~T.; Zurek,~E.;
  Hutchinson,~G.~R. Avogadro: An advanced semantic chemical editor,
  visualization, and analysis platform. \emph{Journal of Cheminformatics}
  \textbf{2012}, \emph{4}, 17, DOI: \doi{10.1186/1758-2946-4-17}\relax
\mciteBstWouldAddEndPuncttrue
\mciteSetBstMidEndSepPunct{\mcitedefaultmidpunct}
{\mcitedefaultendpunct}{\mcitedefaultseppunct}\relax
\EndOfBibitem
\bibitem[avo()]{avogadro-site}
Avogadro: an open-source molecular builder and visualization tool. Version
  1.2.0. \url{http://avogadro.cc/}\relax
\mciteBstWouldAddEndPuncttrue
\mciteSetBstMidEndSepPunct{\mcitedefaultmidpunct}
{\mcitedefaultendpunct}{\mcitedefaultseppunct}\relax
\EndOfBibitem
\bibitem[Hiemenz and Lodge(2007)Hiemenz, and Lodge]{hiemenzlodge}
Hiemenz,~P.~C.; Lodge,~T. \emph{Polymer Chemistry}, 2nd ed.; CRC Press,
  2007\relax
\mciteBstWouldAddEndPuncttrue
\mciteSetBstMidEndSepPunct{\mcitedefaultmidpunct}
{\mcitedefaultendpunct}{\mcitedefaultseppunct}\relax
\EndOfBibitem
\bibitem[Buchholz \latin{et~al.}(2002)Buchholz, Paul, Varnik, and
  Binder]{buchholz2002}
Buchholz,~J.; Paul,~W.; Varnik,~F.; Binder,~K. Cooling rate dependence of the
  glass transition temperature of polymer melts: Molecular dynamics study.
  \emph{The Journal of Chemical Physics} \textbf{2002}, \emph{117}, 7364--7372,
  DOI: \doi{10.1063/1.1508366}\relax
\mciteBstWouldAddEndPuncttrue
\mciteSetBstMidEndSepPunct{\mcitedefaultmidpunct}
{\mcitedefaultendpunct}{\mcitedefaultseppunct}\relax
\EndOfBibitem
\bibitem[Han \latin{et~al.}(1994)Han, Gee, and Boyd]{han1994}
Han,~J.; Gee,~R.~H.; Boyd,~R.~H. Glass Transition Temperatures of Polymers from
  Molecular Dynamics Simulations. \emph{Macromolecules} \textbf{1994},
  \emph{27}, 7781--7784, DOI: \doi{10.1021/ma00104a036}\relax
\mciteBstWouldAddEndPuncttrue
\mciteSetBstMidEndSepPunct{\mcitedefaultmidpunct}
{\mcitedefaultendpunct}{\mcitedefaultseppunct}\relax
\EndOfBibitem
\bibitem[Rigby and Roe(1990)Rigby, and Roe]{rigby1990}
Rigby,~D.; Roe,~R.~J. Molecular dynamics simulation of polymer liquid and
  glass. 4. Free-volume distribution. \emph{Macromolecules} \textbf{1990},
  \emph{23}, 5312--5319, DOI: \doi{10.1021/ma00228a002}\relax
\mciteBstWouldAddEndPuncttrue
\mciteSetBstMidEndSepPunct{\mcitedefaultmidpunct}
{\mcitedefaultendpunct}{\mcitedefaultseppunct}\relax
\EndOfBibitem
\bibitem[Putta and Nemat-Nasser(2001)Putta, and Nemat-Nasser]{putta2001}
Putta,~S.; Nemat-Nasser,~S. Molecularly-based numerical evaluation of free
  volume in amorphous polymers. \emph{Materials Science and Engineering: A}
  \textbf{2001}, \emph{317}, 70--76, DOI:
  \doi{10.1016/S0921-5093(01)01181-9}\relax
\mciteBstWouldAddEndPuncttrue
\mciteSetBstMidEndSepPunct{\mcitedefaultmidpunct}
{\mcitedefaultendpunct}{\mcitedefaultseppunct}\relax
\EndOfBibitem
\bibitem[Widmer-Cooper and Harrowell(2006)Widmer-Cooper, and
  Harrowell]{widmer-cooper_free_2006}
Widmer-Cooper,~A.; Harrowell,~P. Free volume cannot explain the spatial
  heterogeneity of {Debye}–{Waller} factors in a glass-forming binary alloy.
  \emph{Journal of Non-Crystalline Solids} \textbf{2006}, \emph{352},
  5098--5102, DOI: \doi{10.1016/j.jnoncrysol.2006.01.136}\relax
\mciteBstWouldAddEndPuncttrue
\mciteSetBstMidEndSepPunct{\mcitedefaultmidpunct}
{\mcitedefaultendpunct}{\mcitedefaultseppunct}\relax
\EndOfBibitem
\bibitem[Mei \latin{et~al.}(2022)Mei, Zhuang, Lu, An, and
  Wang]{mei_local-average_2022}
Mei,~B.; Zhuang,~B.; Lu,~Y.; An,~L.; Wang,~Z.-G. Local-{Average} {Free}
  {Volume} {Correlates} with {Dynamics} in {Glass} {Formers}. \emph{The Journal
  of Physical Chemistry Letters} \textbf{2022}, \emph{13}, 3957--3964, DOI:
  \doi{10.1021/acs.jpclett.2c00072}, Publisher: American Chemical Society\relax
\mciteBstWouldAddEndPuncttrue
\mciteSetBstMidEndSepPunct{\mcitedefaultmidpunct}
{\mcitedefaultendpunct}{\mcitedefaultseppunct}\relax
\EndOfBibitem
\bibitem[Liu \latin{et~al.}(2009)Liu, Bedrov, Kumar, Veytsman, and
  Colby]{liu2009}
Liu,~W.; Bedrov,~D.; Kumar,~S.~K.; Veytsman,~B.; Colby,~R.~H. Role of
  Distributions of Intramolecular Concentrations on the Dynamics of Miscible
  Polymer Blends Probed by Molecular Dynamics Simulation. \emph{Physical Review
  Letters} \textbf{2009}, \emph{103}, 037801--1--037801--4, DOI:
  \doi{https://doi.org/10.1103/PhysRevLett.103.037801}\relax
\mciteBstWouldAddEndPuncttrue
\mciteSetBstMidEndSepPunct{\mcitedefaultmidpunct}
{\mcitedefaultendpunct}{\mcitedefaultseppunct}\relax
\EndOfBibitem
\bibitem[Moreno and Colmenero(2008)Moreno, and
  Colmenero]{moreno_entangledlike_2008}
Moreno,~A.~J.; Colmenero,~J. Entangledlike {Chain} {Dynamics} in {Nonentangled}
  {Polymer} {Blends} with {Large} {Dynamic} {Asymmetry}. \emph{Physical Review
  Letters} \textbf{2008}, \emph{100}, 126001, DOI:
  \doi{10.1103/PhysRevLett.100.126001}, Publisher: American Physical
  Society\relax
\mciteBstWouldAddEndPuncttrue
\mciteSetBstMidEndSepPunct{\mcitedefaultmidpunct}
{\mcitedefaultendpunct}{\mcitedefaultseppunct}\relax
\EndOfBibitem
\bibitem[Wu(1987)]{wu1987}
Wu,~S. Entanglement, friction, and free volume between dissimilar chains in
  compatible polymer blends. \emph{Journal of Polymer Science Part B: Polymer
  Physics} \textbf{1987}, \emph{25}, 2511--2529, DOI:
  \doi{10.1002/polb.1987.090251207}\relax
\mciteBstWouldAddEndPuncttrue
\mciteSetBstMidEndSepPunct{\mcitedefaultmidpunct}
{\mcitedefaultendpunct}{\mcitedefaultseppunct}\relax
\EndOfBibitem
\bibitem[Moynihan \latin{et~al.}(1974)Moynihan, Easteal, Wilder, and
  Tucker]{tgrate}
Moynihan,~C.~T.; Easteal,~A.~J.; Wilder,~J.; Tucker,~J. Dependence of the glass
  transition temperature on heating and cooling rate. \emph{The Journal of
  Physical Chemistry} \textbf{1974}, \emph{78}, 2673--2677, DOI:
  \doi{10.1021/j100619a008}\relax
\mciteBstWouldAddEndPuncttrue
\mciteSetBstMidEndSepPunct{\mcitedefaultmidpunct}
{\mcitedefaultendpunct}{\mcitedefaultseppunct}\relax
\EndOfBibitem
\bibitem[Hung \latin{et~al.}(2020)Hung, Patra, and
  Simmons]{R:2020_Hung_Forecasting}
Hung,~J.-H.; Patra,~T.~K.; Simmons,~D.~S. Forecasting the experimental glass
  transition from short time relaxation data. \emph{Journal of Non-Crystalline
  Solids} \textbf{2020}, \emph{544}, 120205, DOI:
  \doi{10.1016/j.jnoncrysol.2020.120205}\relax
\mciteBstWouldAddEndPuncttrue
\mciteSetBstMidEndSepPunct{\mcitedefaultmidpunct}
{\mcitedefaultendpunct}{\mcitedefaultseppunct}\relax
\EndOfBibitem
\bibitem[Fernandes \latin{et~al.}(1986)Fernandes, Barlow, and
  Paul]{fernandes1986}
Fernandes,~A.~C.; Barlow,~J.~W.; Paul,~D.~R. Blends containing polymers of
  epichlorohydrin and ethylene oxide. {P}art {I}: {P}olymethacrylates.
  \emph{Journal of Applied Polymer Science} \textbf{1986}, \emph{32},
  5481--5508, DOI: \doi{10.1002/app.1986.070320618}\relax
\mciteBstWouldAddEndPuncttrue
\mciteSetBstMidEndSepPunct{\mcitedefaultmidpunct}
{\mcitedefaultendpunct}{\mcitedefaultseppunct}\relax
\EndOfBibitem
\bibitem[Liberman \latin{et~al.}(1984)Liberman, Gomes, and
  Macchi]{liberman1984}
Liberman,~S.~A.; Gomes,~A. D.~S.; Macchi,~E.~M. Compatibility in poly(ethylene
  oxide)–poly(methyl methacrylate) blends. \emph{Journal of Polymer Science:
  Polymer Chemistry Edition} \textbf{1984}, \emph{22}, 2809--2815, DOI:
  \doi{10.1002/pol.1984.170221107}\relax
\mciteBstWouldAddEndPuncttrue
\mciteSetBstMidEndSepPunct{\mcitedefaultmidpunct}
{\mcitedefaultendpunct}{\mcitedefaultseppunct}\relax
\EndOfBibitem
\bibitem[Lodge \latin{et~al.}(2006)Lodge, Wood, and Haley]{lodge2006}
Lodge,~T.~P.; Wood,~E.~R.; Haley,~J.~C. Two calorimetric glass transitions do
  not necessarily indicate immiscibility: The case of {PEO}/{PMMA}.
  \emph{Journal of Polymer Science Part B: Polymer Physics} \textbf{2006},
  \emph{44}, 756--763, DOI: \doi{10.1002/polb.20735}\relax
\mciteBstWouldAddEndPuncttrue
\mciteSetBstMidEndSepPunct{\mcitedefaultmidpunct}
{\mcitedefaultendpunct}{\mcitedefaultseppunct}\relax
\EndOfBibitem
\bibitem[Brodeck \latin{et~al.}(2009)Brodeck, Alvarez, Arbe, Juranyi, Unruh,
  Holderer, Colmenero, and Richter]{brodeck2009}
Brodeck,~M.; Alvarez,~F.; Arbe,~A.; Juranyi,~F.; Unruh,~T.; Holderer,~O.;
  Colmenero,~J.; Richter,~D. Study of the dynamics of poly(ethylene oxide) by
  combining molecular dynamic simulations and neutron scattering experiments.
  \emph{The Journal of Chemical Physics} \textbf{2009}, \emph{130}, 094908,
  DOI: \doi{10.1063/1.3077858}\relax
\mciteBstWouldAddEndPuncttrue
\mciteSetBstMidEndSepPunct{\mcitedefaultmidpunct}
{\mcitedefaultendpunct}{\mcitedefaultseppunct}\relax
\EndOfBibitem
\bibitem[Agapov and Sokolov(2010)Agapov, and Sokolov]{agapov2010}
Agapov,~A.; Sokolov,~A.~P. Size of the dynamic bead in polymers.
  \emph{Macromolecules} \textbf{2010}, \emph{43}, 9126--9130, DOI:
  \doi{10.1021/ma101222y}\relax
\mciteBstWouldAddEndPuncttrue
\mciteSetBstMidEndSepPunct{\mcitedefaultmidpunct}
{\mcitedefaultendpunct}{\mcitedefaultseppunct}\relax
\EndOfBibitem
\bibitem[Maitra and Heuer(2007)Maitra, and Heuer]{maitra2007}
Maitra,~A.; Heuer,~A. Understanding Segmental Dynamics in Polymer Electrolytes:
  A Computer Study. \emph{Macromolecular Chemistry and Physics} \textbf{2007},
  \emph{208}, 2215--2221, DOI: \doi{10.1002/macp.200700265}\relax
\mciteBstWouldAddEndPuncttrue
\mciteSetBstMidEndSepPunct{\mcitedefaultmidpunct}
{\mcitedefaultendpunct}{\mcitedefaultseppunct}\relax
\EndOfBibitem
\bibitem[Paul \latin{et~al.}(1998)Paul, Smith, Yoon, Farago, Rathgeber, Zirkel,
  Willner, and Richter]{R:1998_Paul_Chain}
Paul,~W.; Smith,~G.~D.; Yoon,~D.~Y.; Farago,~B.; Rathgeber,~S.; Zirkel,~A.;
  Willner,~L.; Richter,~D. Chain Motion in an Unentangled Polyethylene Melt: A
  Critical Test of the Rouse Model by Molecular Dynamics Simulations and
  Neutron Spin Echo Spectroscopy. \emph{Physical Review Letters} \textbf{1998},
  \emph{80}, 2346--2349, DOI: \doi{10.1103/physrevlett.80.2346}\relax
\mciteBstWouldAddEndPuncttrue
\mciteSetBstMidEndSepPunct{\mcitedefaultmidpunct}
{\mcitedefaultendpunct}{\mcitedefaultseppunct}\relax
\EndOfBibitem
\bibitem[Krushev \latin{et~al.}(2002)Krushev, Paul, and
  Smith]{R:2002_Krushev_Role}
Krushev,~S.; Paul,~W.; Smith,~G.~D. The Role of Internal Rotational Barriers in
  Polymer Melt Chain Dynamics. \emph{Macromolecules} \textbf{2002}, \emph{35},
  4198--4203, DOI: \doi{10.1021/ma0115794}\relax
\mciteBstWouldAddEndPuncttrue
\mciteSetBstMidEndSepPunct{\mcitedefaultmidpunct}
{\mcitedefaultendpunct}{\mcitedefaultseppunct}\relax
\EndOfBibitem
\bibitem[Ngai and Roland(1993)Ngai, and Roland]{R:1993_Ngai_Chemical}
Ngai,~K.~L.; Roland,~C.~M. Chemical structure and intermolecular cooperativity:
  dielectric relaxation results. \emph{Macromolecules} \textbf{1993},
  \emph{26}, 6824--6830, DOI: \doi{10.1021/ma00077a019}\relax
\mciteBstWouldAddEndPuncttrue
\mciteSetBstMidEndSepPunct{\mcitedefaultmidpunct}
{\mcitedefaultendpunct}{\mcitedefaultseppunct}\relax
\EndOfBibitem
\end{mcitethebibliography}
%%%%%%%%%%%%%%%%%%%%%%%%%%%%%%%%%%%%%%%%%%%%%%%%%%%%%%%%%%%%%%%%%%%%%

\end{document}

\tableofcontents
\setcounter{tocdepth}{1}

\clearpage
\section{Calculation of apparent $T_\text{g}$} \label{sec:tg_si}
\begin{figure}[h]
    \centering
    \includegraphics{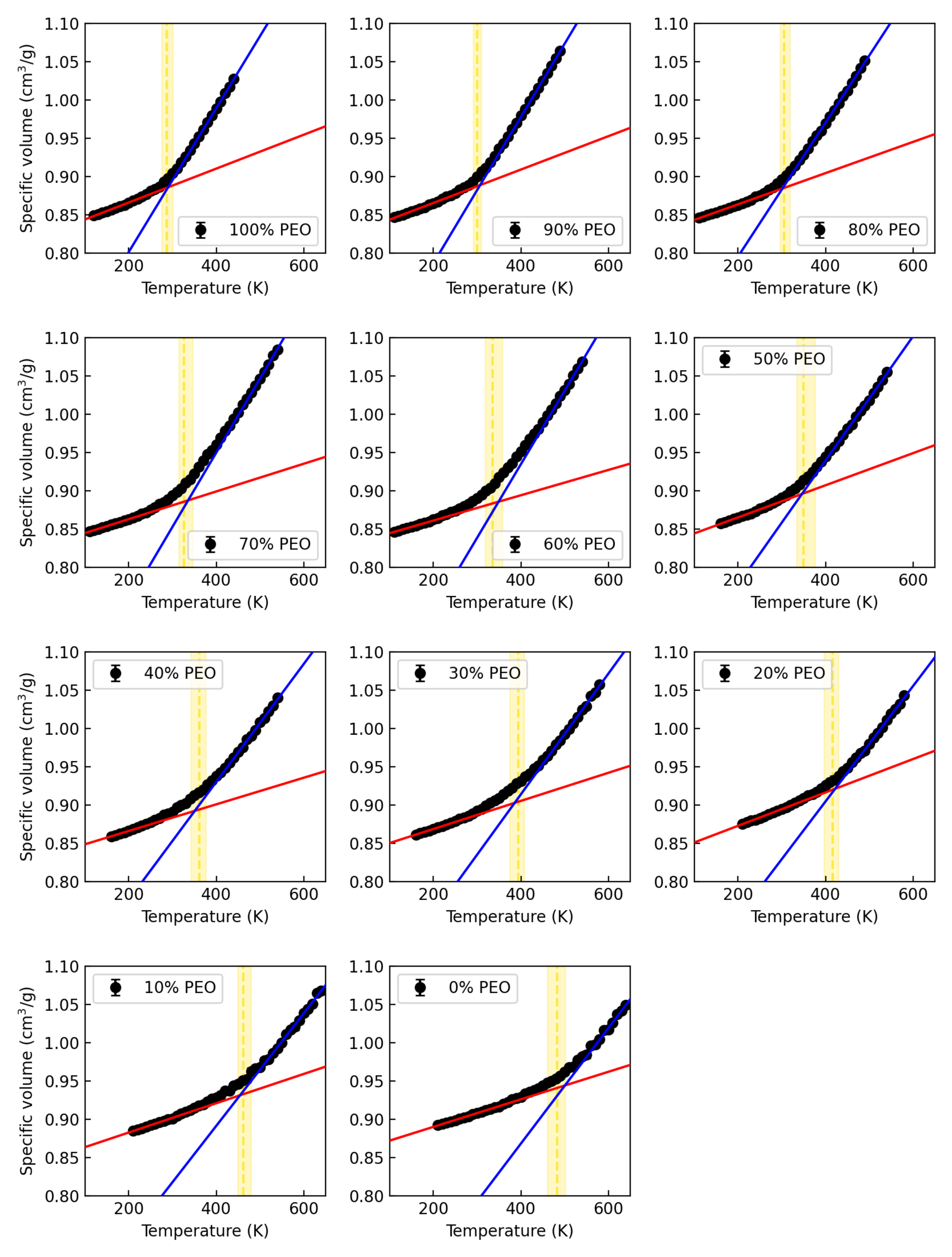}
    \caption{Change in specific volume with temperature used to calculate $T_\text{g}$ for PEO/PMMA blends over blend composition. Each marker and error bar is an average and standard error, respectively, of the specific volume of three independent system configurations at a given temperature. The vertical dashed yellow line is the final calculated $T_\text{g}$. The shaded yellow region is the range of all $T_\text{g}$ values calculated in the bootstrapping process. The red and blue lines are example linear regressions of one instance of the bootstrapping process for the high and low temperature regions, respectively. }
    \label{fig:tg_si}
\end{figure}

The simulated $T_\text{g}$ is calculated using a bootstrapping process that is repeated 10,000 times. The start and end points of the high and low temperature regions selected for each bootstrapping process are given in Table \ref{tab:tgfit}. 
This procedure will only yield a single apparent $T_\text{g}$ representative of the system; different fitting procedures would be required in an attempt to identify multiple $T_\text{g}$ signatures.
The simulated specific volume versus temperature data used in this calculation, final $T_\text{g}$ values, and example linear regressions in the high and low temperature regions for all blends studied are given in Figure \ref{fig:tg_si}.

\begin{table}[ht]
\caption{Start and end points and ranges of temperature used to calculate linear regressions in the high and low temperature regions of specific volume versus temperature data for $T_\text{g}$ calculations of PEO/PMMA blends.}
\centering
\begin{tabular}{ c|c|c } 
Composition & Glassy region fit & Liquid region fit \\
\hline
100 PEO\% & 100-[150,250] & [300,400]-450 \\ 
90 PEO\% & 100-[150,250] & [350,450]-500 \\ 
80 PEO\% & 100-[150,250] & [350,450]-500 \\ 
70 PEO\% & 100-[150,250] & [400,500]-550 \\ 
60 PEO\% & 100-[150,250] & [400,500]-550 \\ 
50 PEO\% & 150-[200,300] & [400,500]-550 \\ 
40 PEO\% & 150-[200,300] & [400,500]-550 \\ 
30 PEO\% & 150-[200,300] & [440-540]-590 \\ 
20 PEO\% & 200-[250,350] & [440-540]-590 \\ 
10 PEO\% & 200-[250,350] & [500,600]-650 \\ 
0 PEO\% & 200-[250,350] & [500,600]-650 \\ 
\end{tabular}
\label{tab:tgfit}
\end{table}

\clearpage
\section{Calculation of free volume}
\begin{figure}[h]
    \centering
    \includegraphics{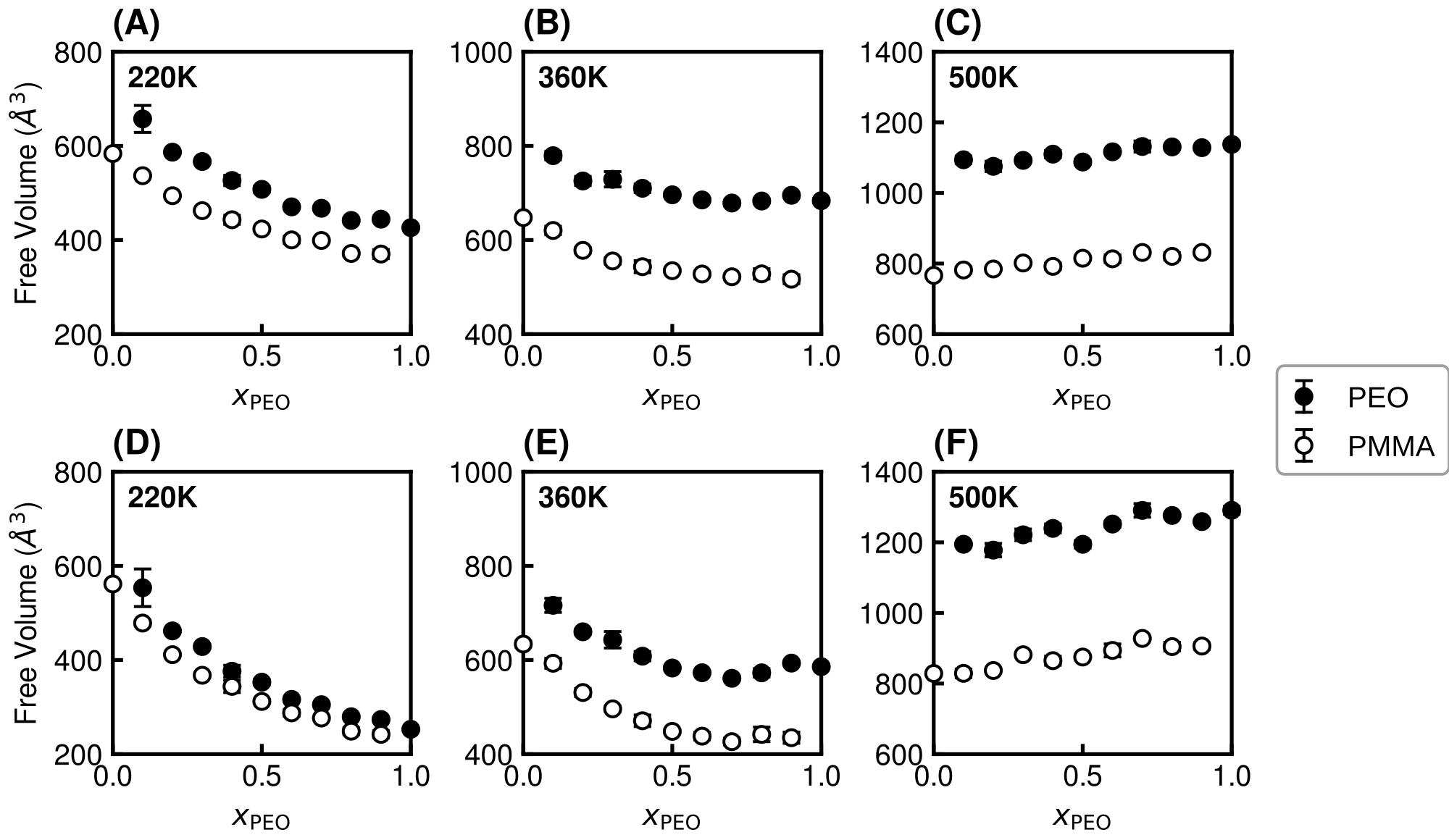}
    \caption{Local free volume of PEO and PMMA chains versus blend composition calculated using different probe radii. Free volume is calculated using probes of radius (A-C) 0.25 {\AA} and (D-F) 0.5 {\AA}. Error bars are standard errors calculated for three independent system configurations. }
    \label{fig:fv_si}
\end{figure}

The size of the probe used in the free volume calculation dictates the magnitude of free volume calculated for a polymer chain. Probes that are too large will underestimate free volume while probes that are too small will overestimate. However, we are interested in the qualitative change in free volume with $x_\text{PEO}$. To ensure that probe size does not affect our analysis, we calculated free volume using probes of radius 0.25 {\AA} and 0.5 {\AA}. These values are smaller than the size of the smallest atom in the polymer, hydrogen, which has a $\sigma=1.008$ {\AA}. As shown in Figure \ref{fig:fv_si}, we observe that adjusting the probe size affects the magnitude of free volume surrounding the chain, as expected, but does not change the qualitative trends in free volume versus $x_\text{PEO}$. Thus, using any of these probe sizes would be adequate for the analysis presented in the main text. We proceed using probes of radius 0.5 {\AA}. 

\clearpage
\section{Comparison of neat PEO and neat PMMA free volume}
\begin{figure}[h]
    \centering
    \includegraphics{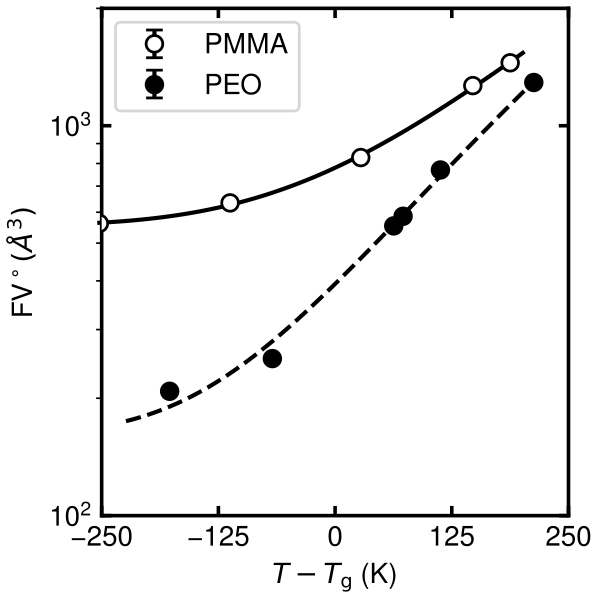}
    \caption{Local free volume of neat PEO and neat PMMA chains versus $T-T_\text{g}$. Free volume is calculated using probes of radius 0.5~\AA. The solid black line represents a fit to the neat PMMA free volume and the dashed line represents a fit to the neat PEO free volume. Error bars are included as standard errors calculated for three independent system configurations but are generally smaller than the symbol size.}
    \label{fig:neat_fv_si}
\end{figure}

Changes in segmental mobilities of PEO and PMMA upon blending directly correlate with changes in free volume. We suggest that the unexpected behavior of segmental mobility at 220 K, where mobility for both PEO and PMMA decrease in blends with more PEO despite PEO being the more mobile polymer, is caused by the mobilities of both species being comparably low at 220 K and the less efficient packing of PMMA compared to PEO. We show that PMMA packs less efficiently than PEO in Figure \ref{fig:neat_fv_si}, where the free volume per chain of neat PMMA is confirmed to be greater than the free volume per chain of neat PEO at all values of $T-T\text{g}$. 

\clearpage 
\section{Relative segmental mobility enhancement as a function of temperature}

\begin{figure}[h]
    \centering
    \includegraphics{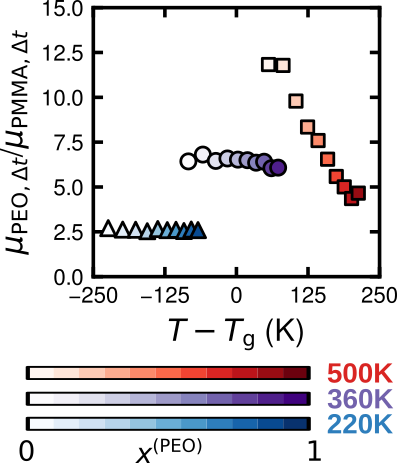}
    \caption{Relative mobility enhancement of PEO over PMMA as a function of temperature. The color gradient corresponds to a gradient in blend composition containing the most PEO (dark) to the least PEO (light). Error bars are standard errors for three independent system configurations at a given composition and temperature and are generally smaller than the symbol size.}
    \label{fig:relmob_si}
\end{figure}

Local segmental mobility as a function of temperature in Section \ref{sec:localdyn} reveals unexpected behaviors at 220 K which oppose the behaviors at 360 K and 500 K. This is due to the low relative mobility of PEO over PMMA at 220 K (below the blend $T_\text{g}$). As such, the dominant factor controlling dynamics is free volume. At 360 K and 500 K, the relative mobility of PEO over PMMA is much higher, as shown in Figure \ref{fig:relmob_si}. 

\clearpage
\section{Influence of local environment on normalized segmental mobility}

\begin{figure}[h]
    \centering
    \includegraphics[width=0.95\linewidth]{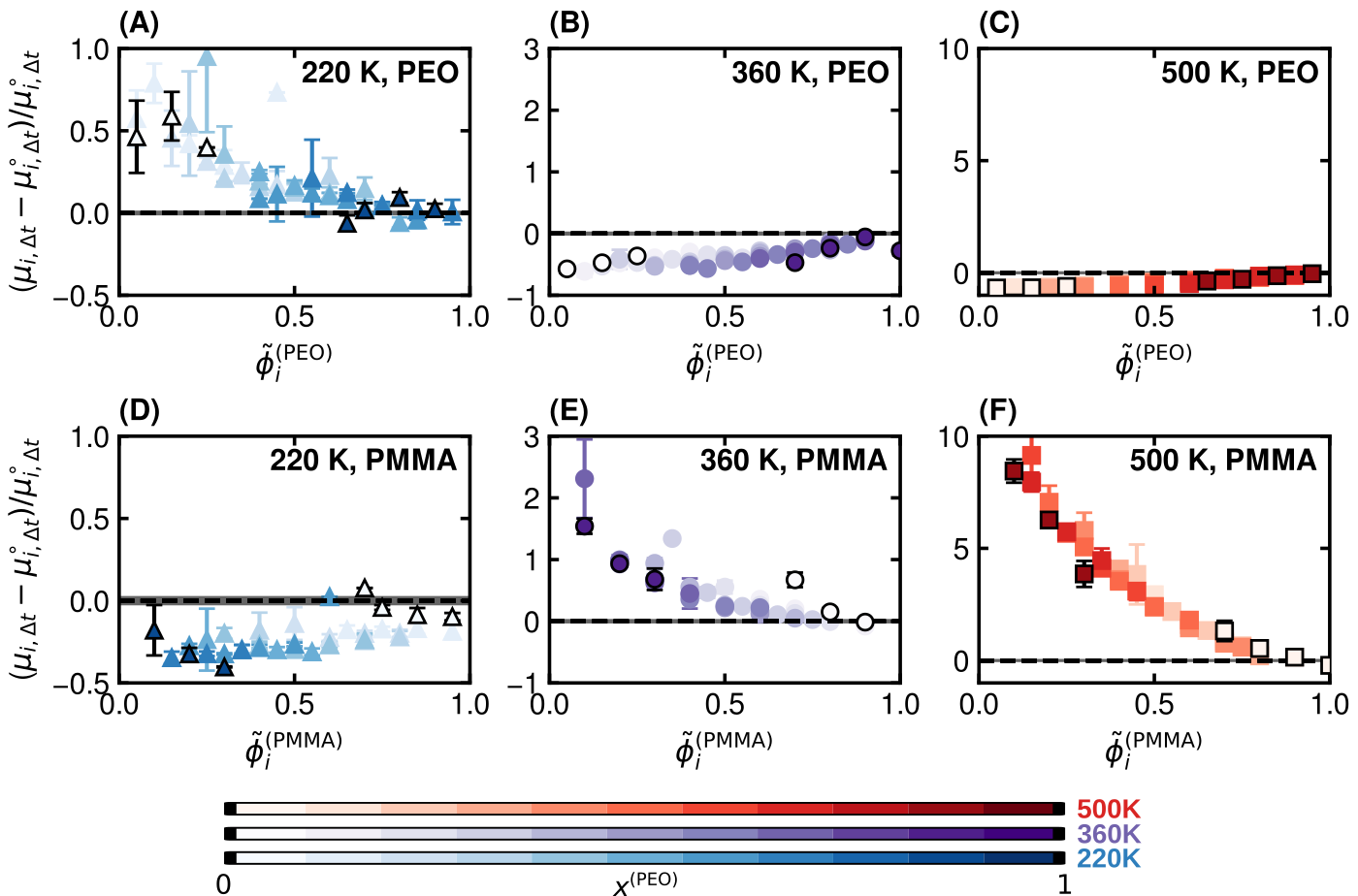}
    \caption{Variation in normalized relative segmental mobility over local environment. Results are shown for PEO at (A) 220 K, (B) 360 K, and (C) 500 K and PMMA at (D) 220 K, (E) 360 K, and (F) 500 K. Segmental mobility is calculated relative to and normalized by $\mu_{i,\Delta t}$ of the neat polymers, denoted as $\mu_{i,\Delta t}^\circ$. The color gradient corresponds to a gradient in blend composition containing the most PEO (dark) to the least PEO (light). Results for chains in blends with the most extreme compositions ($x^\text{(PEO)}=0.1$ and $0.9$) are outlined in black as a visual aid. Error bars are standard errors for three independent system configurations at a given composition and temperature. Horizontal dashed lines correspond to the relative segmental mobility of (A-C) pure PEO and (D-F) pure PMMA. The gray area around the dashed lines are standard deviations calculated from three independent configurations of the neat systems. }
    \label{fig:msfspnorm_si}
\end{figure}

Analysis of the segmental mobility as a function of local mobility in Section \ref{sec:local_absenv} reveals that changes to the local environment lead to larger magnitudes of $\mu_{i,\Delta t}$ suppression than $\mu_{i,\Delta t}$ enhancement. However, $\mu_{i,\Delta t}$ normalized by the local mobility of each neat species ($\mu_{i,\Delta t}^\circ$), shown in Figure \ref{fig:msfspnorm_si}, reveals that the local environment actually has a much larger relative effect on the enhancement of normalized $\mu_{i,\Delta t}$ than the suppression of normalized $\mu_{i,\Delta t}$. 

\clearpage
\section{Influence of the distribution of local environment on segmental mobility}

The distribution of local environment is found to correlate with the change in $\mu_{i,\Delta t} - \mu_{i,\Delta t}^\circ$. This is concluded by analyzing the distribution of intermolecular PMMA around PEO and PMMA and its normalized Shannon entropy ($\tilde{H}$) at 500 K in Section \ref{sec:distenv}. We have additionally calculated the distributions of intramolecular units and intermolecular PEO around PEO and PMMA at 220 K, 360 K, and 500 K, along with the distribution of intermolecular PMMA at 220 K and 360 K, and their normalized Shannon entropies. 

\begin{figure}[h]
    \centering
    \includegraphics[width=0.95\linewidth]{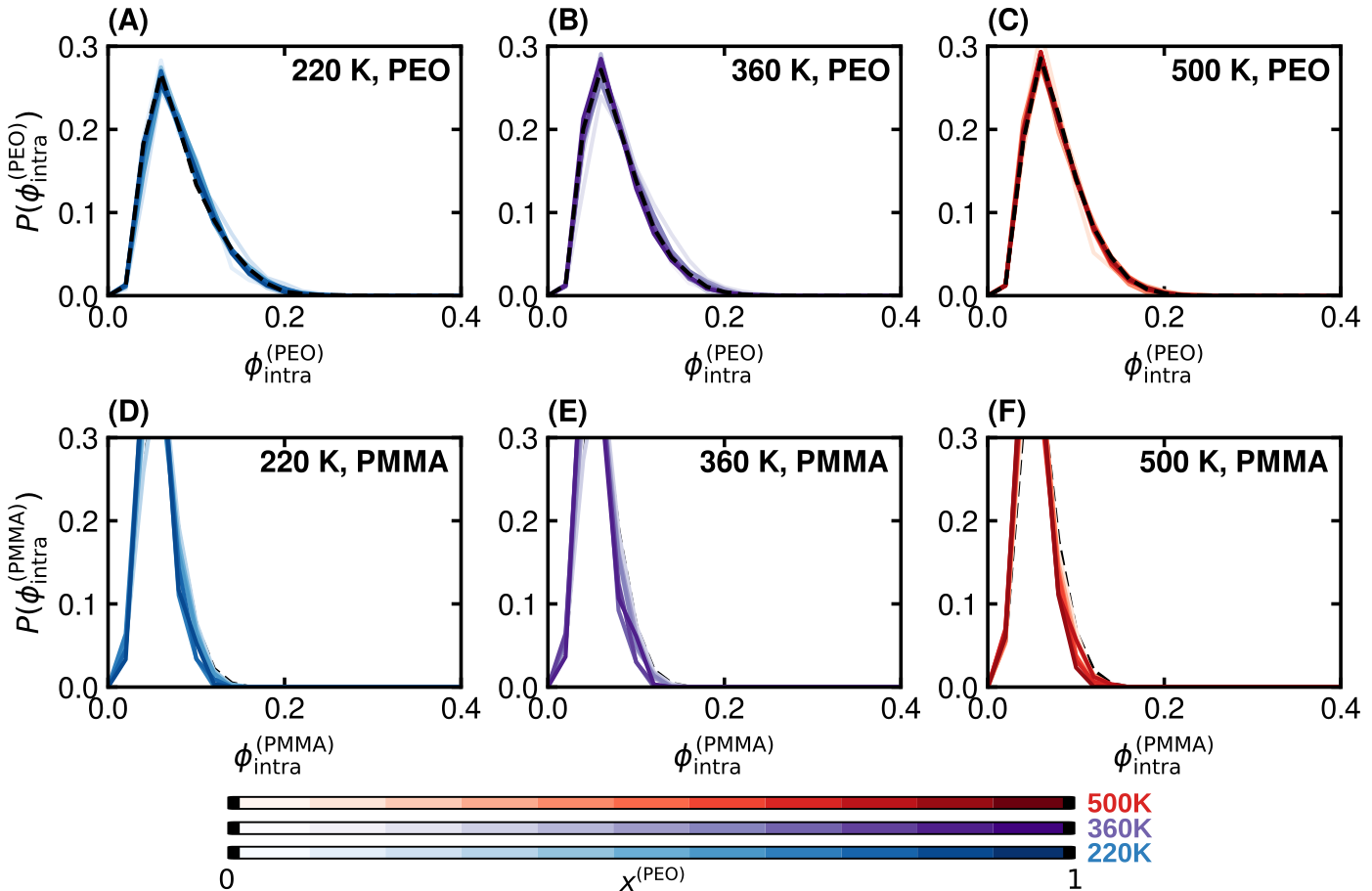}
    \caption{Probability distributions of the volume fraction of intramolecular (A,B,C) PEO and (D,E,F) PMMA atoms within a Kuhn length of each backbone carbon. Results are shown for PEO at (A) 220 K, (B) 360 K, and (C) 500 K and PMMA at (D) 220 K, (E) 360 K, and (F) 500 K. The color gradient corresponds to a gradient in blend composition containing the most PEO (dark) to the least PEO (light). Results for pure systems are shown as black dashed lines. }
    \label{fig:intra_si}
\end{figure}

\begin{figure}[h]
    \centering
    \includegraphics[width=0.95\linewidth]{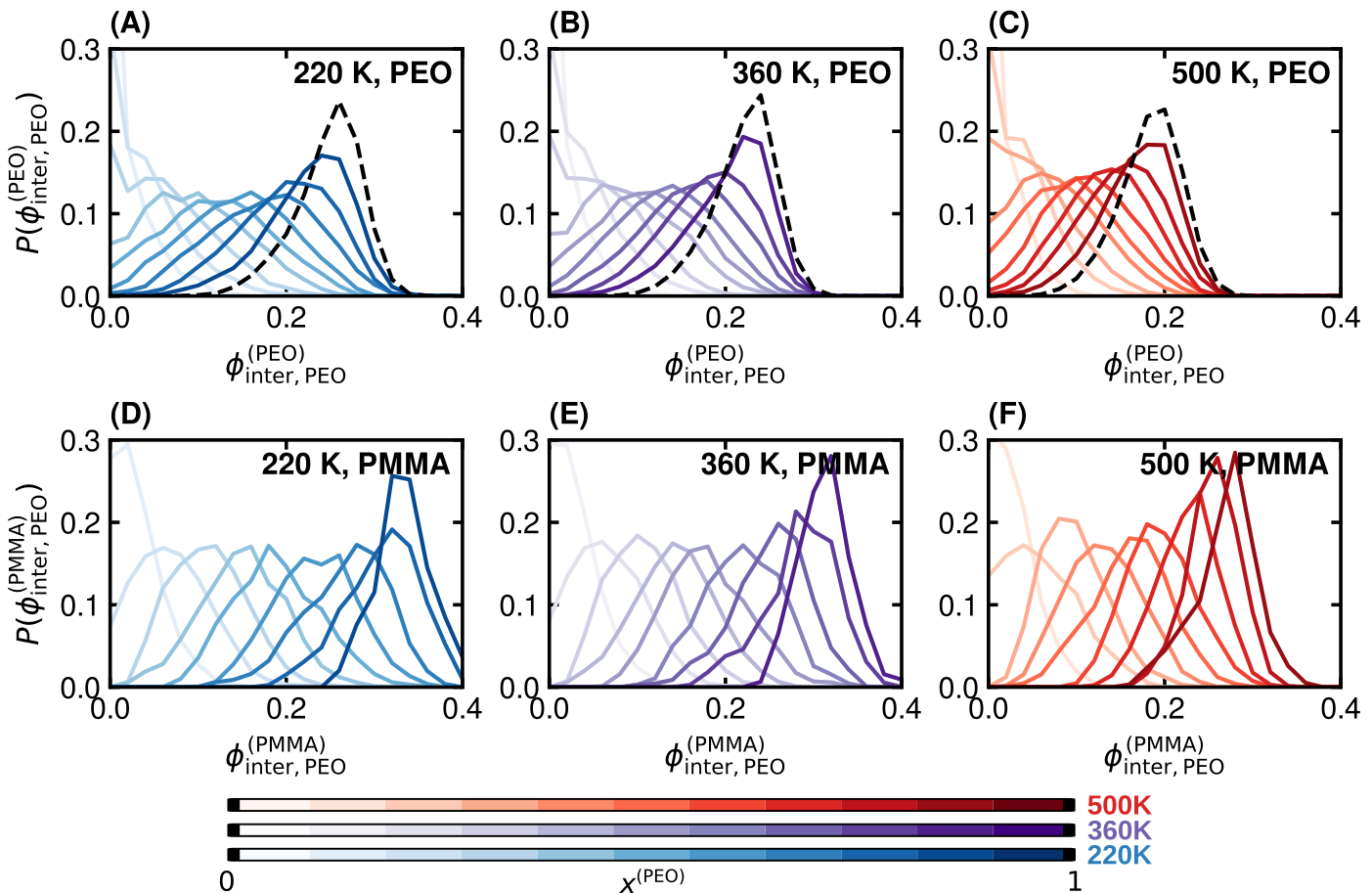}
    \caption{Probability distributions of the volume fraction of intermolecular PEO within a Kuhn length of backbone carbons on (A,B,C) PEO chains and (D,E,F) PMMA chains. Results are shown for PEO at (A) 220 K, (B) 360 K, and (C) 500 K and PMMA at (D) 220 K, (E) 360 K, and (F) 500 K. The color gradient corresponds to a gradient in blend composition containing the most PEO (dark) to the least PEO (light). Results for pure systems are shown as black dashed lines. }
    \label{fig:interpeo_si}
\end{figure}

\begin{figure}[h]
    \centering
    \includegraphics[width=0.95\linewidth]{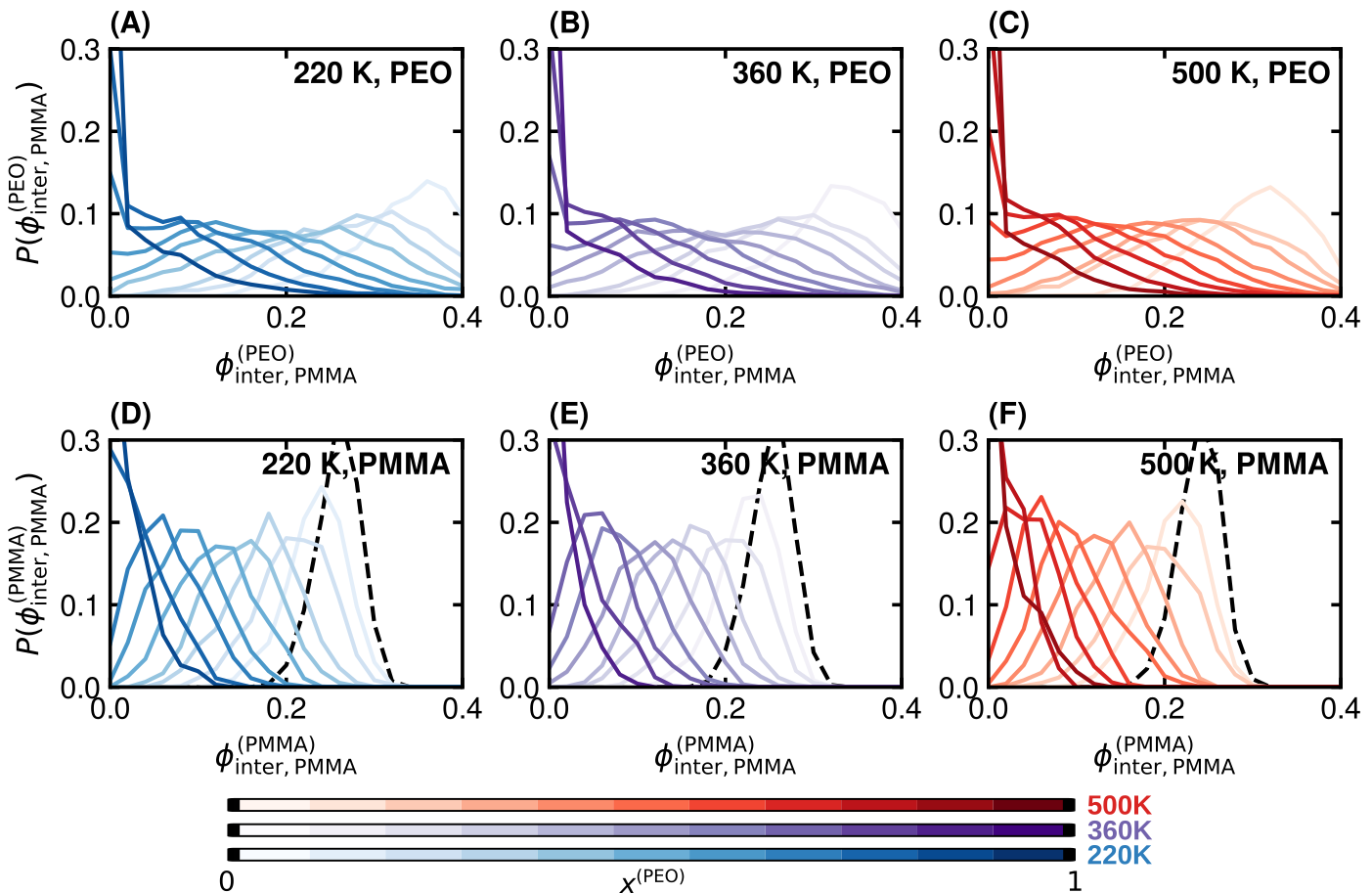}
    \caption{Probability distributions of the volume fraction of intermolecular PMMA within a Kuhn length of backbone carbons on (A,B,C) PEO chains and (D,E,F) PMMA chains. Results are shown for PEO at (A) 220 K, (B) 360 K, and (C) 500 K and PMMA at (D) 220 K, (E) 360 K, and (F) 500 K. The color gradient corresponds to a gradient in blend composition containing the most PEO (dark) to the least PEO (light). Results for pure systems are shown as black dashed lines. }
    \label{fig:interpmma_si}
\end{figure}

\begin{figure}[h]
    \centering
    \includegraphics{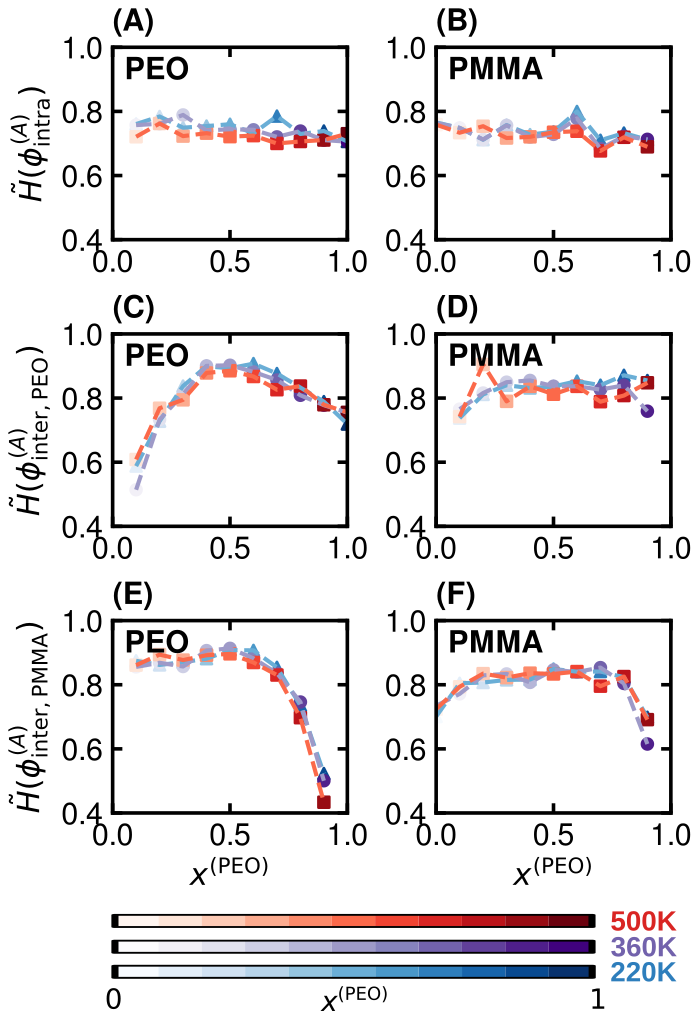}
    \caption{Normalized Shannon entropy of distributions of intra and intermolecular species around backbone carbons of (A,C,E) PEO and (B,D,F) PMMA versus blend composition. The entropies are shown for distributions of (A) intramolecular PEO, (B) intramolecular PMMA, (C,D) intermolecular PMMA, and (E,F) intermolecular PEO within one Kuhn length of the backbone carbons of each species at (red squares) 500 K, (purple circles) 360 K, and (blue triangles) 220 K. Dashed lines provide a guide to the eye. }
    \label{fig:entropy_si}
\end{figure}

Distributions of intramolecular PEO and PMMA are shown in Figure \ref{fig:intra_si} and their $\tilde{H}$ in Figure \ref{fig:entropy_si}A-B. All intramolecular distributions are found to be temperature and blend composition invariant. This is expected due to the small size of the cooperative sphere used in these calculations (the radii of the spheres are Kuhn lengths). Distributions of intermolecular PEO are shown in Figure \ref{fig:interpeo_si} and their $\tilde{H}$ in Figure \ref{fig:entropy_si}C-D. Although these distributions exhibit similar behaviors to the distributions of intermolecular PMMA that are discussed in Section \ref{sec:distenv}, the diversity in local PEO around PEO is less than the diversity in local PMMA around PEO, particularly at low $x^\text{(PEO)}$. We conclude this because the distributions in \ref{fig:interpeo_si}A-C are more narrow than the distributions in \ref{fig:interpmma_si}A-C. This is also evident from the low values of $\tilde{H}$ for low $x^\text{(PEO)}$ observed in Figure \ref{fig:interpeo_si}. We attribute the reduction in distribution diversity to the asymmetric influence of local environment that are echoed throughout the text, wherein proximity to PMMA induces stronger dynamical changes than proximity to PEO. Distributions of intermolecular PMMA are shown in Figure \ref{fig:interpmma_si} and their $\tilde{H}$ are shown in Figure \ref{fig:entropy_si}E-F. Results are shown to be independent of temperature, as stated in Section \ref{sec:distenv}. 

\clearpage
\section{Rouse mode analysis} \label{sec:rouse_si}
\begin{figure}[h]
    \centering
    \includegraphics{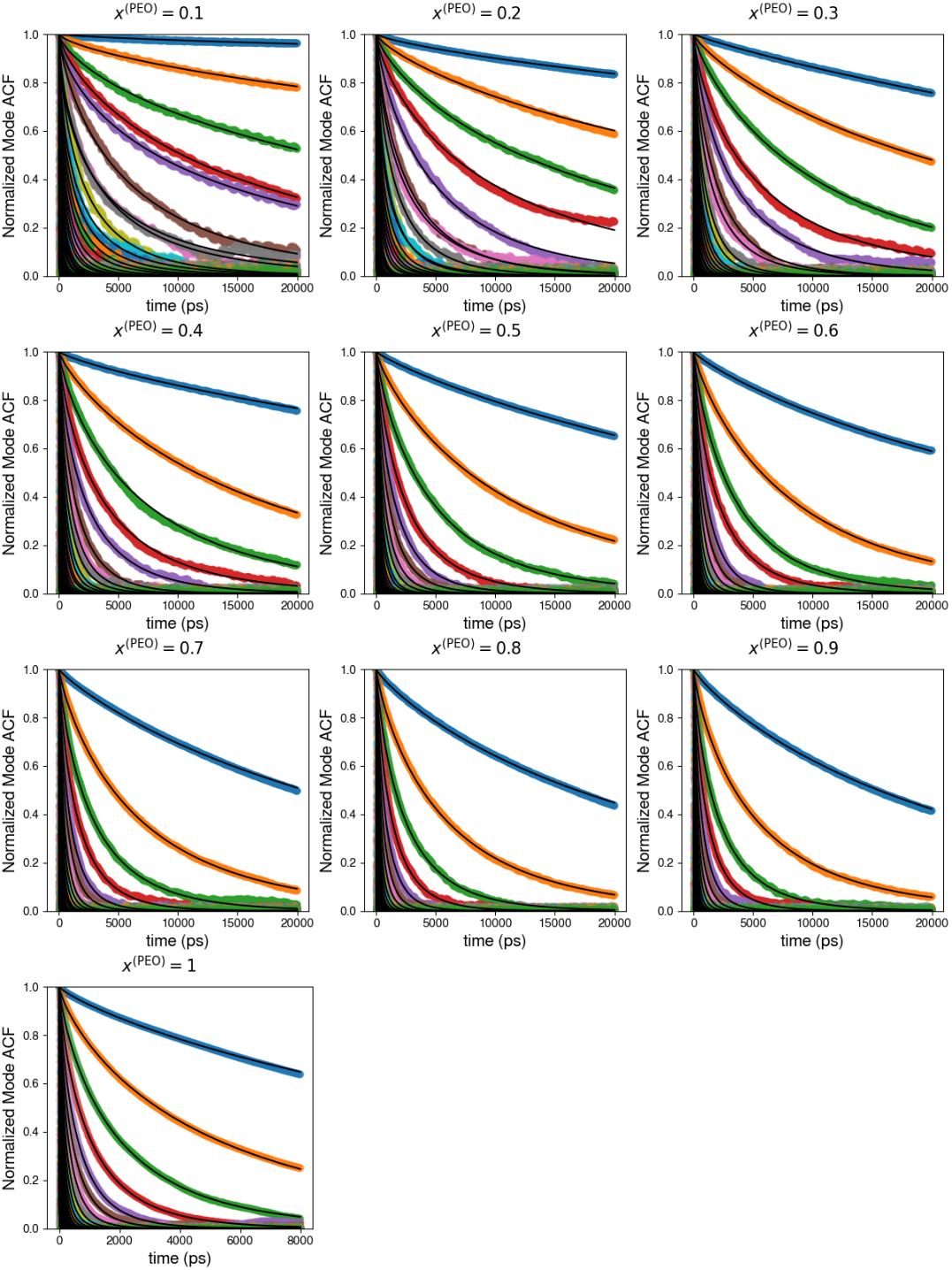}
    \caption{Normalized Rouse mode autocorrelation functions (ACFs) and their fits for PEO chains in the first run of each blend composition. Each ACF shown in each panel is for one Rouse mode. The topmost ACF is the ACF for the $p=0$th Rouse mode, the ACF below it is for the $p=1$st Rouse mode, and so on. The ACFs of all 74 Rouse modes for PEO are plotted. Black solid lines are fits to each ACF. }
    \label{fig:rouse_peo_si}
\end{figure}

\begin{figure}[h]
    \centering
    \includegraphics{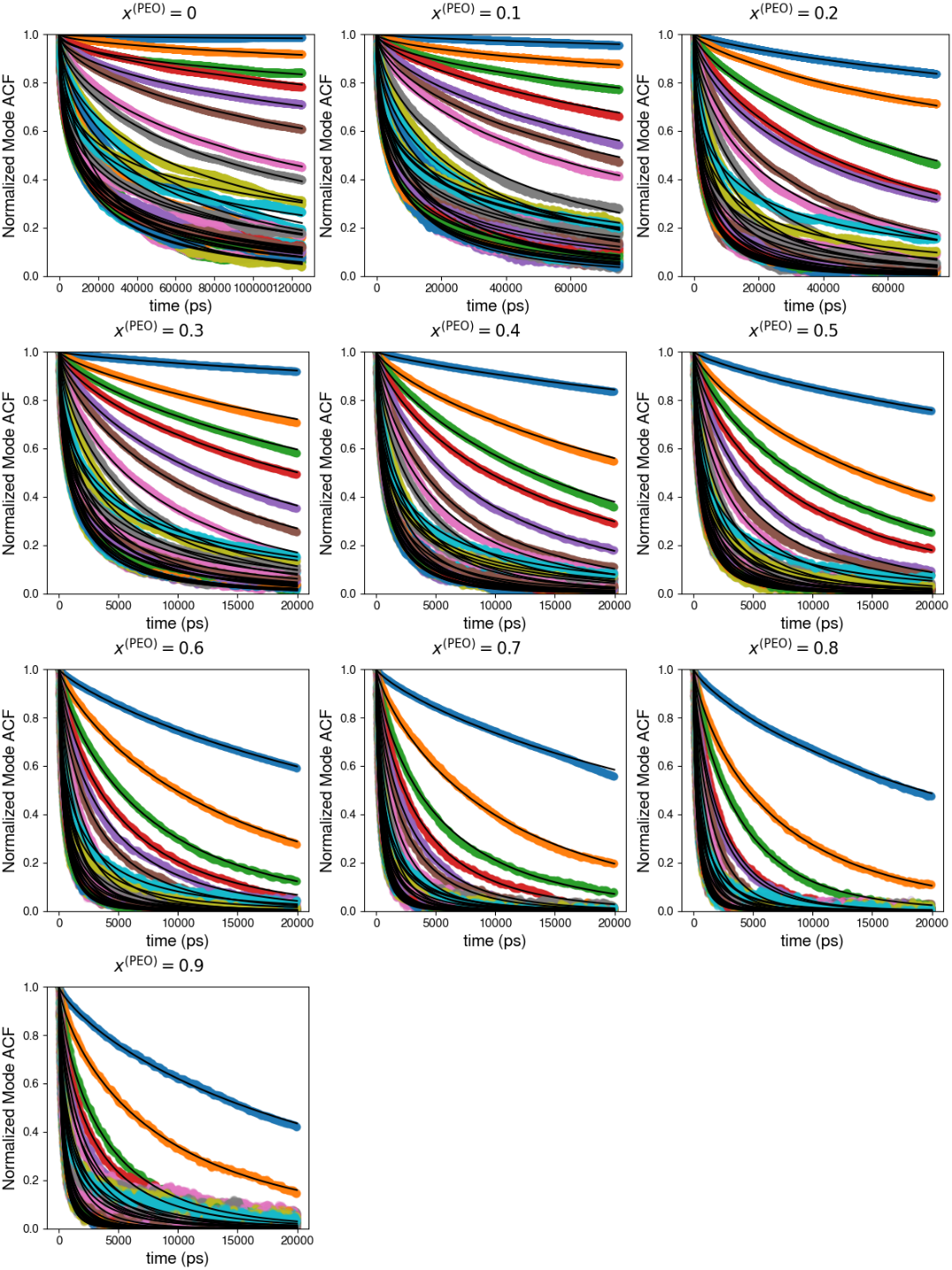}
    \caption{Normalized Rouse mode autocorrelation functions (ACFs) and their fits for PMMA chains in the first run of each blend composition. Each ACF shown in each panel is for one Rouse mode. The topmost ACF is the ACF for the $p=0$th Rouse mode, the ACF below it is for the $p=1$st Rouse mode, and so on. The ACFs of all 32 Rouse modes for PMMA are plotted. Black solid lines are fits to each ACF. }
    \label{fig:rouse_pmma_si}
\end{figure}

Effective relaxation times and stretching parameters used in the Rouse mode analysis are extracted from normalized Rouse mode ACFs of PEO and PMMA. Representative ACFs from one run per $x^\text{(PEO)}$ at 500~K and their fits using values of $\tau_p^\text{eff}$ and $\beta_p$ calculated from the linear regression method described in Section \ref{sec:Rouse} and allowing $\beta$ to vary with both $p$ and $x^\text{(PEO)}$ without L2 regularization are shown for PEO and PMMA in Figures \ref{fig:rouse_peo_si} and \ref{fig:rouse_pmma_si}, respectively. 


\begin{figure}[h]
    \centering
    \includegraphics[width=0.5\linewidth]{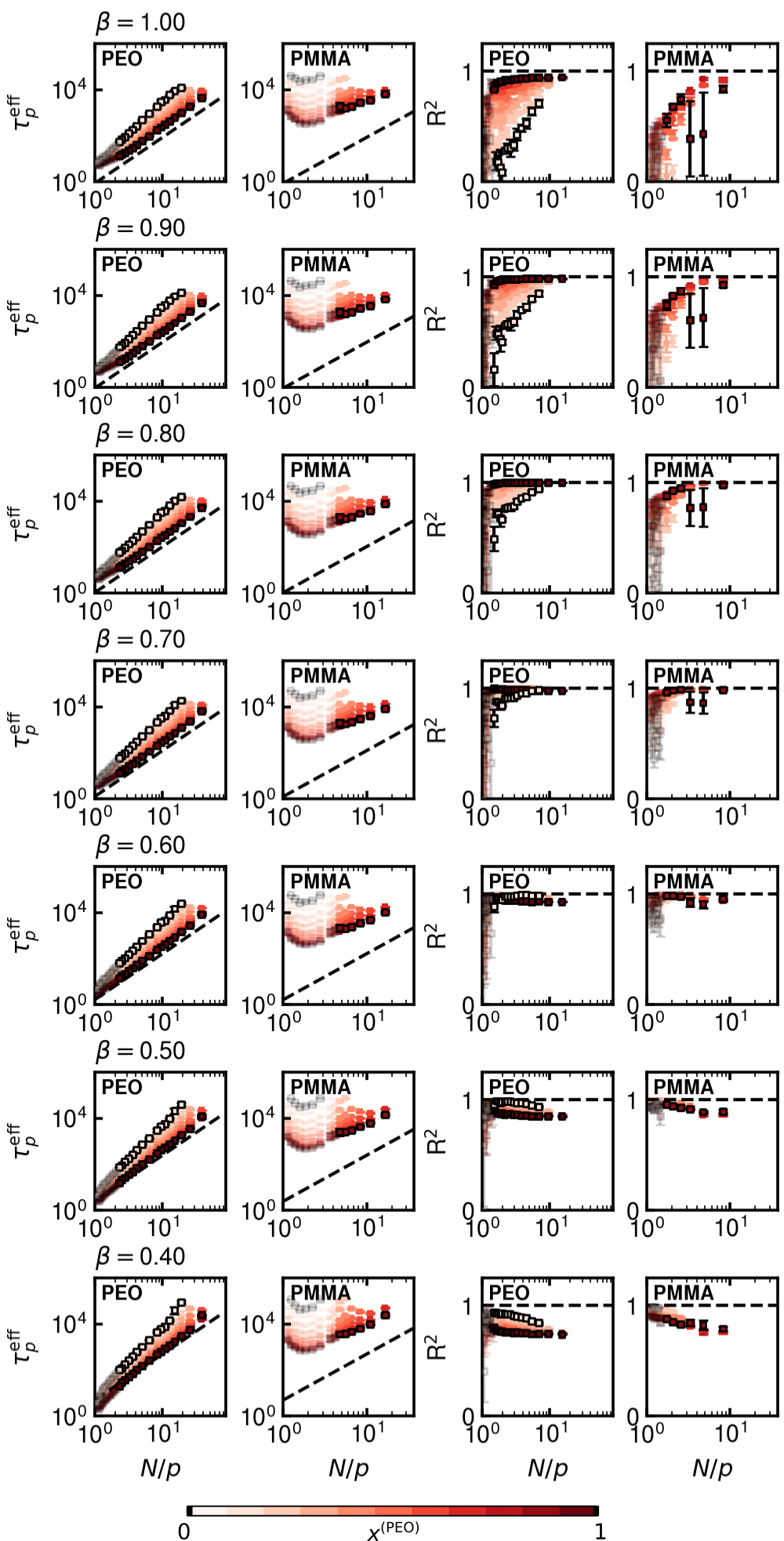}
    \caption{Variations in fit to Rouse mode autocorrelation functions quantified as the R$^2$ coefficient of determination and resulting relaxation times $\tau_p^\text{eff}$ for PEO and PMMA using a range of fixed $\beta$ values. Data shown is for systems at 500 K and color gradient corresponds to 
    The two leftmost columns show the $\tau_p^\text{eff}$ for PEO and PMMA versus sub-chain length $N/p$ while the two rightmost columns show the corresponding R$^2$ values versus $N/p$. 
    Data shown is at 500 K. 
    For $p>8$, symbols for data are only shown for every third value of $p$ for visual clarity. 
    Dashed black lines in the two leftmost columns are a guide to the eye to indicate the expected ideal scaling of $\tau_p \sim p^{-2}$. 
    Results for chains in blends with the most extreme compositions are outlined in black for visual clarity of trends. Error bars reflect standard errors from three independent systems and are generally smaller than the symbol size.}
    \label{fig:rouse_fixedbeta_si}
\end{figure}

Stretching parameters $\beta$ are fit to the ACFs using three different methods to show that the extracted $\tau_p^\text{eff}$ values are independent of fitting method. In each method, we quantify the adequacy of the fit using the R$^2$ coefficient of determination. In the first method, we fit the ACFs using fixed values of $\beta$ that vary from 1 to 0.4. Results are shown in Figure \ref{fig:rouse_fixedbeta_si} We find that for $\beta=1$, adequate fits (R$^2>0.90$) are achieved readily for pure PEO for $p<40$ ($N/p>1.875$) but deviate steadily for lower $x^\text{(PEO)}$ and smaller $N/p$. Adequate fits for PMMA are only achieved for $x^\text{(PEO)}=0.9$ for $p<6$ ($N/p>5.5$) and similarly deviate at low $x^\text{(PEO)}$ and small $N/p$. The fixed values of $\beta$ with the best fits are $\beta=0.6$ for PEO and $\beta=0.5$ for PMMA. 

\begin{figure}[h]
    \centering
    \includegraphics{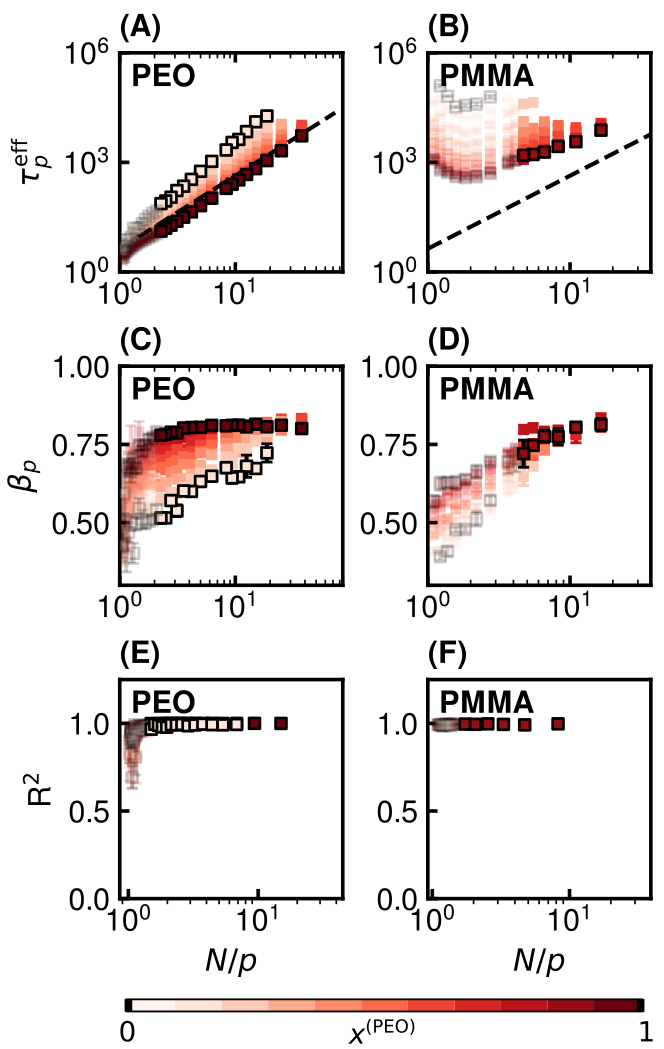}
    \caption{Rouse mode analysis at 500 K for chains in blends of varying composition where $\beta_p$ is allowed to vary with both $p$ and $x^\text{(PEO)}$. The effective Rouse relaxation time $\tau^{\text{eff}}_p$ as a function of sub-chain length $N/p$ for (A) PEO and (B) PMMA.
    The mode-dependent stretching exponent $\beta_p$ for (C) PEO and (D) PMMA. 
    R$^2$ coefficient of determination for the fit of $\tau_p$ and $\beta_p$ to the Rouse mode autocorrelation functions for (E) PEO and (F) PMMA. 
    For $p>8$, symbols for data are only shown for every third value of $p$ for visual clarity. Transparent data points correspond to Rouse modes that correspond to length scales smaller than the Rouse bead size. 
    Dashed black lines in panels (A) and (B) are a guide to the eye to indicate the expected ideal scaling of $\tau_p \sim p^{-2}$. The position of the line is the same across panels and is set to align with the behavior of neat PEO. 
    Results for chains in blends with the most extreme compositions are outlined in black for visual clarity of trends. Error bars reflect standard errors from three independent systems and are generally smaller than the symbol size.}
    \label{fig:rouse_varbeta_si}
\end{figure}

Next, we fit the ACFs by allowing the $\beta$ to vary with $p$ and $x^\text{(PEO)}$. Results are shown in Figure \ref{fig:rouse_varbeta_si} In this case, R$^2$ is closest to 1 for all values of $p$ and $x^\text{(PEO)}$ but, as the reviewer has pointed out, $\beta$ varies significantly across these parameters. 

\begin{figure}[h]
    \centering
    \includegraphics{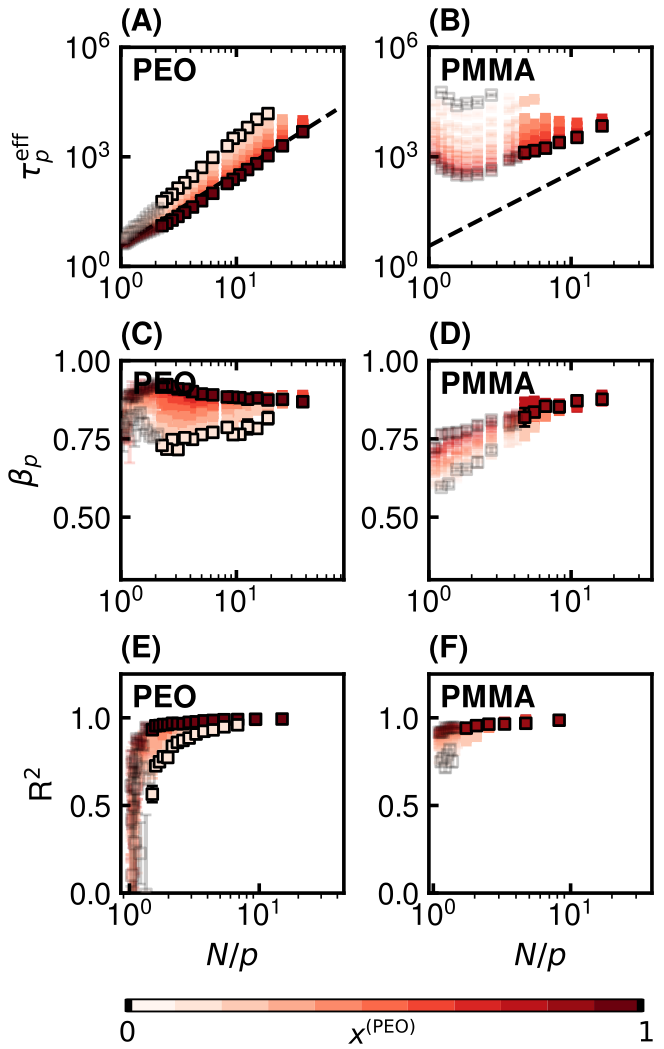}
    \caption{Rouse mode analysis at 500 K for chains in blends of varying composition where $\beta_p$ is allowed to vary with both $p$ and $x^\text{(PEO)}$ but is discouraged from deviation from 1 using L2 regularization. The effective Rouse relaxation time $\tau^{\text{eff}}_p$ as a function of sub-chain length $N/p$ for (A) PEO and (B) PMMA.
    The mode-dependent stretching exponent $\beta_p$ for (C) PEO and (D) PMMA. 
    R$^2$ coefficient of determination for the fit of $\tau_p$ and $\beta_p$ to the Rouse mode autocorrelation functions for (E) PEO and (F) PMMA. 
    For $p>8$, symbols for data are only shown for every third value of $p$ for visual clarity. Transparent data points correspond to Rouse modes that correspond to length scales smaller than the Rouse bead size. 
    Dashed black lines in panels (A) and (B) are a guide to the eye to indicate the expected ideal scaling of $\tau_p \sim p^{-2}$. The position of the line is the same across panels and is set to align with the behavior of neat PEO. 
    Results for chains in blends with the most extreme compositions are outlined in black for visual clarity of trends. Error bars reflect standard errors from three independent systems and are generally smaller than the symbol size.}
    \label{fig:rouse_varbetal2_si}
\end{figure}

Finally, we fit the ACFs using L2 regularization weighted to discourage $\beta$ from deviating from 1. Since the number of data points varies dramatically across $p$, a $\lambda$ value of $\alpha/2$ is used, where $\alpha$ is the number of data points available for fitting. Results are shown in Figure \ref{fig:rouse_varbetal2_si} and show R$^2$ values that are worse than the fits for a fully variable $\beta$ but are better than the fits for $\beta=1$. They are similar to the values we get from the best-fit fixed $\beta$ fits. 

In each of these three cases, we observe that while $\beta$ varies widely with fit method, the approximate shapes of the mostly-decayed relaxation times $\tau_p^\text{eff}$ (opaque points) are the same. As a result, any analysis of $\beta$ would be confounded by the method of the fitting. In Section \ref{sec:rouseanalysis}, we thus avoid avoid analysis of $\beta$ and restrict the Rouse analysis to the behavior of $\tau_p^\text{eff}$, which is not dependent on fitting method. 


\begin{figure}[h]
    \centering
    \includegraphics{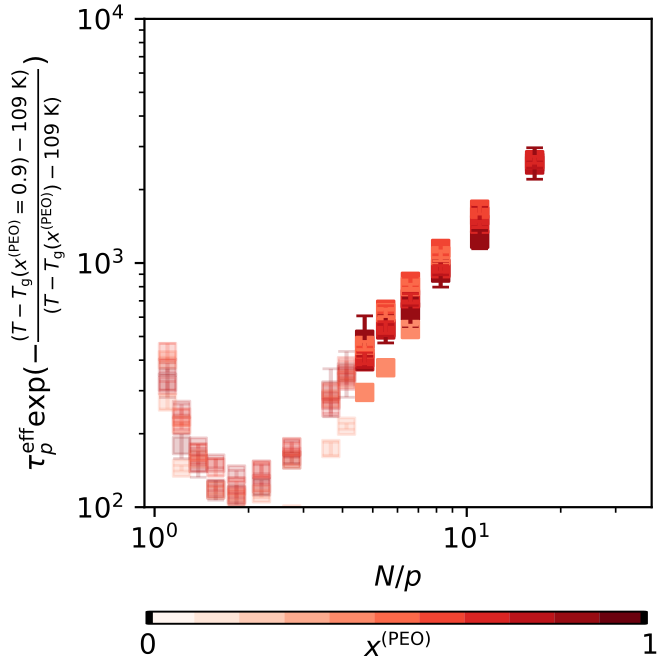}
    \caption{Shifted relaxation times of PMMA Rouse modes. Results are shown only for $x^\text{(PEO)}\geq0.5$, as the shifted relaxation times for $x^\text{(PEO)}<0.5$ deviate from the master curve.
    For $p>8$, symbols for data are only shown for every third value of $p$ for visual clarity. Transparent data points correspond to Rouse modes that correspond to length scales smaller than the Rouse bead size. Error bars reflect standard errors from three independent systems and are generally smaller than the symbol size.}
    \label{fig:rouse_pmma_collapse}
\end{figure}

Finally, the collapse of the relaxation times of PMMA Rouse modes onto a master curve is shown in Figure \ref{fig:rouse_pmma_collapse} to highlight that collective PMMA dynamics upon blending are shifted by a constant, composition-dependent rescaling factor governed by the distance from $T_\text{g}$. The relaxation times in Figure \ref{fig:rouse_pmma_collapse} are shifted by a factor of $\exp{\left(-\frac{T-T_\text{g}(x^\text{(PEO)}=0.9)-109~\text{K}}{T-T_\text{g}(x^\text{(PEO)})-109~\text{K}}\right)}$ to collapse onto $\tau_p^\text{eff}(x^\text{(PEO)}=0.9)$ where $T$=500 K. The constant factor of 109~K is used in the shift factor to minimize the mean-squared error of the collapsed relaxation times with $\tau_p^\text{eff}(x^\text{(PEO)}=0.9)$. For $x^\text{(PEO)}=0$ to $0.4$, the blend $T_\text{g}$ values are $\geq393$~K causing the shifted relaxation times to deviate because $T-T_\text{g}(x^\text{(PEO)})-109~\text{K} < 0~\text{K}$. Deviations in collective dynamics are expected close to the blend $T_\text{g}$, so the shifted data points for $x^\text{(PEO)}$ between and including 0 and 0.4 are not shown. 

\clearpage

%\bibliography{ref}